\documentclass[11pt]{article}
\usepackage{amsmath}
\usepackage{amsfonts}
\usepackage{amssymb}
\usepackage{bbm}

\usepackage{color}
\usepackage[usenames,dvipsnames]{xcolor}

\usepackage{setspace}
\usepackage[margin=1in]{geometry}
\usepackage{authblk}
\usepackage{graphicx}

\usepackage[sort, round, longnamesfirst]{natbib} 

\newtheorem{theorem}{Theorem}

\newtheorem{corollary}[theorem]{Corollary}

\newtheorem{lemma}[theorem]{Lemma}

\usepackage{bm}

\title{
Dark Matter in (Volatility and) Equity Option Risk Premiums \\  \textbf{(\underline{Operations Research} December 2022)} }

\author[1]{Gurdip Bakshi}
\affil[1]{Fox School of Business, Temple University,~~~gurdip.bakshi@temple.edu}
\author[2]{John Crosby}
\affil[2]{Strome College of Business, Old Dominion University,~~johnc2205@yahoo.com}
\author[3]{Xiaohui Gao}
\affil[3]{Fox School of Business, Temple University,~~~xiaohui.gao.bakshi@temple.edu}
\date{}

\begin{document}
\maketitle
\thispagestyle{empty}

\vspace{-6mm}
\begin{abstract}

\noindent Emphasizing the statistics of jumps crossing the strike and local time, we develop a
decomposition of
equity option risk premiums. Operationalizing this theoretical treatment, we equip the pricing kernel process with unspanned risks,
embed (unspanned) jump risks, and allow equity return volatility to contain unspanned risks. Unspanned risks are
consistent with negative risk premiums for jumps crossing the strike and local
time and imply negative risk premiums for out-of-the-money call options
and straddles. The empirical evidence from weekly and farther-dated index options is supportive of
our theory of economically relevant unspanned risks and reveals ``dark matter" in option risk premiums.



\end{abstract}

\begin{flushleft}
\vspace{2 mm}\\

\vspace{3mm}

\end{flushleft}

\vspace{13mm}
\thanks{
\noindent
We are extremely grateful for the feedback of Steven Kou (the editor), the
associate editor,
and the two referees.
First posted on SSRN: November 23, 2020. Participants at the
Canadian Derivatives Institute Conference (September 2020),
Midwest Finance Association (March 2021),
EFA (April 2021),
ITAM conference (May 2021),
Sofie conference,
the Econometrics Society meeting (Melbourne, Australia),
the University of Maryland,
Temple,
Case Western University, and
University of Manitoba (Asper School) provided many insightful
comments that we incorporated in the paper.
The paper has improved
from the feedback of Winston Dou (discussant),
Jeroen Dalderop (discussant),
Olivier Scaillet (discussant),
and Dmitry Makarov (discussant).
We are indebted to Francis Longstaff for urging us to explore the connection of local time risk premiums to gamma risk premiums. Peter Carr suggested that we make precise how the variance/dispersion risk premium is related to the local time risk premium.
Liuren Wu alerted us to a result that implied variance equates to the expected time integral of instantaneous variance weighted by dollar gamma.
Zhenzhen Fan,
Bjorn Eraker,
Steve Heston,
Pete Kyle,
Bingxin Li,
Mark Loewenstein,
Dilip Madan,
Peter Ritchken,
Oleg Rytchkov,
Ivan Shaliastovich,
Tobias Sichert,
Jinming Xue,
Justin Vitanza,
Li Wang,
Liuren Wu,
and Wei Zhou provided comments that improved our paper.
The authors are grateful to colleagues at Temple University and the University of
Maryland for
discussions.
An earlier version was circulated under the
title ``The Dark Matter in Equity Index Volatility Dynamics: Assessing
the Economic Rationales for Unspanned Risks."} 

%

\newpage
\doublespacing
\setcounter{page}{1}


\section{Introduction}
\label{sec:introduction}


Is there dark matter embedded in volatility and equity options?
We present a semimartingale theoretical approach that allows
us to introduce the constructs of risk premiums on \emph{jumps crossing}
the strike (from
above and below
(details later)) and on \emph{local time}. A semimartingale is the most general
type of process suitable for modeling equity prices.

The treatment of jumps crossing the strike and local time is integral to our
theory, because their absence would be counterfactual from an empirical standpoint. We label such
abstract uncertainties --- driven by unspanned risk components --- ``dark matter," as they can
be hard to identify, but their presence is implied in options data,
and the workings of dark matter can be economically influential.

Through our theoretical characterizations, we reveal the manner in which call option risk premiums can be decomposed into
dark matter risk premiums and \emph{upside} equity risk premiums.
Our empirical exercises are based on weekly equity index options (the ``weeklys"), in addition to the farther-dated (index and futures) options up to 88 days maturity.


\textbf{Elements of our approach.}  We propose a theory with three tenets. First, equity volatility is impacted
by both spanned and unspanned risks. Unspanned risks refer to uncertainties not spanned by equity futures but possibly
spanned by
options.

Second, the jump structure is unspecified, and no stance is taken about the exact nature of discontinuities
(e.g., \citet[Figure 7]{AitsahaliaJacod:2012}). Akin to \citet*{Merton:76} and \citet*{Kou:2002}, the jumps
constitute unspanned risks that are unhedgeable.

Third, we highlight pricing kernel evolution
that incorporates both unspanned and spanned risks. Essential to our decompositions is Tanaka's formula
for semimartingales, which gives rise to the analytical forms of (i) jumps crossing the strike and (ii) local time.

Rooted in our theory is the notion that unspanned risks differentially impact the physical and risk-neutral expectations of (i) jumps
crossing the strike and (ii) local time. To reproduce data traits, the properties of unspanned risks
in the pricing kernel, price jumps, and volatility dynamics must be such that the
risk premiums for jumps crossing the strike and local time are
negative. We formalize
how the concept of local time risk premium is distinct from volatility risk premium.

\textbf{Implications of a theory with unspanned risks and dark matter.} The implications of dark matter permeate
the spectrum of claims on equity and volatility, on both the downside
and the upside. For instance, risk premiums of out-of-the-money (OTM) calls can only be negative, as
supported by our empirical work from weeklys,
if the dark matter risk premiums --- the sum of risk premiums for jumps crossing the strike
and local time --- are negative.
Negative dark matter risk premiums stem from
unspanned risks impacting the pricing kernel, volatility dynamics, and price jumps.\footnote{Our investigation favors return volatility dynamics that cohabit
with unspanned risks. To our knowledge, the
scope of this feature has not been appreciated in the theoretical and empirical
equity pricing
literature.}


\textbf{Relation to the dark matter literature.} The work of \citet*{Chen_Dou_Kogan:JF2020} formalizes
a theory for measuring dark matter in asset pricing models.
Their
approach
is founded in the observation
that some
models rely on a form of dark matter,
by which they mean
economic
components or parameters that are
difficult to measure directly.
Complementing,
we depict
dark matter as
variables whose
dominant source is unspanned risks
in volatility and (price) jumps crossing the strike, and we use it to
summarize the properties of option returns.
We additionally show that dark matter risk premiums
take
center stage in the construction of the volatility
risk premium.

Paving the way for a better appreciation of
dark matter uncertainties,
\citet*{Cheng_Dou_Liao:ECMTA2021}
develop
model evaluation procedures for testing asset pricing models.
Their
proposed
econometric
methodology,
while not implemented in this paper,
can be adapted to
probe 
the dark matter restrictions
of option pricing models with unspanned volatility and jump risks.

The subject of our paper invites
connections
with \citet*{Chen_Dou_Kogan:JF2020} and
\citet*{Cheng_Dou_Liao:ECMTA2021}. Like them, we utilize the dark matter link, consistent with the notion
from cosmology:
The dynamics of the local time, the jumps crossing the strike, and the properties of the
pricing kernel may be hard
to identify directly using equity index returns. Instead,
their relevance can
be inferred only from option returns through the standpoint of the
model-implied restrictions.


Our contributions complement, yet differ from,
\citet*{Chen_Dou_Kogan:JF2020} and  \citet*{Cheng_Dou_Liao:ECMTA2021}.
First, they consider dark matter as the degree of fragility for
potentially
misspecified models
formulated under the data-generating
measure $\mathbb{P}$,
whereas our usage pertains to local time and
jumps crossing the strike under
both $\mathbb{P}$ and an
equivalent martingale measure $\mathbb{Q}$.
Second, we
develop the notion of risk premiums for dark matter and
economically isolate
their sign by taking cues from
option excess returns differentiated by
strikes and maturities. Third, we employ option data
to analyze
the presence of
dark matter --- specifically, to unravel
the workings of unspanned risks in the pricing kernel.

\textbf{Empirical takeaways informed by option excess returns.}
Although we do not observe dark matter,
we can infer the effect
of negative dark matter risk premiums from call risk
premiums
getting more negative deeper
OTM.
The empirical setting of weeklys
aids
in decoupling the effect of jumps crossing the strike from that of
local time.
Our bootstrap exercises show
that risk premiums
for jumps crossing the strike are equally pronounced on the upside as
they are on the downside. With
weeklys,
the dark matter and its risk premium are shaped by jumps crossing the strike. This is gauged by the size
of the risk premiums for puts, and calls, deeper OTM.

Our evidence from negative straddle risk premiums
undermines
the ``no unspanned risks" hypothesis. We infer negative risk
premiums for local time from farther-dated options. Our findings are consistent with a dislike for jumps
crossing the strike (as inferred from weeklys)
and a dislike for unspanned volatility risks (as
inferred from farther-dated options).
A rationale is
that jump movements across
actively traded strike thresholds are pertinent to
traders
and to the exposures of option writers.
Our conclusions stem from the behavior of option returns and
do not hinge on parametric assumptions about the evolution of the pricing kernel,
returns, and volatility.

\textbf{Theoretical and empirical context for why our approach
is relevant.}
We present an explanation
that conforms with
data features from
the equity options market.
If there were no unspanned jump risks in the pricing kernel,
then no risk premium would be elicited for jumps crossing the strike,
refuting
empirical evidence.
Imparting
direct
theoretical and empirical content, our predictions are devised using Tanaka's formula for
semimartingales.
This framework
gives economic footings to the concepts of jumps
crossing the strike terms and local time
and yields the context for the salient
data features of option returns.

Efforts to understand options data are ongoing.
\citet*{AndersenBollerslevDieboldLabys:ecma2003} present
a theory in which the
price process can be decomposed
into a continuous-sample path part and a jump part.
Essential to \citet*{CarrWu:2003bJF}
is the question of what type of risk-neutral processes underlie options, and
they discern
the relevance
of both
continuous and jump components.
The treatment of \citet*{Bollerslev_todorov:JF_2011} shows that
the compensation for rare events accounts for a large fraction of the
equity and variance
risk premiums.
\citet*{Todorov_Tauchen:2011JBES} favor a volatility process with jumps of infinite variation.
Using high-frequency data,
\citet*{AitsahaliaJacod:2012}
show that models are amiss if they fail
to simultaneously incorporate the continuous, small, and large jump
components of
returns.
\citet*{Andersen_fusari_todorov:2015_JFE} identify a factor driving the left jump tail of the risk-neutral
distribution. They show
that option markets embody
critical
information about the
risk premium and its dynamics.

Our approach is about distilling the effects of
unspanned
risks relevant to
trading
options. Our interest is not modeling the
volatility or
price
jump distributions but rather, it is on
uncovering the
properties that unspanned risks --- in the pricing kernel, price jumps, and volatility --- must possess
to be compatible with option returns.
While dissecting the channel of unspanned risks, we propose model-free
characterizations.
All in all,
we offer
differentiation
by framing theoretical and empirical
questions using the constructs
of local time and jumps crossing the strike and synthesizing
economic mechanisms by
combining short- and farther-dated
option prices. \vspace{-4mm}

\section{Dark matter, unspanned risks, and
option risk premiums} 
\label{sec-GeneralDiffusionDynamics}


Consider a theoretical framework
in which an equity index is tradeable and written upon
it is a futures contract.
Essential for interpretations in the market for equities,
we consider the setting of
a general \emph{semimartingale}
(that encompasses diverse forms of discontinuities (jumps)).

There are certain risks that are spanned by equity index futures and risks that are, by definition,
unspanned. The sources of risks are allied to movements in volatility as
well as jump discontinuities. 

In what follows, let $(\Omega,\mathcal{F},(\mathcal{F}_t)_{0 \leq t \leq \mathfrak{T}},\mathbb{P})$ be a filtered probability space, with $\mathfrak{T}$ being
a fixed
finite time.
The filtration $(\mathcal{F}_t)_{0 \leq t \leq \mathfrak{T}}$ satisfies the usual
conditions.
Stochastic processes are assumed to be right continuous with left limits.

Let $\mathbb{P}$ denote the physical probability measure.
Since markets are not complete,
there is neither a unique equivalent martingale measure
nor a unique pricing
kernel.
We consider an equivalent martingale measure $\mathbb{Q}$ and a pricing kernel $M_t$
consistent with the absence of arbitrage.

Additionally, we assume that $M_t$ is a \emph{semimartingale}.
Fixing notation,
$\mathbb{E}^{\mathbb{P}}_{t}( \bullet ) \equiv  \mathbb{E}^{\mathbb{P}}( \bullet | \mathcal{F}_t )$ (respectively, $\mathbb{E}^{\mathbb{Q}}_{t}( \bullet ) \equiv  \mathbb{E}^{\mathbb{Q}}( \bullet | \mathcal{F}_t )$) is the expectation under $\mathbb{P}$ (respectively, $\mathbb{Q}$), \emph{conditional} on $\mathcal{F}_t$. Furthermore,  $r$ is the spot interest-rate, assumed constant.

\noindent \textbf{Equity
premium.}
The (cum dividend) equity index price, at time $t$,
is denoted by $S_t$,
and is a semimartingale.
We maintain that
the time
$t$
conditional equity
premium is positive over any holding period ${T}_O-t$; that is,
$\mathbb{E}_{t}^{\mathbb{P}}( \frac{S_{{T}_O}}{S_t} ) - e^{r ({T}_O-t)}>0$.

\noindent \textbf{Gross equity futures return.}
We denote the time $t$ equity futures price by $F_{t}^{T_F}$, where $T_F$ denotes the expiration date of the futures contract.
It holds that
\begin{align}
F_t^{T_F} &~=
\mathbb{E}_{t}^{\mathbb{P}} ( \frac{M_{\ell}}{M_{t} e^{ -r (\ell-t)}}  F_{\ell}^{T_F} ) \, = \mathbb{E}_{t}^{\mathbb{Q}}( F_{\ell}^{T_F} ),
&~\mbox{ \, \, for all $t$ and $\ell$ satisfying $t \leq \ell \leq T_F$},\mbox{ \, \, }~~
\label{eq:RelationsInGeneralDynamics}  \\
 &~=  S_t \, e^{r (T_F - t)},
 &\mbox{ \, \,(i.e., cost of carry with $S_{T_F} = F_{T_F}^{T_F}$), \, \, }~
 \label{fuut}
\end{align}
where
$\frac{M_{\ell}}{M_{t} e^{ -r (\ell-t)}}$ represents the Radon-Nikodym derivative. Hence, the
process $(G_\ell)$ defined by 
\begin{eqnarray}
G_\ell \equiv \frac{F_{\ell}^{T_F}}{F_{t}^{T_F}},~\mbox{ }~\mathrm{represents~the~\emph{gross~futures~return},~from}~t~\mathrm{to}~\ell,~\mathrm{for}~\ell~\mathrm{satisfying}~t \leq \ell \leq T_F.
\label{eq:DefinitionOfYProcess}
\end{eqnarray}

\noindent \textbf{Futures risk premium on the downside and upside.}
The 
futures risk premium, with $G_t=1$, is given by
$\mathbb{E}_{t}^{\mathbb{P}}(\frac{F_{T_O}^{T_F}}{F_{t}^{T_F}}) -
\mathbb{E}_{t}^{\mathbb{Q}}(\frac{F_{T_O}^{T_F}}{F_{t}^{T_F}}) =
\mathbb{E}_{t}^{\mathbb{P}}(\frac{F_{T_O}^{T_F}}{F_{t}^{T_F}}) -
1 =  \mathbb{E}_{t}^{\mathbb{P}}( G_{T_O}) - G_{t} = \mathbb{E}_{t}^{\mathbb{P}}( \int_{t}^{T_O} dG_{\ell} )$. Define $k$ as
\begin{equation}
 k \, \equiv \, \frac{K}{F_{t}^{T_F}} \, \in \, (0,\infty),
~
\text{\small
which~is~the~option~moneyness~for~strike~price}~K. \mbox{ \, } ~
\end{equation}
In light of their connection to option risk premiums,
we define the following futures risk premiums:
\begin{align}
&\mathbb{E}_{t}^{\mathbb{P}}( \int_{t+}^{{T}_O} \mathbbm{1}_{\{G_{\ell-} < k\}} \,dG_{\ell} )&
&&
&\mbox{(downside~risk premium, $k<1$)}~~~\mathrm{and}& \label{a.x2} \\
&\mathbb{E}_{t}^{\mathbb{P}}( \int_{t+}^{{T}_O} \mathbbm{1}_{\{G_{\ell-} > k\}} \,dG_{\ell} )&
&&
&\mbox{(upside~risk premium, $k>1$).}~& \label{a.x1}
\end{align}
Additionally, $\mathbbm{1}_{\{G_{\ell-} > k\}}=1$ if $G_{\ell-} > k$ and is zero otherwise.
In equations (\ref{a.x2})--(\ref{a.x1}), $G_{\ell-}$ can be thought of as the value ``just an instant before time $\ell$."

Both of the terms in
equations (\ref{a.x2})--(\ref{a.x1}) reflect risk premiums since
$\mathbb{E}_{t}^{\mathbb{Q}}( \int_{t+}^{{T}_O} \mathbbm{1}_{\{G_{\ell-} < k\}} \, dG_{\ell} )=0$ and $\mathbb{E}_{t}^{\mathbb{Q}}( \int_{t+}^{{T}_O} \mathbbm{1}_{\{G_{\ell-} > k\}} \,dG_{\ell} )=0$.
This is because
$(F_{\ell}^{T_F})$ and $(G_\ell)$ are martingales under
$\mathbb{Q}$.


\noindent \textbf{Options on the equity futures price with moneyness $k$.}
Consider an option written on the equity futures price over
$t$ to ${T}_O$ with strike price $K$ (or moneyness $k$).
Therefore,
\begin{equation}
\mathrm{for ~ OTM ~ (at\mbox{-}the\mbox{-}money) ~calls} ~ k \, > \, 1 ~ ~ (k \, = \, 1) ~ ~ \mathrm{and ~ for ~ OTM ~ puts}, ~ k \, < \, 1. ~ ~ ~ ~ ~
~ \mbox{ \, \quad \, }
\end{equation}
It is understood that $t \leq {T}_O \leq T_F$, where $T_O$ is the maturity of the option.
The expected return of holding a call option on equity futures over $t$ to ${T}_O$ with moneyness $k$,
denoted $\mu^{{T}_O}_{t,{\tiny \mathrm{call}}}[k]$, satisfies
\begin{eqnarray}
1 + \mu^{{T}_O}_{t,{\tiny \mathrm{call}}}[k]  &\equiv&  \frac{\mathbb{E}_{t}^{\mathbb{P}}( \max (F_{{T}_O}^{T_F} - K, 0) )}{ e^{-r ({T}_O - t)}\,  \mathbb{E}_{t}^{\mathbb{Q}}(
\max(F_{{T}_O}^{T_F} - K, 0) )}
~=~  \, \frac{\mathbb{E}_{t}^{\mathbb{P}}( \max(G_{{T}_O} - k,0) )}
{ e^{-r ({T}_O - t)}\, \mathbb{E}_{t}^{\mathbb{Q}}(
\max(G_{{T}_O} - k,0)  )}.~~~\mbox{ \, \, \, \, }
\label{eq:ExpectedHoldingReturn1GneralDynamics}
\end{eqnarray}


\noindent \textbf{Tanaka's formula for semimartingales.}
Our Theorem~\ref{claimm:claim1call_jump}
will rely
on Tanaka's formula for (general) semimartingales.
Specifically (and relevant for call option payoffs),
Tanaka's formula for semimartingales --- as in \citet*[Theorem 68, page 216]{Protter:2013} ---
implies (mapping
his notation of $x^{+}=\max(x,0)$ and $x^{-}=-\min(x,0)=\max(-x,0)$)
\begin{eqnarray}
\max( G_{T_O} - k, 0 ) ~-~ \overbrace{ {\underbrace{\max( G_{t} - k, 0 )}_{\tiny =0,~\mbox{for~OTM~calls}}}}^{\tiny \mbox{intrinsic~value}} & = &
\int_{t+}^{T_O} \mathbbm{1}_{\{G_{\ell -} > k\}} dG_{\ell} ~+~
\overbrace{\mathbb{L}^{T_O}_t[k]}^{\tiny \mbox{local~time}}~~\mbox{ \, \, } \nonumber \\
& & +~~ \underbrace{\sum_{t < \ell \leq T_O} \mathbbm{1}_{\{G_{\ell -} \leq k\}} \, \max( G_{\ell} - k, 0 )}_{~\equiv~a_t^{T_O}[k]~~\tiny \mbox{(jumps~crossing~the~strike~from~below)}} \nonumber \\
& & +~~~ \underbrace{\sum_{t < \ell \leq T_O} \mathbbm{1}_{\{G_{\ell -} > k\}} \, \max( k - G_{\ell}, 0 ).}_{~\equiv~b_t^{T_O}[k]~~\tiny \mbox{(jumps~crossing~the~strike~from~above)}}
~~\mbox{ \quad}~~
\label{eq:TanakaJumps}
\end{eqnarray}

The summand terms on the second and third lines characterize \emph{large deviations} or significant events
and do not appear in the absence of jumps. We interpret
them as follows (presuming $k>1$):
\begin{description}
\item $\mathbbm{1}_{\{G_{\ell -} \leq k\}} \max( G_{\ell} - k, 0 )$
is only nonzero when $G_{\ell -} \leq k$ and $G_{\ell} > k$ --- loosely speaking, when
a jump at time $\ell$ results in $G$ jumping from below $k$ to above $k$ (i.e., the equity futures price jumps
upward
and crosses the level of the strike).

\item $\mathbbm{1}_{\{G_{\ell \, -} > k\}} \max( k - G_{\ell}, 0 )$
is only nonzero when $G_{\ell -} > k$ and $G_{\ell} < k$ --- loosely speaking, when
a jump at time $\ell$ results in $G$ jumping from above $k$ to below $k$.
\end{description}
In a continuous semimartingale setting, $a_t^{T_O}[k]$ and $b_t^{T_O}[k]$ vanish.
Finally, the term $\int_{t+}^{T_O} \mathbbm{1}_{\{G_{\ell -} > k\}} \,dG_{\ell}$ is a stochastic integral representing the gains/losses to a dynamic trading strategy that takes a
long position of magnitude $\frac{1}{F_{t}^{T_F}}$ at time $\ell$, in the futures, if, and only if, $G_{\ell  -} > k$ (i.e., $F_{\ell -}^{T_F} > K$).

Likewise, and relevant for put option payoffs, Tanaka's formula for semimartingales implies
\begin{eqnarray}
\max( k - G_{T_O}, 0 ) -
\overbrace{ {\underbrace{\max( k- G_{t}, 0 )}_{\tiny =0,~\mbox{for~OTM~puts}}}}^{\tiny \mbox{intrinsic~value}}
  & = & ~ - ~
\int_{t+}^{T_O} \mathbbm{1}_{\{G_{\ell -} < k\}} dG_{\ell} ~+~
\overbrace{\mathbb{L}^{T_O}_t[k]}^{\tiny \mbox{local~time}}~~\mbox{ \, \, } \nonumber \\
& &+~~ \underbrace{\sum_{t < \ell \leq T_O} \mathbbm{1}_{\{G_{\ell  -} \geq k\}} \, \max( k - G_{\ell}, 0 )}_{~\equiv~c_t^{T_O}[k]~\tiny~\mbox{(jumps~crossing~the~strike~from~above)}} \nonumber \\
& & +~~~ \underbrace{\sum_{t < \ell \leq T_O} \mathbbm{1}_{\{G_{\ell -} < k\}} \, \max(  G_{\ell}- k, 0 ).}_{~\equiv~d_t^{T_O}[k]~\tiny~\mbox{(jumps~crossing~the~strike~from~below)}}
~~\mbox{ \quad}~~ \label{eq:TanakaPutCaseJumps}
\end{eqnarray}
We interpret the terms in our context as follows (presuming~$k <1$):
\begin{description}
\item $\mathbbm{1}_{\{G_{\ell -} \geq k\}} \max( k-G_{\ell}, 0 )$
is only nonzero when $G_{\ell -} \geq k$ and $G_{\ell} < k$ --- loosely speaking, when
a jump at time $\ell$ results in $G$ jumping from above $k$ to below $k$.

\item $\mathbbm{1}_{\{G_{\ell -} < k\}} \max( G_{\ell} - k, 0 )$
is only nonzero when $G_{\ell -} < k$ and $G_{\ell} > k$ --- loosely speaking, when
a jump at time $\ell$ results in $G$ jumping from below $k$ to above $k$.

\end{description}
In a continuous semimartingale setting, $c_t^{T_O}[k]$ and $d_t^{T_O}[k]$
are identically zero.
The
term $-\int_{t+}^{T_O} \mathbbm{1}_{\{G_{\ell -} < k\}} \,dG_{\ell}$ reflects
the
gains/losses to a dynamic trading strategy
that takes a short futures  position of magnitude
$\frac{1}{F_{t}^{T_F}}$ at time $\ell$, if and only if, $G_{\ell  -} < k$ (i.e., $F_{\ell -}^{T_F} < K$).

\noindent \textbf{Local time and
risk premiums on local time.}
In
equations
(\ref{eq:TanakaJumps}) and (\ref{eq:TanakaPutCaseJumps}),
the term
\begin{eqnarray}
\mathbb{L}^{T_O}_t[k]=\frac{1}{2} \int_{t}^{T_O} \delta_{\{G_\ell ~-~ k\}}\,d [ G^\mathrm{c}, G^\mathrm{c} ]_{\ell} ~~ \mathrm{is~the~\emph{local~time}.}
~~~\text{\small (\citet*[Theorem~71, page~221]{Protter:2013})}  ~ ~
\label{ltt.1}
\end{eqnarray}
In
(\ref{ltt.1}), $\delta_{\{\bullet\}}$ is the Dirac delta function,
and $[ G^\mathrm{c}, G^\mathrm{c} ]_{\ell}$ denotes the path-by-path  continuous part of
the
quadratic variation, defined
(see \citet*[page~70]{Protter:2013})
as
\begin{equation}
[ G^\mathrm{c}, G^\mathrm{c} ]_{\ell} ~~ \equiv ~ \underbrace{~ [ G, G ]_{\ell} ~ }_{\tiny \mbox{quadratic~variation}} ~ - ~ \underbrace{\sum_{t \leq h \leq \ell} (G_{h} - G_{h -})^2.}_{\tiny \mbox{sum~of~squares~of~the~jumps}} ~~ \mbox{ \, \quad } \label{eq:RelationQVToPathByPathContinuousQV}
\end{equation}

Intuitively,
$\mathbb{L}^{T_O}_t[k]$ captures the slice of uncertainty associated with the
time that $G_{\ell}$ spends at the level $k$. In economic terms, one may contemplate $\mathbb{L}^{T_O}_t[k]$ as a form of
volatility
uncertainty. Continuous semimartingales
imply $\sum_{t \leq h \leq \ell} (G_{h} - G_{h -})^2  = 0$, for all $h$,
so, in this case, one
may view \emph{local time} as a measure of integrated variance over $T_O-t$ computed
when $(G_{\ell})$ is
\emph{exactly} $k$.

The local time reflects sample path properties that do not vary according to the
measures $\mathbb{P}$ or $\mathbb{Q}$.
At the same time, the expectations of $\mathbb{L}^{T_O}_t[k]$ under $\mathbb{P}$ and $\mathbb{Q}$ may differ. We define
\begin{equation}
\mathbb{E}^{\mathbb{P}}_{t}( \mathbb{L}^{T_O}_t[k] ) ~-~ \mathbb{E}^{\mathbb{Q}}_{t}( \mathbb{L}^{T_O}_t[k] )
~\mbox{ \, } ~\mathrm{as~the~\emph{local~time~risk~premium}~for~moneyness}~k. ~~~ \mbox{ \,  \, }~
\end{equation}
We interpret the local time risk premium, between $t$ and $T_O$, as conveying the risk premium for the strip of
volatility
uncertainty associated with $k$.

Local time risk premiums corresponding to the downside ($k <1$) can be distinct from those to the upside
($k >1$). We will show the manner in which the local time risk premium at $k=1$ associates with
the risk premium on straddles (under some mild assumptions). This analytical association is concrete for continuous semimartingales.

\noindent \textbf{Dark matter, unspanned risks, and dark matter risk premiums.} Before we present our theoretical results and explore their empirical implications, we emphasize that the dynamics of the pricing kernel
and futures return volatility may contain
both spanned and unspanned diffusive risks as well as
jump risks. In other words,
they may contain risks that are spanned by equity futures
as well as risks that are
not spanned by equity futures but may be spanned by options.

The complexity of local time and of the ``jumps crossing the strike" terms (i.e., $a_t^{T_O}[k]$,
$b_t^{T_O}[k]$, $c_t^{T_O}[k]$, and $d_t^{T_O}[k]$)
gives rise to the following definition of dark matter:
\begin{gather}
\underbrace{\mathrm{Dark~Matter}}_{\tiny(\mbox{over}~t~\mbox{to}~T_{O})}~=~
\begin{cases}
  \mathrm{D}^{d, T_O}_t[k]\equiv\underbrace{\mathbb{L}^{T_O}_t[k]}_{\tiny \mbox{local~time}} ~+~
\underbrace{c_t^{T_O}[k] + d_t^{T_O}[k],}_{\tiny \mbox{jumps~crossing~the~strike~terms~(eq.~(\ref{eq:TanakaPutCaseJumps}))}}
 & \text{for } k<1, \\
  \mathrm{D}^{\tiny \mbox{atm}, T_O}_t[1]\equiv~
\mathbb{L}^{T_O}_t[1] ~~+~~ a_t^{T_O}[1] + b_t^{T_O}[1]
 & \text{for } k=1, \\
  \mathrm{D}^{u, T_O}_t[k] ~\equiv~ \underbrace{\mathbb{L}^{T_O}_t[k]}_{\tiny \mbox{local~time}} ~+~
\underbrace{a_t^{T_O}[k]+b_t^{T_O}[k],}_{\tiny \mbox{jumps~crossing~the~strike~terms~(eq.~(\ref{eq:TanakaJumps}))}}
 & \text{for } k>1.
\end{cases}
\label{darkk.1}
\end{gather}
Then, we can define as follows:
\begin{gather}
\underbrace{\mathrm{Dark~Matter~Risk~Premium}}_{\tiny (\mbox{over}~t~\mbox{to}~T_{O})}
~\equiv~
\begin{cases}
\mathbb{E}^{\mathbb{P}}_{t}( \mathrm{D}^{d, T_O}_t[k] ) ~-~ \mathbb{E}^{\mathbb{Q}}_{t}( \mathrm{D}^{d, T_O}_t[k] ), & \text{for } k<1,
~ \mbox{ \, \, } ~ \\
\mathbb{E}^{\mathbb{P}}_{t}( \mathrm{D}^{\tiny \mbox{atm}, T_O}_t[1] ) ~-~ \mathbb{E}^{\mathbb{Q}}_{t}( \mathrm{D}^{\tiny \mbox{atm}, T_O}_t[1] ), & \text{for } k=1,~~\mathrm{and} ~ \mbox{ \, \, } ~ \\
\mathbb{E}^{\mathbb{P}}_{t}( \mathrm{D}^{u, T_O}_t[k] ) ~-~ \mathbb{E}^{\mathbb{Q}}_{t}( \mathrm{D}^{u, T_O}_t[k] ), & \text{for } k>1. ~ \mbox{ \, } ~ \mbox{ \, \, } ~
\end{cases}
\label{darkk.rp}
\end{gather}

We note that, due to the convexity of $a_t^{T_O}[k]$, $b_t^{T_O}[k]$, $c_t^{T_O}[k]$, and $d_t^{T_O}[k]$ in
$G_\ell$, we have $\mathbb{E}^{\mathbb{Q}}_{t}(a_t^{T_O}[k])>0$,
$\mathbb{E}^{\mathbb{Q}}_{t}(b_t^{T_O}[k])>0$,
$\mathbb{E}^{\mathbb{Q}}_{t}(c_t^{T_O}[k])>0$, and
$\mathbb{E}^{\mathbb{Q}}_{t}(d_t^{T_O}[k])>0$.

The source of risk premiums
on $a_t^{T_O}[k]$, $b_t^{T_O}[k]$, $c_t^{T_O}[k]$, and $d_t^{T_O}[k]$ is, by definition,
unspanned jump risks (one
may not be able to trade during a jump). In other words, the risk associated with jumps crossing the strike cannot be
eliminated.
Now we state: \vspace{-3mm}
\setcounter{theorem}{0}
\begin{theorem}[Negative risk premiums for dark matter]
\label{claimm:claim1call_jump}
The call risk premium at $k>1$ \emph{can} be negative only
if the dark matter risk premium at $k>1$, as defined in (\ref{darkk.rp}),
is negative. The
straddle risk premium is negative only if the dark matter risk premium at
$k=1$
is negative.
\vspace{-3mm}
\end{theorem}
\noindent {\bf Proof:} See Appendix~\ref{appsec:jumppps}. $\blacksquare$

Using Tanaka's formula for semimartingales (details in
Appendix~\ref{appsec:jumppps}), we derive
the following expression for
the \emph{call risk premium} (for $k>1$) as follows:
\begin{equation}
\underbrace{1 + \mu^{{T}_O}_{t,{\tiny \mathrm{call}}}[k] - e^{r ({T}_O - t)}}_{\tiny \mbox{expected~excess~return~of~calls}}
 =  \underbrace{\frac{e^{r ({T}_O - t)} }{ \mathbb{E}_{t}^{\mathbb{Q}}( \mathrm{D}^{u, T_O}_t[k] ) }}_{>0}
\{
\underbrace{ \mathbb{E}_{t}^{\mathbb{P}}( \mathrm{D}^{u, T_O}_t[k] )
~-~ \mathbb{E}_{t}^{\mathbb{Q}}( \mathrm{D}^{u, T_O}_t[k] )}_{\tiny \mbox{risk~premium~for~dark~matter}~(k>1)}
~+~
\underbrace{ \mathbb{E}_{t}^{\mathbb{P}}( \int_{t+}^{{T}_O} \mathbbm{1}_{\{G_{\ell-} > k\}} \,dG_{\ell} )}_{\tiny \mbox{upside~risk~premium}}
 \}.
\label{eq:excess_mucall1x}
\end{equation}
Theorem~\ref{claimm:claim1call_jump} establishes when call risk premiums can be negative. Negative
call (or straddle) risk premiums imply the relevance of unspanned risks. 


The \emph{put risk premium} (for $k<1$) can be determined (details in
Appendix~\ref{appsec:jumppps}) as follows:
\begin{equation}
\underbrace{1 + \mu^{{T}_O}_{t,{\tiny \mathrm{put}}}[k] - e^{r ({T}_O - t)}}_{\tiny \mbox{expected~excess~return~of~puts}}
 =  \underbrace{\frac{e^{r ({T}_O - t)} }{ \mathbb{E}_{t}^{\mathbb{Q}}( \mathrm{D}^{d, T_O}_t[k] ) }}_{ >0}
\{ \underbrace{ \mathbb{E}_{t}^{\mathbb{P}}( \mathrm{D}^{d, T_O}_t[k] )
~-~ \mathbb{E}_{t}^{\mathbb{Q}}( \mathrm{D}^{d, T_O}_t[k] )}_{\tiny \mbox{risk~premium~for~dark~matter}~(k<1)}
~-~ \underbrace{ \mathbb{E}_{t}^{\mathbb{P}}( \int_{t+}^{{T}_O} \mathbbm{1}_{\{G_{\ell-} < k\}} \,dG_{\ell} )}_{\tiny \mbox{downside~risk~premium}}
\}.
\label{eq:excess_muputs}
\end{equation}
If $\mathbb{E}_{t}^{\mathbb{P}}( \mathrm{D}^{d, T_O}_t[k] )
- \mathbb{E}_{t}^{\mathbb{Q}}( \mathrm{D}^{d, T_O}_t[k] ) < 0$, then
the
put risk premium is
negative. This implication is empirically supported in return data of OTM
puts.


\noindent \textbf{Dark matter risk premium ($k=1$) and straddle risk premium.}
In
Appendix~\ref{appsec:jumppps} (equation (\ref{eq:StraddleInterim1Jumps})),
we develop the link of the local time risk
premium (when $k=1$) and risk premium for jumps crossing the strike
(from below and above $k=1$)
to the straddle risk premium.
The latter risk premium effect can be traced
to the quantity $a_t^{T_O}[1] + b_t^{T_O}[1]=\sum_{t < \ell \leq T_O} \{\mathbbm{1}_{\{G_{\ell  -} < 1\}}  \max(  G_{\ell}- 1, 0 )+  \mathbbm{1}_{\{G_{\ell  -} > 1\}}  \max( 1 - G_{\ell}, 0 )\}$, which
represents jumps that cross $k=1$ in either direction. Importantly, the existence and relevance of
dark matter can be detected from straddle risk premiums.

\noindent \textbf{Linking dark matter risk premiums
to the risk premium on volatility uncertainty.}
To formalize this notion, suppose $\{\log \frac{F_{T_O}^{T_F}}{F_t^{T_F}}\}^2$ represents uncertainty
related to the volatility
of futures returns over $t$ to $T_O$.
Then the risk premium on dark matter
is a building block for constructing the risk premium on
volatility uncertainty.
It is seen that (Internet Appendix~(Section~\ref{appsec:dispersion})) \small
\begin{eqnarray}
\underbrace{\mathbb{E}_t^{\mathbb{P}} ( \big\{\log \frac{F_{T_O}^{T_F}}{F_t^{T_F}}\big\}^2 )
- \mathbb{E}_t^{\mathbb{Q}} ( \big\{\log \frac{F_{T_O}^{T_F}}{F_t^{T_F}}\big\}^2 )}_{\mathrm{risk~premium~on~squared~log~contract}}
& = & -\mathrm{e}_t^{\mathbb{P}} ~+~ \int\limits_{0}^\infty ~ \omega[k] \, \underbrace{\{ \mathbb{E}_t^{\mathbb{P}} ( \mathbb{L}^{T_O}_t[k] )-\mathbb{E}_t^{\mathbb{Q}} ( \mathbb{L}^{T_O}_t[k] )\}}_{\mathrm{risk~premium~for~local~time}}\,dk~\mbox{ \, \, \, } \nonumber \\
&+& \int\limits_{0}^1 \omega[k]\, \underbrace{ \{
\mathbb{E}_t^{\mathbb{P}}( c_t^{T_O}[k] + d_t^{T_O}[k] )
- \mathbb{E}_t^{\mathbb{Q}}( c_t^{T_O}[k] + d_t^{T_O}[k] ) \}}_{\mathrm{risk~premium~for~jumps~crossing~the~strike}~(k<1)} \, dk  \nonumber \\
&+& \int\limits_{1}^\infty \omega[k]\, \underbrace{\{
\mathbb{E}_t^{\mathbb{P}}( a_t^{T_O}[k]+ b_t^{T_O}[k] )
- \mathbb{E}_t^{\mathbb{Q}}( a_t^{T_O}[k]+ b_t^{T_O}[k]) \}}_{\mathrm{risk~premium~for~jumps~crossing~the~strike}~(k>1)} \, dk,
~~\mbox{ \, \, }\label{eq:LogFuturesSqasb11InResult}  \nonumber\\
&&~~~~\mbox{with}~ \omega[k] ~ \equiv ~ \frac{2}{k^2} ( 1 - \log k ),
~~\mbox{ \, \, }
~\mbox{ \, \, }
~~\mbox{ \, }
\end{eqnarray}  \normalsize
where $\mathrm{e}_t^{\mathbb{P}}$ has the economic interpretation of the expected total gain/loss, over $t$ to $T_O$, from
a  dynamic equity futures trading strategy (details in Internet Appendix~\ref{appsec:dispersion} (equation (\ref{eq:DefMuP}))).

\noindent \textbf{Absence of unspanned risks in the pricing kernel and a continuous semimartingale model setting
with stochastic return volatility.}
The final question is: Is it possible to obtain negative risk premiums for OTM calls if there
are unspanned diffusive risks in volatility dynamics but not in the pricing kernel?
This continuous semimartingale
environment is revealing for two reasons. First, the jumps crossing the strike terms --- $a_t^{T_O}[k]$, $b_t^{T_O}[k]$, $c_t^{T_O}[k]$, and $d_t^{T_O}[k]$ --- \emph{vanish}. Second, one can delineate the distinction between spanned and unspanned \emph{diffusive} risks.

Reconciling intuition,
we establish the takeaway that
OTM call option risk premiums will be positive if
there are no unspanned risks in
the pricing kernel.\footnote{This analysis is presented in
Internet Appendix (Section~\ref{app:LTRPSpannedDiffusiveVolRisk}) to save on space.}
The model studied in Section~\ref{sec:furtherrr} ascribes
clear-cut roles for spanned and unspanned risks,
and we show that unspanned risks can generate negative local time risk premiums and negative risk premiums of
calls and straddles.
\vspace{-4mm}

\section{Supportive empirical evidence on dark matter}
\label{sec:EmpiricalDataSupportUnspannedIndexVol}



Suppose there is potential for jumps crossing the strike
and
the pricing kernel contains risks that are not spanned by equity futures but do
correlate with risks that intersect local time,
then this attribute may give rise to negative call
option
risk premiums. Such a feature speaks to the relevance of dark matter.

The risk premium on dark matter is implied to
be negative if the straddle risk premium
is negative or if the call option risk premiums are negative at some $k>1$.
Our goal is to detect dark matter and probe its workings.



\noindent \textbf{A. Implication-rich weeklys (short-dated
options).}
Notably, weeklys
are considered gamma plays, whereas long-dated options are vega plays.
With no more than eight days to maturity,
the delta
of such options can move quickly along directional movement.


In conjunction with shrinking
time value for weeklys,
the insight to exploit
is that the source of dark matter risk premiums is
predominantly risk premiums for jumps crossing the strike, pertinently so for deep OTM options.
Complementing this channel, straddle risk premiums are linked to risk  premiums on price
jumps without regard to
their direction.



\noindent \textbf{B. Framing the theoretical predictions.} Our theory allows us to
formulate the following predictions about equity option risk premiums: \vspace{-2mm}
\begin{enumerate}

\item[\textbf{H1.}] \textbf{No unspanned risks hypothesis.}
If there are no unspanned risks in the pricing kernel,
the risk premium of OTM calls is  \emph{positive} and the risk premium of
straddles is \emph{zero}.

\item[\textbf{H2.}] \textbf{Negative risk premiums for \emph{jumps} crossing the strike hypothesis for
\emph{short-dated} options.}
Deep OTM weekly
options exhibit negative risk premiums, in line with negative risk premiums for jumps
crossing the strike.



\item[\textbf{H3.}] \textbf{Negative risk premiums on dark matter hypothesis.} If there are unspanned risks,
the risk premium on dark matter
(for moneyness $k$) can be negative. Then, the risk premiums of straddles and OTM calls can be \emph{negative}.

\end{enumerate}
We examine these predictions using option returns computed over expiration cycles.
Our focus is on option maturities that are actively traded: weeklys
(eight days), 28 days, and 88 days.

\noindent \textbf{C. Excess returns of weeklys}. Weekly options are instrumental
in identifying
and isolating
the jumps crossing the strike component of dark matter.
Motivated by questions concerning our hypotheses, we first construct the time-series of excess returns of options on the
S\&P 500 index over the \emph{weekly} expiration cycles.

Specifically, for $T_{O}-t=$ 8 days (on average), and setting $k= \frac{K}{S_t}$, 
\begin{equation}
{q_{t,{\tiny \mathrm{call}} }^{{T}_O}[k]} ~=~ \frac{ \max( S_{{T}_O} - k \,S_t,0)}{\mathrm{call}_{t}[k \,S_{t}]}~-~e^{r(T_{O}-t)},
~~~\mathrm{where}~\log(k)~\mathrm{is}~1\%, 2\%,~\mathrm{and}~3\%~\mathrm{OTM},
\end{equation}
and $\mathrm{call}_{t}[k \, S_{t}]$ is the
ask price of an OTM call with strike $K=k \, S_{t}$.
Anchoring our discussions,
the selected
$\log(k)$
are allied
to a delta of 27, 12, and  6 (in \%, likewise for puts).
The  straddle excess return is
\begin{equation}
q_{t,{\tiny \mathrm{straddle}} }^{{T}_O} ~=~\frac{\max( S_t - S_{{T}_O},0) ~+~ \max(S_{{T}_O} - S_t,0)}{
\mathrm{put}_{t}[S_{t}]
~+~ \mathrm{call}_{t}[S_{t}]  }~-~e^{r(T_{O}-t)},
\end{equation}
where $\mathrm{put}_{t}[S_{t}]$ is the ask price of an at-the-money (ATM) put with strike $K= S_{t}$.

Weekly options initiate on a Thursday and expire on the Friday of the
following week.
The first (final) expiration cycle is 1/13/2011
(12/20/2018). Hence, our analysis covers 415 weekly expiration cycles. These weekly options are
associated with sizable open interest and volume.


%

\noindent \textbf{D. Drawing inferences from empirical measures of
option risk premiums.}
Our theoretical results pertain to the expectation of option returns conditional on the
filtration
$\mathcal{F}_t$; that is, $\mathbb{E}^{\mathbb{P}}_{t}( \bullet ) =
\mathbb{E}^{\mathbb{P}}( \bullet | \mathcal{F}_t )$. To measure this
object empirically,
we construct
average excess option returns (over expiration cycles)
conditional on $\{{\cal F}_t \in \mathfrak{s}\}$, for some
variable $\mathfrak{s}$.

We are guided by the implication that historically generated excess returns conform with ex-ante
expected excess returns.
Our criteria for $\mathfrak{s}$ are that they connect to
time $t$ information
tracked by market participants.
Each $\mathfrak{s}$ is  arranged so as to be in one of the following three categories:
\begin{gather}
{\cal F}_t \in \mathfrak{s}=
\begin{cases}
\mathfrak{s}_{\tiny \mbox{bad}}& \mathrm{(when~the~equity~premium~is~presumably~high)}, \\
\mathfrak{s}_{\tiny \mbox{normal}}&\mathrm{(when~the~equity~premium~is~presumably~normal)},~~\mathrm{and} \\
\mathfrak{s}_{\tiny \mbox{good}}& \mathrm{(when~the~equity~premium~is~presumably~low)}. \\
\end{cases}
\label{filt.1}
\end{gather}
Thus, we draw inferences based on partitioned average excess option returns.
Pertinent to our exercise for \emph{weekly option returns}, we
consider the following variables
to surrogate $\mathfrak{s}$:

\begin{enumerate}

\item \textbf{Change in the Weekly Economic Index}$_{t}$. Reflects the weekly innovation in the WEI
index (source: New York Fed). A decline indicates a weakening economy.

\item \textbf{Quadratic Variation}$_{t}$. Sum of daily squared (log) returns over
the \emph{prior} expiration cycle (eight days). A high $\mathrm{QV}_{t}$ corresponds
to unfavorable economic states.

\item \textbf{Risk Reversal}$_t$. The negative skew, reflected in $\log(\frac{\mathrm{IV_t^{\tiny \mbox{put}}}[k]}{\mathrm{IV_t^{\tiny \mbox{call}}}[k]}$), mirrors
    downside protection concerns. The implied volatility (IV$_{t}$) for puts (calls) uses $\log(k)$ equal
    to $-2\%$ (2\%).

\item \textbf{Change in Volatility}$_{t}$ ($\log(\frac{\mathrm{IV}^{\tiny \mbox{atm}}_{t}}{\mathrm{IV}^{\tiny \mbox{atm}}_{t-1}})$).
A positive change in ATM implied volatility, over the prior expiration cycle, coincides
with rising market uncertainty (and wary investors). The implied volatility is the average across
ATM puts and calls of weekly options.


\item \textbf{Recent Market}$_{t}$: Log relative of the S\&P 500 index over the prior
expiration cycle.
 \vspace{-3mm}





\end{enumerate}

Our rationale for considering these variables is that they may be correlated with subsequent variation in
equity premiums
and may influence dark matter risk premiums.

\noindent \textbf{E. Support for our predictions about dark matter from \emph{weeklys}}.
We consider a regression framework, where excess returns of calls
is the dependent variable (likewise for
straddles and
puts), as follows:
\begin{equation}
q_{t,{\tiny \mathrm{call}} }^{{T}_O}[k]
=
\underbrace{\mu_{\{ {\cal F}_{t} \in \mathfrak{s}_{\tiny \mbox{bad}} \} } \mathbbm{1}_{\{ {\cal F}_{t} \in\mathfrak{s}_{\tiny \mbox{bad}} \}}
+ \mu_{\{ {\cal F}_{t} \in \mathfrak{s}_{\tiny \mbox{normal}} \} }
\mathbbm{1}_{\{ {\cal F}_{t} \in\mathfrak{s}_{\tiny \mbox{normal}} \}}
+ \mu_{\{ {\cal F}_{t} \in \mathfrak{s}_{\tiny \mbox{good}} \} }
\mathbbm{1}_{\{ {\cal F}_{t} \in\mathfrak{s}_{\tiny \mbox{good}} \}}}_{\tiny \mbox{Dichotomizing~expected~excess~returns~across~economic~states}} +
\underbrace{\epsilon_{T_{O}}.}_{\tiny \mbox{error~term}}
\label{eq:regress_three_states}
\end{equation}

Table~\ref{tab:weekly} reports
the
estimates of partitioned average excess returns of puts, straddles, and calls,
without
making
distributional assumptions about $\epsilon_{T_{O}}$.
For instance, $\mu_{\{ {\cal F}_{t} \in \mathfrak{s}_{\tiny \mbox{bad}} \} }$ reflects the call
risk premium in bad economic states, which, in turn, tends to be associated with higher equity
premiums.
The
presence of $\epsilon_{T_{O}}$
recognizes the departures between observed option excess
returns and ex-ante expected option excess returns.

The superscripts ***, **, and * on estimates indicate statistical significance
at 1\%, 5\%, and 10\%, respectively. We rely on the HAC estimator
of \citet*{NeweyWest:87} with the lag
selected automatically. The reported partitioned average weekly
option returns are \emph{not} annualized.

Having laid the groundwork, we have hypothesized
that the local time component of the dark matter risk premium for $k<1$ and $k>1$
will be negligible in the case of weeklys. This is because, for small $T_{O}-t$, concerns about jump risks outweigh concerns about
volatility risks.\footnote{The size of the local time risk premiums for very short horizon options can also be understood
from the standpoint of
\citet*{Andersen_Fusari_Todorov:2017JFweekly}.
They suggest
the possibility that the variance of the continuous
component of equity returns is effectively almost constant over small $T_{O}-t$.}

Mindful of these
considerations, \emph{for~small}~$T_O - t$, we, hence, posit 
\begin{gather*}
\underbrace{\mathrm{Dark~Matter}}_{\tiny \mathrm{for~weeklys}}~\approx \sum_{t < \ell \leq T_O}
\begin{cases}
\mathbbm{1}_{\{G_{\ell  -} \geq k\}}  \max( k - G_{\ell}, 0 ) +
\mathbbm{1}_{\{G_{\ell  -} < k\}}  \max(  G_{\ell}- k, 0 )
&~~~~\text{puts}, k<1 \\
\mathbbm{1}_{\{G_{\ell  -} \leq k\}} \max( G_{\ell} - k, 0 ) +
\mathbbm{1}_{\{G_{\ell -} > k\}} \max( k - G_{\ell}, 0 )
 &~~~~\text{calls},~k>1.
\end{cases}
\label{emp.1}
\end{gather*}

Viewed through the prism of our theory, what do the weekly options data tell us? The empirical pattern
that emerges from Table~\ref{tab:weekly} is fourfold.
First,
the
partitioned
average excess returns of
straddles are negative (14 out of 15 estimates).
The weekly straddle return is $-10\%$ unconditionally.

Second, the partitioned average excess returns of
3\% OTM calls are
negative.  Consistent with our predictions, the negative effect of the risk premium for
$\sum_{t < \ell \leq T_O} \{ \mathbbm{1}_{\{G_{\ell  -} \leq k\}} \max( G_{\ell} - k, 0 ) +
\mathbbm{1}_{\{G_{\ell -} > k\}} \max( k - G_{\ell}, 0 )\}$
dominates
the
effect of
$\mathbb{E}_{t}^{\mathbb{P}}( \int_{t+}^{{T}_O} \mathbbm{1}_{\{G_{\ell-} > k\}}\, dG_{\ell} )$ at
high $k>1$.
The OTM call excess return is $-59\%$ unconditionally. The upshot from
the model-derived restrictions is that
the risk premiums for jumps crossing the strike are implied to
be negative at high $k>1$.

Third, the
difference in the partitioned average excess returns  of 3\% and 1\% OTM calls
are significantly negative. Our bootstrap-based exercise (Table~\ref{tab:bootstrap} (Panel A)) furnishes a finding that
the associated 95\% lower and upper confidence intervals do not tend to bracket zero.

Fourth, all estimates of partitioned average excess returns of OTM puts are negative.
The unconditional return
of $-59\%$ for 3\% OTM call, as opposed to $-61\%$ for 3\% OTM put, with the same
absolute delta, is revealing. Based on the 95\% bootstrap confidence intervals shown in
Table~\ref{tab:bootstrap} (Panel B), the risk premium for jumps crossing the strike for
$k>1$ (i.e., on the upside) is statistically
at par with that for $k<1$ (i.e., on the downside).
This finding stands out across three bootstrap procedures (IID, stationary, and circular block)
that we employ to safeguard inference.

Our Theorem~\ref{claimm:claim1call_jump}, in conjunction with the analytical link
in equation
(\ref{eq:StraddleInterim1Jumps}), \emph{for~small}~$T_O-t$, can
be considered
as a form of
specification test for the absence of unspanned risks. This is because of the correspondence between
the straddle risk premium and the risk premium for jumps crossing the strike
and local time. Stated differently,
the negative partitioned average weekly excess straddle returns mimic the sign and magnitude of the
dark matter risk premium at $k=1$.\footnote{The dichotomy
observed between partitioned average excess call returns for $\mathfrak{s}_{\tiny \mbox{bad}}$ and
$\mathfrak{s}_{\tiny \mbox{good}}$ can be understood in the context of our theory. To be specific, $\mathfrak{s}_{\tiny \mbox{bad}}$
may reflect \emph{high} prevailing $\mathbb{E}_{t}^{\mathbb{P}}( \int_{t+}^{{T}_O} \mathbbm{1}_{\{G_{\ell-} > k\}}\, dG_{\ell} )$, which
translates into positive partitioned average excess returns for
1\% OTM calls. This
dimension
may further help to explain the outcome that partitioned average excess call returns for $\mathfrak{s}_{\tiny \mbox{bad}}$ are
typically higher compared to those in $\mathfrak{s}_{\tiny \mbox{good}}$.}



Reinforcing the view that jumps crossing
the strike are a pertinent component of dark matter, we report the returns of crash-neutral
straddles 
in Table~\ref{tab:weekly} (final column).
Our treatment of the short put position
accounts for the posting of required collateral as per \citet*[page 22]{CBOE:2000}.
The salient finding is that average returns of crash-neutral straddles are small (and close to zero).
This outcome
supports a view that the risk premium for the jumps crossing
the strike component of shorting puts
balances out
the negative risk premium component for
long straddle positions.
%

%

The
negative average option excess returns for ultra-short maturities
further corroborate the relevance of jumps crossing the strike
(as noted in Internet Appendix (Table~\ref{tab:2and3day})).
These maturities of two- and three-day
manifest option prices that are higher
than the minimum tick size and
reflect positive likelihood of expiring in-the-money (i.e.,
$\mathbbm{1}_{\{ q_{t, {T}_O} >0 \}}$).
In sum, our evidence highlights hurdles
facing option models looking to match the behavior of ultra-short maturity option payoffs under both
$\mathbb{P}$ and $\mathbb{Q}$.


In what
ways
could
liquidity
considerations, margin requirements, and heterogeneous trading
contribute to outcomes of
negative returns of deep OTM calls? We address this issue from
three 
angles.
First, alleviating concerns that lack of liquidity may overly influence option returns, we
report (i) open interest and (ii) trading volume in Table~\ref{tab:weekly}
(and also Tables~\ref{tab:equity_options},
\ref{tab:table2}, and \ref{tab:opretspfutx}). In line with \citet*{Muravyev_Pearson:RFS2020},
deep OTM options do not appear to come with sharply lower open interest or
thin trading volume.

Second, we
consider OTM
calls with as small as
\textit{1 delta} and these strikes
maintain
positive
open interest and trading volume.\footnote{This feature is noted in Internet Appendix~(Table~\ref{tab:deep_weekly}).}
Our evidence indicates
that call option risk premiums are
negative
at progressively higher strikes.


Third,  it is plausible that bid-ask spreads widen
when market participants are adversely exposed to large price jumps.
Taking cues from \citet*{Christoffersen_Goyenko_Jacobs_Karoui:RFS2018},
we recompute option returns
using the midpoint of bid and ask prices.
The pattern of negative returns to buying deep OTM call options
remains qualitatively unchanged.\footnote{We display this evidence in Internet Appendix
(Table~\ref{tab:bid-ask}).}

\noindent \textbf{F. Composition of dark matter from farther-dated options.} To study
the nature of
dark matter risk premiums,
we examine
evidence from farther-dated options, with
${T}_O-t$ equal to 28 and 88 days (on average).
Farther-dated options can highlight
the relevance
of local time risk premiums because concerns
associated with jumps crossing the strike may be, relatively speaking, lessened.

Tables \ref{tab:equity_options}, \ref{tab:table2},  and  \ref{tab:opretspfutx} uncover negative partitioned average
excess returns of straddles. The uniformly negative estimates, in particular, for 88-day options, attests to the
notion of negative local time risk premiums.
Essentially, this exercise identifies the dark matter risk premium (for $k=1$) as being negative and
significant.
Our findings are an acknowledgment of
a viewpoint that aversion to unspanned risks
is implied within farther-dated options.

Consistent with our theoretical predictions, the negative effect of $\mathbb{E}_{t}^{\mathbb{P}}( \mathrm{D}^{u,T_O}_t[k] ) - \mathbb{E}_{t}^{\mathbb{Q}}( \mathrm{D}^{u, T_O}_t[k] )$ overcomes
the
effect of $\mathbb{E}_{t}^{\mathbb{P}}( \int_{t+}^{{T}_O} \mathbbm{1}_{\{G_{\ell-} > k\}}\, dG_{\ell} )$ at high $k>1$.
Specifically, based on Tables \ref{tab:equity_options}
and \ref{tab:table2} --- which cover 28-day options --- we garner that risk premiums for
5\% OTM calls exhibit partitioned average excess returns that are negative in 21 out of 30 entries.

Complementary evidence comes from Table~\ref{tab:opretspfutx}, which covers 88-day options, and shows that
partitioned average excess returns of 12\% (i.e.,
6 delta) OTM calls are  negative.
These estimates are statistically significant in 10 out of 15 entries. Additionally, the unconditional call
risk premiums get more negative deeper OTM (i.e., going from 32 to
6
delta). This outcome reflects
the interaction
between dark matter risk premiums --- which may get more negative with higher $k>1$ ---
and upside equity risk premiums.

Our theoretical results were designed in terms of equity futures, and the expected returns of their options, to exploit the
analytical convenience of the property that the futures price is a martingale under the $\mathbb{Q}$-measure.
This
aspect
is not essential, as noted in the context of Tables \ref{tab:equity_options} and~\ref{tab:table2}
and
because $F_{T_{O}}^{T_{O}}=S_{T_{O}}$.
First, there is agreement on negative straddle risk premiums and negative risk premiums for calls 5\% OTM. Second,
the evidence for negative put risk premiums is mutually consistent. Taken together, our
evidence favors dark matter risk premiums that tend to be more pronounced at both low $k<1$ and high $k>1$.

\noindent \textbf{G. Reconciling the various pieces of evidence and our hypotheses.}
The implication from straddle risk premiums
across the three maturities is that one can reject
the ``No unspanned risks" hypothesis.
Also, essential is
the data outcome
that the partitioned average excess returns of calls,
which depict call risk premiums, are negative at high $k>1$,
which is indicative of dark matter.

What is the foundation
of these findings? Connecting to
equation (\ref{eq:excess_mucall1x}),
$\mathbb{E}_{t}^{\mathbb{P}}( \int_{t+}^{{T}_O} \mathbbm{1}_{\{G_{\ell-} > k\}} \,dG_{\ell} )$
is likely to be small at higher $k$
and is conceivably dominated by the magnitude of the dark matter risk premium.
Accompanying
these effects across option maturities,
the straddle risk premiums being negative is
a further indication that  dark matter is relevant. 
The negative dark matter risk premiums --- imputed from traded options ---
support our theory that there are unspanned risks  and that they are economically pertinent.

Our theoretical predictions are free of parametric
assumptions about the evolution of the pricing kernel, price jumps, and return volatility.
Dark matter is needed to explain the behavior of call option risk premiums.
Although
we do not observe dark
matter directly, we are able to detect the workings of dark matter, and its risk premium, in the turning point of the call risk premiums
computed at rising $k$, as reflected in partitioned average excess return of calls \emph{switching sign} from positive to negative.

The
consequences
of our approach are
compatible with unspanned volatility risks
being disliked and jumps crossing the strike being disliked. The latter finding is informed by our
evidence from the weeklys. It is with the OTM weeklys that we can decouple the effects of jumps crossing the strike
from local time. These effects would otherwise be blended
within dark matter.

\vspace{-3mm}

\section{Dark matter in option pricing models}
\label{sec:furtherrr}

The distinguishing feature of our theory is that it maps
option risk premiums to the risk premiums for dark matter while
emphasizing
the statistics
of jumps crossing the strike and local time. What
are the consequences of dark matter embedded in an option pricing model? We explore
the dark matter property, as elaborated
in \citet*{Chen_Dou_Kogan:JF2020},
by parameterizing
uncertainties related to
unspanned diffusive risks and price and volatility jump risks
in option pricing models.
\vspace{2mm}


Consider a
parametric option pricing model that arises from the following setup under $\mathbb{P}$:
\begin{eqnarray}
\underbrace{\frac{dM_t}{M_{t-}}}_{\underset{\tiny \mbox{kernel}}{\tiny \mbox{pricing}}} & = & -r\, dt
+ \eta[t,\mathrm{v}_t] \underbrace{d z_t^{\mathbb{P}}}_{\underset{\tiny \mbox{risks}}{\tiny \mbox{spanned}}}
+~ \theta[t,\mathrm{v}_t] \underbrace{du_t^{\mathbb{P}}}_{\underset{\tiny \mbox{risks}}{\tiny \mbox{unspanned}}}
+ \underbrace{(e^{\mathbbm{x}_m} - 1) \, d \mathbb{N}_t^{\mathbb{P}}
}_{\underset{\tiny \mbox{risks}}{\text{\tiny unspanned~jump}}}
- {{\bm \lambda}^{\mathbb{P}}_{\tiny \mbox{jump}}} \, \mathbb{E}^{\mathbb{P}}( e^{\mathbbm{x}_m} - 1 ) \, dt,
\label{eq:dou1}\\
& &
\eta[t,\mathrm{v}_t] \, = \, - \frac{1}{ \sqrt{\mathrm{v}_t}}(
\alpha_{\tiny \mbox{vol}} +\lambda_{\tiny \mbox{vol}} \, \mathrm{v}_t),
~~~~~~~~
\theta[t,\mathrm{v}_t]  =  - \theta_{\mathrm{LT}}\, \sqrt{\mathrm{v}_t},
~\mbox{ \, }
\label{eq:dou2}
\\
\frac{d F_{t}^{T_F}}{F_{t-}^{T_F}} & = &  \overbrace{
(\alpha_{\tiny \mbox{vol}} + \lambda_{\tiny \mbox{vol}} \, \mathrm{v}_t +
\mu_{\tiny \mbox{jump}})}^{\text{\tiny futures~risk~premium}}  \, dt
+\sqrt{\mathrm{v}_t}
d z^{\mathbb{P}}_t
~+~\underbrace{(e^{\mathbbm{x}_s} - 1) \, d \mathbb{N}_t^{\mathbb{P}}}_{\underset{\tiny \mbox{jump~risks}}{\text{\tiny unspanned~price}}}
 -
{\bm \lambda}^{\mathbb{P}}_{\tiny \mbox{jump}} \, \mathbb{E}^{\mathbb{P}}( e^{\mathbbm{x}_s} - 1 ) \, dt,
\label{eq:dou3} \\
\underbrace{d\mathrm{v}_t}_{\tiny \mbox{variance}}  &=&
 ( \phi_{\tiny \mbox{vol}}^{\mathbb{P}} - \kappa_{\tiny \mbox{vol}}^{\mathbb{P}} \,\mathrm{v}_t )\, dt +
  \sigma_{\tiny \mbox{vol}} \, \sqrt{\mathrm{v}_t} \,\rho_{\tiny \mbox{vol}}
\underbrace{d z_t^{\mathbb{P}}}_{\underset{\tiny \mbox{risks}}{\tiny \mbox{spanned}}}
+~ \sigma_{\tiny \mbox{vol}} \sqrt{\mathrm{v}_t} \, \sqrt{1-\rho^2_{\tiny \mbox{vol}}} \,
\underbrace{du_t^{\mathbb{P}}}_{\underset{\tiny \mbox{risks}}{\tiny \mbox{unspanned}}} +
\underbrace{\mathbbm{x}_{\mathrm{v}} \, d \mathbb{N}^{\mathbb{P}}_{t}}_{\underset{\tiny \mbox{(additive)}}{\tiny \mbox{jumps~in}~\mathrm{v}_t}},~~ \mbox{ \, \,  }~ \label{eq:dou4}\\
\underbrace{d \mathbb{N}^{\mathbb{P}}_{t}}_{\tiny \mbox{Poisson~jump}} &=& \left \{
\begin{array}{ll}
1
& \hspace{10 mm} \mbox{with~probability}~{\bm \lambda}^{\mathbb{P}}_{\tiny \mbox{jump}}\,dt \\
0
& \hspace{10 mm} \mbox{with~probability}~1- {\bm \lambda}^{\mathbb{P}}_{\tiny \mbox{jump}}\,dt \\
\end{array}
\right.  \label{eq:dou5} \\
\mathbbm{x}_{\mathrm{v}} &&\text{variance~jumps~follow~spectrally~positive~i.i.d.~distribution~under}~\mathbb{P} ~ ~~ ~~~ ~ \label{eq:dou6a} \\
(\mathbbm{x}_m, \mathbbm{x}_s) &&\text{jumps in $M_t$ and $F_{t}^{T_F}$ have i.i.d. distributions~under}~\mathbb{P}.
\label{eq:dou6}
\end{eqnarray}


In this model, $\mathrm{v}_t$ denotes
the variance of the diffusive component of the
equity (futures) return,
and  $z_t^{\mathbb{P}}$ and $u_t^{\mathbb{P}}$ are each
independent
standard Brownian motions.
Unspanned risks are commingled with spanned risks in both the $M_t$ and $\mathrm{v}_t$ dynamics.


How does this model ---
which
traverses  the dimension of unspanned diffusive volatility risks and unspanned price and volatility jump risks --- fare in summarizing
option risk premiums? First,
the risk premiums associated with jumps crossing the strike vary across alternative
jump specifications under $\mathbb{P}$ and $\mathbb{Q}$.
Our model analysis, pursued in Internet Appendix (Section~\ref{app:jumps_across}),
shows that the risk premiums for jumps
crossing the strike can
rationalize negative risk premiums
of OTM calls.
We establish this attribute for
jump specifications of \citet*{Merton:76}, \citet*{Kou:2002}, and \citet*{DuffiePanSingleton:2000}.
However, akin to the dark matter property, reconciliation
between option models and data requires
a stand on the
properties of jumps under $\mathbb{P}$ and $\mathbb{Q}$.\footnote{Our approach
aims to understand
the  differences in option risk
premiums
across strikes. Additionally,
we emphasize weekly options, which allow us to draw the distinctions between
risk premiums for jumps crossing the strike on the downside versus on the
upside. While \citet*{Merton:76} emphasizes
\emph{downward} jumps in equities, \citet*{Kou:2002} presents a model with
\emph{both upward} and downward jumps.
See also \citet*{Aitsahalia_Yacine:2004}.
We refer the reader to works
that
consider
models with jumps
(in price and volatility) and/or
stochastic volatility. See, among others,
\citet*{BakshiCaoChen:97},
\citet*{Bates:2000},
\citet*{Pan:2002},
\citet*{ErakerJohannesPolson:2003},
\citet*{Eraker:2004},
\citet*{Kou_Wang:2004},
\citet*{BroadieChernovJohannes:2007},
and \citet*{Cai_Kou:2011}.}



Three sources contribute to local time risk premiums in this model: (i) unspanned diffusive risks,
(ii) unspanned volatility jump risks, and (iii) spanned diffusive risks. This
analysis is rather lengthy
and is presented in Internet Appendix (Section~\ref{app:var_jumps}).

Notably, we show
that the parameter $\theta_{\mathrm{LT}}$  --- introduced in (\ref{eq:dou2}) --- controls the contribution of priced unspanned diffusive volatility risks over $T_{O}-t$ to local time risk
premiums.\footnote{We show this in Internet Appendix (Section~\ref{app:unspannedDiffusiveVolRisk}).} The overall consequence is that the
local time risk premiums for unspanned diffusive risks can be negative
(provided $\theta_{\mathrm{LT}} < 0$), which contributes to negative call option risk premiums.

Additionally, the local time risk premiums due to unspanned jump volatility risks
can be negative.\footnote{These restrictions are identified in Internet Appendix (Section~\ref{app:unspannedVolJumpRisk}).}
The takeaway is that any potential
misspecification of models with $\theta_{\mathrm{LT}} \equiv 0$, or absent of unspanned
volatility jump risks, may be hard to disentangle without data on option returns.

Finally, if spanned risks were the only source of uncertainty in the pricing kernel (i.e., if $\mathbbm{x}_m \equiv 0$ and
$\theta_{\mathrm{LT}} \equiv 0$), then
local time risk premiums are such that the risk premiums for OTM calls
would be \emph{positive}. The
implication is that unspanned risks, and hence, dark matter,
are relevant to capturing realities of option risk premiums.

Our decomposition of option risk premiums
provides additional perspectives. First, option models rely upon
variables and parameters that may be hard to reliably extract from equity and volatility dynamics. Second, some parameter restrictions required for empirical consistency may not be directly verifiable.
For example,
to align negative local time risk premiums for volatility jumps, we deduce that option
model parameterizations must be such that large positive jumps in volatility associate with large positive
jumps in the pricing kernel. However, the pricing kernel is not a directly inferable quantity.

\noindent \textbf{Connections with other option modeling frameworks.}
Through Tanaka's formula, we emphasize the analyticity of local time and jumps crossing the strike
and this angle deviates from others.

\citet*{CovalShumway:2001}
feature a theory in which the call option risk premium is \emph{positive and increasing} in the strike price. Our
prediction, with unspanned risks, with or without jumps, is for the opposite, when the dark matter risk premium is sufficiently negative, and we pose this as a testable implication at high $k>1$
(i.e., farther OTM
calls).
The work of \citet*{Christoffersen_Jacobs_Heston:2013RFSa}
considers
a log stochastic discount factor (SDF),
affine in the return of the equity
and its variance, but the SDF's projection onto
returns is nonmonotonic.
Their
framework does not
formalize a theory of option risk premiums across strikes.



On the other hand, {\citet*{BakshiMadanPanayotov:2010JFE}}
consider a model with heterogeneity in beliefs with personalized
change of measure
for investors, long and short
equity.
In this setting, it is shown
that the risk premium of
OTM calls
can be negative when the SDF
admits an increasing region to the
upside.
The approach in our paper relies
on a dynamic model with unspanned risks,
and it does
not take a stand on whether the
SDF is nonmonotonic.

\citet*{Andersen_Fusari_Todorov:2017JFweekly} explore the merits of using weekly options.
They formalize the argument that the jump intensity rate of the discontinuous component and
the return variance of the continuous component will vary little for short-dated options.
In particular, the variance of the continuous component can be regarded as a constant
over very short horizons.
Complementing their approach,
we uncouple, using weeklys (analogous to small $T_{O}-t$), the
effects of risk premiums on local time from risk premiums on
jumps crossing the strike.

Our perspective about local time risk premiums --- gleaned from option returns --- intersects with work
on volatility.
\citet*{Carr_Wu:2016_JFE} model
implied volatility dynamics and then derive implications for the shape of the
volatility surface.
\citet*{Eraker_Wu:2017} show
negative average returns to holding volatility products.
What emerges from the analysis of \citet*{Aitsahalia_Karaman_Mancini:JOE2019} is
that variance swap rates incorporate
a significant price jump component.

The driving mechanism of our theory of
option risk premiums is dark matter.
\citet*{Jonesc:2006} considers
factor models of index option returns but
does not emphasize jumps, and
the setup does not offer differentiation
between diffusive and discontinuous return components. 
\citet*{Broadie_Chernov_Ghysels:2009RFS} explore option mispricing and examine unconditional
returns to writing puts on the S\&P 500 index futures.
Essential to \citet*{Bollerslev_Todorov_Xu:2015_JFE} is that the variance risk premium helps predict
market returns and that much of this predictability arises from the part of the variance risk premium associated with tail risk.
\vspace{-4mm}

\section{Conclusion}
\label{eq:concluding_remarks}

Is there dark matter embedded in volatility and in equity options?
That is, are
there unspanned risks
that
are hard to observe but elicit
risk premiums on equity options?
Building on this question,
our answer is ``yes," and we provide supportive empirical evidence.


We present a semimartingale theoretical approach that allows us to study
the constructs of \emph{jumps crossing the strike}
(from below and above) and of \emph{local time}. Our treatment of jumps crossing
the strike and of local time is essential to our theory, because their
absence would go against our empirical evidence.
We label such abstract uncertainties dark matter,
as they can be hard to identify, but their presence
is inferred
in options data.
Dark matter generates statistically significant risk premiums,
and the workings of dark matter can be economically influential.

Developing this line of inquiry, we reveal the manner in which call option risk premiums can be decomposed into
dark matter risk premiums and \emph{upside} equity risk premiums.
Our theoretical
treatment predicts negative call option risk premiums and a negative straddle risk premium \emph{only if}
there are unspanned risks in the pricing kernel.
Our empirical findings are
consistent with
the relevance of unspanned risks and dark matter in option risk premiums.

We develop theoretical results with testable implications.
The key to attaining consistency with data attributes lies in equipping
the pricing kernel dynamics, and the
volatility dynamics with unspanned risks (the jump risks are intrinsically unspanned),
which ends up inducing negative
dark matter risk premiums. What stands out from our analysis is the compatibility between
negative dark matter risk premiums and negative risk premiums of straddles and deep out-of-the-money call options.
Our empirical investigation substantiates these implications,
thus, aligning with our theory of unspanned risks and dark matter in equity markets.

\newpage
\clearpage
                                            \bibliographystyle{abbrvnat}


\bibliography{masterbib_gao}

\newpage
\appendix

\begin{center}
{\Large \bf Appendix}
\end{center}
\vspace{-3mm}
\setcounter{equation}{0}
\renewcommand{\theequation}{A\arabic{equation}}

\section{\bf \small Appendix A: Proof of Theorem~\ref{claimm:claim1call_jump} ((general) semimartingales)} 
\label{appsec:jumppps}

Suppose $(F_{\ell}^{T_F})$, and thus $(G_\ell)$, for $\ell \geq t$, are semimartingales.
This theoretical
environment
allows for the possibility of jumps in
the futures price, as well as for stochastic volatility effects (including accommodating jumps in volatility).

Henceforth, the
term $\mathbb{L}^{T_O}_t[k]$ is local time (as defined in  (\ref{ltt.1})).

By construction,  $G_t=1$.
Since the stochastic processes $(F_{\ell}^{T_F})$ and $(G_\ell)$ are $\mathbb{Q}$ martingales,
\begin{align}
&\mathbb{E}_{t}^{\mathbb{Q}}( \int_{t+}^{T_O} \mathbbm{1}_{\{G_{\ell \, -} < k\}} dG_{\ell} ) ~=~0 &
&\mathrm{and}&
&\mathbb{E}_{t}^{\mathbb{Q}}( \int_{t+}^{T_O} \mathbbm{1}_{\{G_{\ell \, -} > k\}} dG_{\ell} ) ~=~0.&
\end{align}
%

\noindent \textbf{I. OTM call option risk premium.}
We employ Tanaka's formula in
(\ref{eq:TanakaJumps}).

Using
the definition of the expected return of a call option, the fact that $(F_{\ell}^{T_F})$ is a martingale
under $\mathbb{Q}$, and considering OTM calls, that is $k> 1$, so that $\max( G_{t} - k, 0 ) = 0$, we obtain
\begin{equation}
1 + \mu^{{T}_O}_{t,{\tiny \mathrm{call}}}[k] \, = \,
\frac{\mathbb{E}_{t}^{\mathbb{P}}( \int_{t+}^{T_O} \mathbbm{1}_{\{G_{\ell  -} > k\}} dG_{\ell})
~+~ \mathbb{E}_{t}^{\mathbb{P}}( \mathbb{L}^{T_O}_t[k])
~+~
\mathbb{E}_{t}^{\mathbb{P}}(a_t^{T_O}[k]) + \mathbb{E}_{t}^{\mathbb{P}}(b_t^{T_O}[k]) }
{ e^{-r ({T}_O - t)} \{ \mathbb{E}_{t}^{\mathbb{Q}}( \mathbb{L}^{T_O}_t[k] )
~+~ \mathbb{E}_{t}^{\mathbb{Q}}( a_t^{T_O}[k] )
~+~ \mathbb{E}_{t}^{\mathbb{Q}}( b_t^{T_O}[k]) \} }.~~~\mbox{ \, \, }
\label{eq:ExpectedHoldingReturn1GneralDynamicsWithJumps}
\end{equation}
From the definition of $\mathrm{D}^{u, T_O}_t[k] = \mathbb{L}^{T_O}_t[k] + a_t^{T_O}[k] + b_t^{T_O}[k]$ in
(\ref{darkk.1}), we note that
\begin{equation}
\mathbb{E}_{t}^{\mathbb{Q}}( \mathrm{D}^{u, T_O}_t[k]) ~=~ \mathbb{E}_{t}^{\mathbb{Q}}(\mathbb{L}^{T_O}_t[k] + a_t^{T_O}[k] + b_t^{T_O}[k])~>0,~~~~\mathrm{for}~k>1.
\end{equation}
This follows, since
\begin{align}
&\mathbb{E}_{t}^{\mathbb{Q}}( \mathbb{L}^{T_O}_t[k] ) > 0 &(\mathbb{L}^{T_O}_t[k]~\mathrm{is~a~nonnegative~random~variable).} \\
&\mathbb{E}_{t}^{\mathbb{Q}}( a_t^{T_O}[k]) > 0~~\mathrm{and}~~\mathbb{E}_{t}^{\mathbb{Q}}( b_t^{T_O}[k]) > 0
&(a_t^{T_O}[k]~\mathrm{and}~b_t^{T_O}[k]~\mbox{are~each~{convex in $G_\ell$).}}
\end{align}

Subtracting $e^{r ({T}_O - t)}$ from both sides of
(\ref{eq:ExpectedHoldingReturn1GneralDynamicsWithJumps}),
we obtain the following: 
\begin{equation*}
1 + \mu^{{T}_O}_{t,{\tiny \mathrm{call}}}[k] - e^{r ({T}_O - t)} =
\frac{e^{r ({T}_O - t)}}{\mathbb{E}_{t}^{\mathbb{Q}}( \mathrm{D}^{u, T_O}_t[k])  }
\{
\underbrace{\mathbb{E}_{t}^{\mathbb{P}}( \int_{t+}^{T_O} \mathbbm{1}_{\{G_{\ell  -} > k\}} dG_{\ell})}_{\tiny~\mbox{upside~risk~premium}}
~+~ \underbrace{\mathbb{E}_{t}^{\mathbb{P}}( \mathrm{D}^{u, T_O}_t[k] ) - \mathbb{E}_{t}^{\mathbb{Q}}( \mathrm{D}^{u, T_O}_t[k] )}_{\tiny \mbox{risk~premium~for~dark~matter}} \}.
\label{eq:ExpectedHoldingReturn1call}
\end{equation*}

If the upside risk premium $\mathbb{E}_{t}^{\mathbb{P}}( \int_{t+}^{{T}_O} \mathbbm{1}_{\{G_{\ell -} > k\}} dG_{\ell} )$
were
positive, the expected excess return of an OTM call on the equity futures \emph{can} be negative only if
\begin{equation}
\mathbb{E}^{\mathbb{P}}_{t}( \mathrm{D}^{u, T_O}_t[k] ) ~-~ \mathbb{E}^{\mathbb{Q}}_{t}( \mathrm{D}^{u, T_O}_t[k] )~\mathrm{is~negative~for}~k>1.
\end{equation}
The following case is instructive:
\begin{itemize}
\item Suppose $(F_{\ell}^{T_F})$ is a continuous semimartingale. Then  $a_t^{T_O}[k]=b_t^{T_O}[k]=0$ and the source of
the risk premium for dark matter is the risk premium for local time (for $k>1$).

\end{itemize}
We have verified the statement of Theorem~\ref{claimm:claim1call_jump} with respect to OTM calls. $\square$

\noindent \textbf{II. OTM put option risk premium.} With the definition, for $k<1$, in (\ref{darkk.1})
that $\mathrm{D}^{d, T_O}_t[k]= \mathbb{L}^{T_O}_t[k] + c_t^{T_O}[k] + d_t^{T_O}[k]$, and Tanaka's formula in (\ref{eq:TanakaPutCaseJumps}),
we obtain the following:
\begin{equation*}
\underbrace{1 + \mu^{{T}_O}_{t,{\tiny \mathrm{put}}}[k] - e^{r ({T}_O - t)}}_{\tiny \mbox{expected~excess~return~of~puts}} =
\frac{e^{r ({T}_O - t)}}{\mathbb{E}_{t}^{\mathbb{Q}}( \mathrm{D}^{d,T_O}_t[k])  }
\{~
- \underbrace{\mathbb{E}_{t}^{\mathbb{P}}( \int_{t+}^{T_O} \mathbbm{1}_{\{G_{\ell -} < k\}} dG_{\ell})}_{~\tiny \mbox{downside~risk~premium}}
+ \underbrace{\mathbb{E}_{t}^{\mathbb{P}}( \mathrm{D}^{d,T_O}_t[k] ) - \mathbb{E}_{t}^{\mathbb{Q}}( \mathrm{D}^{d,T_O}_t[k] )}_{\tiny \mbox{risk~premium~for~dark~matter}} \}.
\label{eq:ExpectedHoldingReturn1put}
\end{equation*}
If the downside risk premium $\mathbb{E}_{t}^{\mathbb{P}}( \int_{t+}^{{T}_O} \mathbbm{1}_{\{G_{\ell -} < k\}} dG_{\ell} )$
were
positive, the put risk premium is negative when the risk premium for dark matter is negative. $\square$

\noindent \textbf{III. Straddle risk premium.}
Since at $k = 1$, $a_t^{T_O}[1] = d_t^{T_O}[1]$, and
$b_t^{T_O}[1] = c_t^{T_O}[1]$, it holds that
\begin{eqnarray}
&a_t^{T_O}[1] + b_t^{T_O}[1] + c_t^{T_O}[1] + d_t^{T_O}[1] ~ = ~ 2 \,(a_t^{T_O}[1] + b_t^{T_O}[1])~\equiv~ 2 \, \mathbb{A}_t^{T_O}[1], ~ \mbox{ \, \, \, \, \, } ~  &  \\
& \mathrm{where} ~~ ~ \mathbb{A}_t^{T_O}[1] ~ \equiv ~  \sum_{t < \ell \leq T_O} \underbrace{\{
 \mathbbm{1}_{\{G_{\ell \, -} < 1\}} \, \max(  G_{\ell}- 1, 0 ) \, + \,
 \mathbbm{1}_{\{G_{\ell  -} > 1\}} \, \max( 1 - G_{\ell}, 0 )\}}_{\tiny \mbox{jumps~crossing~the~strike~from~below~and~above,}~ k = 1}. ~ \mbox{ \, \, \, \, } ~ &
\end{eqnarray} 

Suppose further that, for $k=1$, the futures risk premium to the
upside is approximately equal to the futures risk premium to the downside.
This is akin to an assumption that return movements
(anchored to $F_{t}^{T_F}$)
to the downside or upside
are equally probable and unforecastable.

Specifically,
\begin{equation}
\overbrace{\mathbb{E}_{t}^{\mathbb{P}}( \int_{t+}^{T_{O}} \mathbbm{1}_{\{G_{\ell -} > 1 \}} dG_{\ell} ) }
^{\tiny \mbox{upside~risk~premium~for}~\tiny {k=1}}
~~ - ~~
\overbrace{\mathbb{E}_{t}^{\mathbb{P}}( \int_{t+}^{T_{O}} \mathbbm{1}_{\{G_{\ell -} < 1 \}} dG_{\ell} )}^{\tiny \mbox{downside~risk~premium~for}~{k=1}}  ~~ \approx ~ 0. ~~ \mbox{ \, \, \, \, \, \, }~~~~ \label{eq:RiskPremiumUpsideEqualsRiskPremiumDownsideJumps}
\end{equation}
Then we have
\begin{eqnarray}
& & \overbrace{1 + \mu^{T_{O}}_{t,{\tiny \mbox{straddle}}} - e^{r (T_{O}-t)}}^{\tiny \mbox{straddle~risk~premium}} \nonumber \\
& & ~ = ~ e^{r (T_{O}-t)} (
\frac{\overbrace{\mathbb{E}_{t}^{\mathbb{P}}( \int_{t+}^{T_{O}} \mathbbm{1}_{\{G_{\ell -} > 1 \}} dG_{\ell}
- \int_{t+}^{T_{O}} \mathbbm{1}_{\{G_{\ell -} < 1 \}} dG_{\ell}}^{\approx 0}
~+~ 2 \, \mathbb{L}_t^{{T}_O}[1] + 2\, \mathbb{A}_t^{T_O}[1] )}{ \mathbb{E}_{t}^{\mathbb{Q}}( 2 \, \mathbb{L}_t^{{T}_O}[1]  + 2\, \mathbb{A}_t^{T_O}[1] )} - 1) ~ \mbox{ \, \, \, } ~ \nonumber \\
& & ~=~\frac{e^{r (T_{O}-t)}}{\mathbb{E}_{t}^{\mathbb{Q}}( \mathbb{L}_t^{{T}_O}[1] +\mathbb{A}_t^{{T}_O}[1] )} ~
\{ \underbrace{\mathbb{E}_{t}^{\mathbb{P}}( \mathbb{L}_t^{{T}_O}[1]) -\mathbb{E}_{t}^{\mathbb{Q}}( \mathbb{L}_t^{{T}_O}[1] )}_{\underset{(k=1)}{\tiny \mbox{local~time~risk~premium}}} ~+~
\underbrace{\mathbb{E}_{t}^{\mathbb{P}}( \mathbb{A}_t^{{T}_O}[1]) -\mathbb{E}_{t}^{\mathbb{Q}}( \mathbb{A}_t^{{T}_O}[1] )}_{\underset{ \mathrm{strike~from~below~and~above}~k=1}{\tiny \mbox{risk~premium~for~jumps~crossing~the}}} \}.
\label{eq:StraddleInterim1Jumps}
\end{eqnarray}
The continuous semimartingale analog of (\ref{eq:StraddleInterim1Jumps}) is obtained by setting
$\mathbb{A}_t^{T_O}[1]=0$ (because, in this case, there are no jumps).  $\square$

Internet Appendix (Section~\ref{app:jumps_acrossPutsStraddles}) further shows that when there are no unspanned risks in the pricing kernel, the straddle risk premium is zero.


We have the proof of Theorem~\ref{claimm:claim1call_jump}. $\blacksquare$ 

\newpage

\newpage
\begin{table}[h!]
\footnotesize
\caption{\textbf{{
Risk premiums for \emph{weekly options} on the S\&P 500 index
}}}
\vspace{2mm}
\label{tab:weekly}
The sample period is 01/13/2011 to 12/20/2018, with 415 weekly option expiration cycles (8 days to maturity (on average)).
The weekly options data on S\&P 500 index is from the CBOE.
We construct the excess return of OTM puts, OTM calls, and straddles (ATM and crash-neutral) over weekly
expiration cycles. These calculations are done at the ask option price. The returns of a crash-neutral
straddle combines a long straddle position and a short 3\% OTM put position.
The following is the regression specification (analogously for puts and straddles):
\begin{align*}
&q_{t,{\tiny \mathrm{call}} }^{{T}_O}[k] =
\mu_{\{ {\cal F}_{t} \in \mathfrak{s}_{\tiny \mbox{bad}} \} } \mathbbm{1}_{\{ {\cal F}_{t} \in\mathfrak{s}_{\tiny \mbox{bad}} \}}
+ \mu_{\{ {\cal F}_{t} \in \mathfrak{s}_{\tiny \mbox{normal}} \} }
\mathbbm{1}_{\{ {\cal F}_{t} \in\mathfrak{s}_{\tiny \mbox{normal}} \}}
+ \mu_{\{ {\cal F}_{t} \in \mathfrak{s}_{\tiny \mbox{good}} \} }
\mathbbm{1}_{\{ {\cal F}_{t} \in\mathfrak{s}_{\tiny \mbox{good}} \}} ~+~ \underbrace{\epsilon_{T_{O}}.}_{\mathrm{error~term}}&
\end{align*}
We use proxies for the variable $\mathfrak{s}$, known at the beginning of the expiration cycle.
The variable construction for this weekly exercise is described
in the text.
For example, WEI is the weekly economic index. 
\\ 

We indicate statistical significance at 1\%, 5\%, and 10\% by the superscripts ***, **, and *, respectively,
where the $p$-values rely on the \citet*{NeweyWest:87} HAC estimator (with the lag selected automatically).
The reported
put (respectively, call) delta is $-{\cal N}(-d_1)$ (respectively, ${\cal N}(d_1)$), where $d_1=  \frac{1}{ \sigma \sqrt{T_O-t}} \{ - \log k + r (T_O-t) + \frac{1}{2} \sigma^2 (T_O-t)\}$. 
SD is the standard deviation, and $\mathbbm{1}_{\{ q_{t, {T}_O} >0 \}}$ is the proportion (in \%) of
option positions that generate positive returns.
We tabulate the average open interest and trading volume, all observed on the first day of the weekly option expiration cycle.
The average number of strikes across puts and calls is 112.
\begin{center}
\setlength{\tabcolsep}{0.06in}
\begin{tabular}{lll c ccc ccc cc ccc} \hline
       &   & &           &           &           &           &           &           &  \\
       &   & &\multicolumn{3}{c}{OTM puts on equity} &           &  \multicolumn{3}{c}{OTM calls on equity}& &\multicolumn{2}{c}{Straddle} \\
       &   & &\multicolumn{3}{c}{$\log(k)\times 100$} &        & \multicolumn{3}{c}{$\log(k)\times 100$} &    &\multicolumn{2}{c}{on equity} \\
          \cline{4-6} \cline{8-10} \cline{12-13}
Moneyness (\%)   &  &      & -3         & -2         & -1         &          & 1         & 2         & 3 &    &  ATM  &     Crash-\\
Delta (\%)       &  & & -6         & -12         & -26         &           & 27         & 12         & 6 &    &   &     Neutral\\ \\
Open Interest ($\times 1,000$)    &  & & 10.2         & 9.3         & 7.4         &           & 9.1         & 7.9         & 6.9 &    &   &    \\
Volume ($\times 1,000)$     &  & & 2.5        & 2.5         & 2.6         &           & 3.1         & 2.4        & 1.8 &    &   &    \\
 &         &       &    &           &           &           &                      &  &    &    &\\ \hline
 &         &       &    &           &           &           &                      &  &    &    &\\

\multicolumn{1}{l}{Change in WEI} & \multicolumn{1}{l}{L} & \multicolumn{1}{l}{$\mathfrak{s}_{\tiny \mbox{bad}}$} & \multicolumn{1}{c}{-44} & \multicolumn{1}{c}{-36} & \multicolumn{1}{c}{-30} &           & \multicolumn{1}{c}{60} & \multicolumn{1}{c}{11} & \multicolumn{1}{c}{-53***} &    &  -2  & 0\\
\multicolumn{1}{l}{ } & \multicolumn{1}{l}{M} & \multicolumn{1}{l}{$\mathfrak{s}_{\tiny \mbox{normal}}$} & \multicolumn{1}{c}{-81***} & \multicolumn{1}{c}{-69***} & \multicolumn{1}{c}{-53***} &           & \multicolumn{1}{c}{-16} & \multicolumn{1}{c}{-46***} & \multicolumn{1}{c}{-64***}  &    & -23***   & -1***\\
\multicolumn{1}{l}{ } & \multicolumn{1}{l}{H} & \multicolumn{1}{l}{$\mathfrak{s}_{\tiny \mbox{good}}$} & \multicolumn{1}{c}{-58***} & \multicolumn{1}{c}{-32} & \multicolumn{1}{c}{-16} &           & \multicolumn{1}{c}{-8} & \multicolumn{1}{c}{-38**} & \multicolumn{1}{c}{-59***}  &    & -5   & 0\\
          &           &           &           &           &           &           &           &           &   &    &    &\\
\multicolumn{1}{l}{Quadratic Variation} & \multicolumn{1}{l}{H} & \multicolumn{1}{l}{$\mathfrak{s}_{\tiny \mbox{bad}}$} & \multicolumn{1}{c}{-42} & \multicolumn{1}{c}{-27} & \multicolumn{1}{c}{-19} &           & \multicolumn{1}{c}{3} & \multicolumn{1}{c}{-2} & \multicolumn{1}{c}{-22} &    &  -7  & 0 \\
\multicolumn{1}{l}{ } & \multicolumn{1}{l}{M} & \multicolumn{1}{l}{$\mathfrak{s}_{\tiny \mbox{normal}}$} & \multicolumn{1}{c}{-50**} & \multicolumn{1}{c}{-24} & \multicolumn{1}{c}{-9} &           & \multicolumn{1}{c}{32} & \multicolumn{1}{c}{7} & \multicolumn{1}{c}{-56***}  &    & 2   & 0\\
\multicolumn{1}{l}{ } & \multicolumn{1}{l}{L} & \multicolumn{1}{l}{$\mathfrak{s}_{\tiny \mbox{good}}$} & \multicolumn{1}{c}{-91***} & \multicolumn{1}{c}{-86***} & \multicolumn{1}{c}{-71***} &           & \multicolumn{1}{c}{0} & \multicolumn{1}{c}{-77***} & \multicolumn{1}{c}{-98***} &    &  -25***  &-2*** \\
          &           &           &           &           &           &           &           &           &  &    &    & \\
\multicolumn{1}{l}{Risk Reversal} & \multicolumn{1}{l}{H} & \multicolumn{1}{l}{$\mathfrak{s}_{\tiny \mbox{bad}}$} & \multicolumn{1}{c}{-70***} & \multicolumn{1}{c}{-51***} & \multicolumn{1}{c}{-35**} &           & \multicolumn{1}{c}{24} & \multicolumn{1}{c}{-53***} & \multicolumn{1}{c}{-92***} &    & -10   & 0\\
\multicolumn{1}{l}{ } & \multicolumn{1}{l}{M} & \multicolumn{1}{l}{$\mathfrak{s}_{\tiny \mbox{normal}}$} & \multicolumn{1}{c}{-94***} & \multicolumn{1}{c}{-75***} & \multicolumn{1}{c}{-56***} &           & \multicolumn{1}{c}{18} & \multicolumn{1}{c}{2} & \multicolumn{1}{c}{-36} &    &  -14**  & 0\\
\multicolumn{1}{l}{ } & \multicolumn{1}{l}{L} & \multicolumn{1}{l}{$\mathfrak{s}_{\tiny \mbox{good}}$} & \multicolumn{1}{c}{-19} & \multicolumn{1}{c}{-10} & \multicolumn{1}{c}{-8} &           & \multicolumn{1}{c}{-6} & \multicolumn{1}{c}{-22} & \multicolumn{1}{c}{-49***} &    & -7   &  -1\\
          &           &           &           &           &           &           &           &           &  &    &    &\\
\multicolumn{1}{l}{Change in Volatility} & \multicolumn{1}{l}{H} & \multicolumn{1}{l}{$\mathfrak{s}_{\tiny \mbox{bad}}$} & \multicolumn{1}{c}{-41} & \multicolumn{1}{c}{-38*} & \multicolumn{1}{c}{-37**} &           & \multicolumn{1}{c}{2} & \multicolumn{1}{c}{-7} & \multicolumn{1}{c}{-55***}  &    & -15**   & -1\\
\multicolumn{1}{l}{ } & \multicolumn{1}{l}{M} & \multicolumn{1}{l}{$\mathfrak{s}_{\tiny \mbox{normal}}$} & \multicolumn{1}{c}{-71***} & \multicolumn{1}{c}{-51***} & \multicolumn{1}{c}{-30*} &           & \multicolumn{1}{c}{49} & \multicolumn{1}{c}{-6} & \multicolumn{1}{c}{-43*}  &    &  -5  &  0\\
\multicolumn{1}{l}{ } & \multicolumn{1}{l}{L} & \multicolumn{1}{l}{$\mathfrak{s}_{\tiny \mbox{good}}$} & \multicolumn{1}{c}{-71***} & \multicolumn{1}{c}{-48***} & \multicolumn{1}{c}{-32**} &           & \multicolumn{1}{c}{-15} & \multicolumn{1}{c}{-59***} & \multicolumn{1}{c}{-78***} &    & -10*   & 0 \\
          &           &           &           &           &           &           &           &           &  &    &    & \\
\multicolumn{1}{l}{Recent Market} & \multicolumn{1}{l}{L} & \multicolumn{1}{l}{$\mathfrak{s}_{\tiny \mbox{bad}}$} & \multicolumn{1}{c}{-45} & \multicolumn{1}{c}{-39*} & \multicolumn{1}{c}{-28*} &           & \multicolumn{1}{c}{9} & \multicolumn{1}{c}{-21} & \multicolumn{1}{c}{-53***} &    & -13*   & -1\\
\multicolumn{1}{l}{ } & \multicolumn{1}{l}{M} & \multicolumn{1}{l}{$\mathfrak{s}_{\tiny \mbox{normal}}$} & \multicolumn{1}{c}{-56***} & \multicolumn{1}{c}{-38*} & \multicolumn{1}{c}{-24} &           & \multicolumn{1}{c}{6} & \multicolumn{1}{c}{-33} & \multicolumn{1}{c}{-59***} &    & -5   & 0\\
\multicolumn{1}{l}{ } & \multicolumn{1}{l}{H} & \multicolumn{1}{l}{$\mathfrak{s}_{\tiny \mbox{good}}$} & \multicolumn{1}{c}{-82***} & \multicolumn{1}{c}{-60***} & \multicolumn{1}{c}{-47***} &           & \multicolumn{1}{c}{21} & \multicolumn{1}{c}{-18} & \multicolumn{1}{c}{-65***} &    & -12*   & -1 \\
          &           &           &           &           &           &           &           &           &  &    &    & \\ \hline
          &           &           &           &           &           &           &           &           &  &    &    &\\
\multicolumn{1}{l}{\textbf{Unconditional}} &           & \multicolumn{1}{l}{Average} & \multicolumn{1}{c}{-61} & \multicolumn{1}{c}{-46} & \multicolumn{1}{c}{-33} &           & \multicolumn{1}{c}{12} & \multicolumn{1}{c}{-24} & \multicolumn{1}{c}{-59} &    &  -10  & -1 \\
\multicolumn{1}{l}{\textbf{Estimates}} &           & \multicolumn{1}{l}{SD} & \multicolumn{1}{c}{240} & \multicolumn{1}{c}{216} & \multicolumn{1}{c}{181} &           & \multicolumn{1}{c}{273} & \multicolumn{1}{c}{255} & \multicolumn{1}{c}{210} &    & 78   & 7 \\
\multicolumn{1}{l}{ } &           & \multicolumn{1}{l}{
$\mathbbm{1}_{\{ q_{t, {T}_O} >0 \}}$}
& \multicolumn{1}{c}{6\%} & \multicolumn{1}{c}{10\%} & \multicolumn{1}{c}{17\%} &           & \multicolumn{1}{c}{26\%} & \multicolumn{1}{c}{12\%} & \multicolumn{1}{c}{5\%} &    &   40\% & 43\%\\
          &           &           &           &           &           &           &           &           & &    &    & \\ \hline
\end{tabular}%
\end{center}
\end{table}

\newpage
\begin{table}[h!]
\footnotesize
\caption{\textbf{Disparities in option risk premiums
with weekly options}}
\vspace{2mm}
\label{tab:bootstrap}
The sample period is 01/13/2011 to 12/20/2018, with 415 weekly option expiration cycles (8 days to maturity (on average)).
We construct the excess return of OTM puts and OTM calls over weekly expiration cycles (as in Table~\ref{tab:weekly}).
These calculations are done at the ask option price.
Then we compute
\begin{align*}
&q_{t,{\tiny \mathrm{call}} }^{{T}_O}[k]{\Big |}_{\log(k)=3\%} ~~-~~ q_{t,{\tiny \mathrm{call}} }^{{T}_O}[k]{\Big |}_{\log(k)=1\%}&
&\mbox{(3\% OTM call minus 1\% OTM call)}& \mathrm{and} &\\
&q_{t,{\tiny \mathrm{call}} }^{{T}_O}[k]{\Big |}_{\log(k)=3\%} ~~-~~ q_{t,{\tiny \mathrm{put}} }^{{T}_O}[k]{\Big |}_{\log(k)=-3\%}.&
&\mbox{(3\% OTM call minus 3\% OTM put)}&  &
\end{align*}
Reported are the option risk premium differentials, partitioned according to ${\cal F}_{t} \in \mathfrak{s}_{\tiny \mbox{bad}}$, ${\cal F}_{t} \in \mathfrak{s}_{\tiny \mbox{normal}}$, and ${\cal F}_{t} \in \mathfrak{s}_{\tiny \mbox{good}}$. We employ
proxies for $\mathfrak{s}$ known at the beginning of the expiration cycle (as outlined in Table~\ref{tab:weekly}).
Reported are the two-sided $p$-values for these option risk premium differentials, relying on
the \citet*{NeweyWest:87} HAC estimator (with the lag selected
automatically).
We jointly bootstrap --- via an i.i.d, stationary, or
circular block bootstrap procedures ---
the returns of the options with
replacement and report the 95\% lower and upper confidence intervals. Bootstrap
confidence intervals --- shown as $\lfloor . \rfloor$ --- that bracket zero
imply that the disparity in the option risk premiums is indistinguishable from zero.
We perform 10,000 bootstraps. \vspace{-2mm}
\begin{center}
\setlength{\tabcolsep}{0.06in}
\begin{tabular}{ll l cc ccc ccc } \hline
          &           &           &           &           &           &           &           &           &           &                    \\
          &           &           &           &           &           \multicolumn{6}{c}{\textbf{Bootstrap procedure}}
 \\
 \cline{6-11}
          &           &           &  Estimate          &   NW[$p$]        &           & \multicolumn{1}{c}{\textbf{IID}} &           & \multicolumn{1}{c}{\textbf{Stationary}} &           & \multicolumn{1}{c}{\textbf{Circular}} \\
          &           &           &           &           &           & \multicolumn{1}{c}{$\lfloor$Lower~Upper$\rfloor$} &           & \multicolumn{1}{c}{$\lfloor$Lower~Upper$\rfloor$} &  & \multicolumn{1}{c}{$\lfloor$Lower~Upper$\rfloor$} \\
          \cline{7-7} \cline{9-9} \cline{11-11}
          &           &           &           &           &           &           &           &           &           &                    \\ \hline
          &           &           &           &           &           &           &           &           &           &                    \\
 & &  &        \multicolumn{8}{l}{\textbf{Panel A: Risk premium differentials}}  \\
 & &  &        \multicolumn{8}{l}{~~~~~~~~~~~~~~\textbf{(3\% OTM call minus 1\% OTM call)}}  \\
\multicolumn{1}{l}{Change in WEI} & L         &  $\mathfrak{s}_{\tiny \mbox{bad}}$         & -113      & \textbf{0.01}&     &$\lfloor$-181~ -51$\rfloor$       & & $\lfloor$-206 ~ -47$\rfloor$ & & $\lfloor$-207 ~ -47$\rfloor$ \\
                                 & M         &   $\mathfrak{s}_{\tiny \mbox{normal}}$        & -48       & \textbf{0.00}      &         & $\lfloor$-78~ -16$\rfloor$       &           & $\lfloor$-73~ -22$\rfloor$      &           & $\lfloor$-73~ -22$\rfloor$\\
          & H         &   $\mathfrak{s}_{\tiny \mbox{good}}$        & -51       & \textbf{0.00}      &           & $\lfloor$-85~ -16$\rfloor$       &           & $\lfloor$-68~ -34$\rfloor$       &           & $\lfloor$-67~  -35$\rfloor$ \\ \hline
\multicolumn{1}{l}{Quadratic Variation} & H         &  $\mathfrak{s}_{\tiny \mbox{bad}}$         & -26       & \textbf{0.09}      &           & $\lfloor$-55 ~ 1$\rfloor$         &           & $\lfloor$-50 ~ -1$\rfloor$        &           & $\lfloor$-51 ~ 0$\rfloor$ \\
          & M         &   $\mathfrak{s}_{\tiny \mbox{normal}}$        & -88       & \textbf{0.00}      &           & $\lfloor$-138 ~ -38$\rfloor$       &           & $\lfloor$-135~ -45$\rfloor$       &           & $\lfloor$-137 ~ -45$\rfloor$ \\
          & L         &    $\mathfrak{s}_{\tiny \mbox{good}}$       & -98       & \textbf{0.00}      &           & $\lfloor$-164 ~ -49$\rfloor$       &           & $\lfloor$-154 ~ -55$\rfloor$      &           & $\lfloor$-154~ -55$\rfloor$ \\ \hline
\multicolumn{1}{l}{Risk Reversal} & H         &    $\mathfrak{s}_{\tiny \mbox{bad}}$       & -116      & \textbf{0.00}      &           & $\lfloor$-179 ~ -65$\rfloor$       &           & $\lfloor$-170 ~ -71$\rfloor$       &           & $\lfloor$-172 ~ -71$\rfloor$ \\
          & M         &   $\mathfrak{s}_{\tiny \mbox{normal}}$        & -54       & \textbf{0.02}      &           & $\lfloor$-97  ~ -14$\rfloor$       &           & $\lfloor$-74 ~ -33$\rfloor$       &           & $\lfloor$-74 ~ -33$\rfloor$ \\
          & L         &    $\mathfrak{s}_{\tiny \mbox{good}}$       & -43       & \textbf{0.03}      &           & $\lfloor$-81 ~ -8$\rfloor$        &           & $\lfloor$-76 ~ -14$\rfloor$      &           & $\lfloor$-75 ~ -14$\rfloor$ \\ \hline
\multicolumn{1}{l}{Change in Volatility} & H         &     $\mathfrak{s}_{\tiny \mbox{bad}}$      & -56       & \textbf{0.00}      &           & $\lfloor$-88 ~ -27$\rfloor$       &           & $\lfloor$-83 ~ -32$\rfloor$       &           & $\lfloor$-83 ~ -32$\rfloor$ \\
          & M         &   $\mathfrak{s}_{\tiny \mbox{normal}}$        & -92       & \textbf{0.04}      &           & $\lfloor$-171 ~ -30$\rfloor$       &           & $\lfloor$-178 ~ -25$\rfloor$       &           & $\lfloor$-179 ~ -24$\rfloor$ \\
          & L         &    $\mathfrak{s}_{\tiny \mbox{good}}$       & -64       & \textbf{0.00}      &           & $\lfloor$-90 ~ -37$\rfloor$       &           & $\lfloor$-84 ~ -43$\rfloor$       &           & $\lfloor$-84 ~ -44$\rfloor$ \\ \hline
\multicolumn{1}{l}{Recent Market} & L         &   $\mathfrak{s}_{\tiny \mbox{bad}}$        & -62       & \textbf{0.03}      &           & $\lfloor$-119 ~ -20$\rfloor$       &           & $\lfloor$-116 ~ -22$\rfloor$       &           & $\lfloor$-115 ~ -23$\rfloor$ \\
          & M         &   $\mathfrak{s}_{\tiny \mbox{normal}}$        & -65       & \textbf{0.00}      &           & $\lfloor$-110 ~ -23$\rfloor$       &           & $\lfloor$-103 ~ -27$\rfloor$       &           & $\lfloor$-103 ~ -27$\rfloor$ \\
          & H         &  $\mathfrak{s}_{\tiny \mbox{good}}$         & -85       & \textbf{0.00}      &           & $\lfloor$-130 ~ -45$\rfloor$       &           & $\lfloor$-123 ~ -50$\rfloor$       &           & $\lfloor$-125 ~ -48$\rfloor$ \\
 \hline \hline
          &           &           &           &           &           &           &           &           &           &            \\
 & &  &        \multicolumn{8}{l}{\textbf{Panel B: Risk premium differentials}}  \\
 & &  &        \multicolumn{8}{l}{~~~~~~~~~~~~~~\textbf{(3\% OTM call minus 3\% OTM put)}}  \\
          \multicolumn{1}{l}{Change in WEI} & L         &  $\mathfrak{s}_{\tiny \mbox{bad}}$         & -9        & \textbf{0.80}      &           & $\lfloor$-85 ~ 54$\rfloor$        &           & $\lfloor$-70 ~ 46$\rfloor$        &           & $\lfloor$-70 ~ 46$\rfloor$ \\
          & M         &  $\mathfrak{s}_{\tiny \mbox{normal}}$         & 17        & \textbf{0.36}      &           & $\lfloor$-18 ~ 54$\rfloor$        &           & $\lfloor$-12 ~ 48$\rfloor$        &           & $\lfloor$-12 ~ 48$\rfloor$ \\
          & H         &  $\mathfrak{s}_{\tiny \mbox{good}}$         & -1        & \textbf{0.97}      &           & $\lfloor$-53 ~ 53$\rfloor$        &           & $\lfloor$-26 ~ 23$\rfloor$        &           & $\lfloor$-25 ~ 22$\rfloor$ \\ \hline
\multicolumn{1}{l}{Quadratic Variation} & H         & $\mathfrak{s}_{\tiny \mbox{bad}}$          & 20        & \textbf{0.58}      &           & $\lfloor$-52 ~ 85$\rfloor$        &           & $\lfloor$-31 ~ 68$\rfloor$        &           & $\lfloor$-31 ~ 68$\rfloor$ \\
          & M         &  $\mathfrak{s}_{\tiny \mbox{normal}}$         & -6        & \textbf{0.85}      &           & $\lfloor$-63 ~ 58$\rfloor$        &           & $\lfloor$-48 ~ 36$\rfloor$        &           & $\lfloor$-49 ~ 35$\rfloor$ \\
          & L         &   $\mathfrak{s}_{\tiny \mbox{good}}$        & -7        & \textbf{0.31}      &           & $\lfloor$-24 ~ 4$\rfloor$         &           & $\lfloor$-20 ~  2$\rfloor$         &           & $\lfloor$-20 ~ 2$\rfloor$ \\ \hline
\multicolumn{1}{l}{Risk Reversal} & H         &    $\mathfrak{s}_{\tiny \mbox{bad}}$       & -22       & \textbf{0.20}      &           & $\lfloor$-56 ~ 9$\rfloor$         &           & $\lfloor$-52 ~ 4$\rfloor$         &           & $\lfloor$-52 ~ 4$\rfloor$ \\
          & M         &   $\mathfrak{s}_{\tiny \mbox{normal}}$        & 58        & \textbf{0.02}      &           & $\lfloor$15 ~ 106$\rfloor$       &           & $\lfloor$21 ~ 99$\rfloor$        &           & $\lfloor$22 ~ 100$\rfloor$ \\
          & L         &  $\mathfrak{s}_{\tiny \mbox{good}}$         & -30       & \textbf{0.40}      &           & $\lfloor$-110 ~ 36$\rfloor$        &           & $\lfloor$-98 ~ 27$\rfloor$        &           & $\lfloor$-97 ~ 29$\rfloor$ \\ \hline
\multicolumn{1}{l}{Change in Volatility} & H         &   $\mathfrak{s}_{\tiny \mbox{bad}}$        & -14       & \textbf{0.66}      &           & $\lfloor$-87 ~ 44$\rfloor$        &           & $\lfloor$-71 ~ 37$\rfloor$        &           & $\lfloor$-69 ~ 34$\rfloor$ \\
          & M         & $\mathfrak{s}_{\tiny \mbox{normal}}$          & 28        & \textbf{0.33}      &           & $\lfloor$-28 ~ 86$\rfloor$        &           & $\lfloor$1 ~ 55$\rfloor$        &           & $\lfloor$3 ~ 54$\rfloor$ \\
          & L         & $\mathfrak{s}_{\tiny \mbox{good}}$          & -8        & \textbf{0.66}      &           &$\lfloor$-43 ~ 27$\rfloor$        &           & $\lfloor$-36 ~ 21$\rfloor$        &           & $\lfloor$-36 ~ 21$\rfloor$ \\ \hline
\multicolumn{1}{l}{Recent Market} & L         &   $\mathfrak{s}_{\tiny \mbox{bad}}$        & -8        & \textbf{0.78}      &           & $\lfloor$-75 ~ 46$\rfloor$        &           & $\lfloor$-65 ~ 40$\rfloor$        &           & $\lfloor$-61 ~ 37$\rfloor$ \\
          & M         &   $\mathfrak{s}_{\tiny \mbox{normal}}$        & -3        & \textbf{0.92}      &           & $\lfloor$-58 ~ 54$\rfloor$        &           & $\lfloor$-28 ~ 23$\rfloor$        &           & $\lfloor$-28 ~ 23$\rfloor$ \\
          & H         &   $\mathfrak{s}_{\tiny \mbox{good}}$        & 18        & \textbf{0.38}      &           & $\lfloor$-16 ~ 57$\rfloor$        &           & $\lfloor$-12 ~ 53$\rfloor$        &           & $\lfloor$-12 ~ 53$\rfloor$ \\
\hline \hline
\end{tabular}%
\end{center}
\end{table}


\newpage
\begin{table}[h!]
\footnotesize
\caption{\textbf{{Risk premiums for \emph{28-day} options on the S\&P 500 \emph{index}}}} \vspace{2mm}
\label{tab:equity_options}
The sample period is 01/22/1990 to 12/24/2018, with 348 option expiration cycles (28 days to maturity (on average)).
The 28-day options data on S\&P 500 index is from the CBOE.
We construct the excess return of OTM puts, OTM calls, and straddles (ATM and crash-neutral) over expiration cycles. These calculations are done at the ask option price.
The returns of a crash-neutral
straddle combines a long straddle position and a short 5\% OTM put position.
The following is the regression specification (analogously for puts and straddles):
\begin{align*}
&q_{t,{\tiny \mathrm{call}} }^{{T}_O}[k] =
\mu_{\{ {\cal F}_{t} \in \mathfrak{s}_{\tiny \mbox{bad}} \} } \mathbbm{1}_{\{ {\cal F}_{t} \in\mathfrak{s}_{\tiny \mbox{bad}} \}}
+ \mu_{\{ {\cal F}_{t} \in \mathfrak{s}_{\tiny \mbox{normal}} \} }
\mathbbm{1}_{\{ {\cal F}_{t} \in\mathfrak{s}_{\tiny \mbox{normal}} \}}
+ \mu_{\{ {\cal F}_{t} \in \mathfrak{s}_{\tiny \mbox{good}} \} }
\mathbbm{1}_{\{ {\cal F}_{t} \in\mathfrak{s}_{\tiny \mbox{good}} \}} +
\epsilon_{T_{O}}.&
\end{align*}
We use the following proxies for the variable $\mathfrak{s}$, known at the beginning of the expiration cycle.
\begin{description}
\item[-] \textit{Dividend Yield}$_{t}$: A high dividend yield (from Robert Shiller's website)
aligns with bad states. 

\item[-] \textit{Quadratic Variation}$_t$. Sum of daily squared (log) returns over
the \emph{prior} expiration cycle.

\item[-] \textit{Risk Reversal$_t$} ($\log(\frac{\mathrm{IV_t^{\mathrm{put}}}[k]}{\mathrm{IV_t^{\mathrm{call}}}[k]}$)). The 28-day implied volatility for puts (calls) uses $\log(k)$ equal to $-3\%$ (3\%).

\item[-] \textit{Change in Volatility$_t$} ($\log(\frac{\mathrm{IV}^{\mathrm{atm}}_{t}}{\mathrm{IV}^{\mathrm{atm}}_{t-1}})$).
The 28-day implied volatility (IV$_{t}$) is the average across
ATM puts and calls.


\item[-] \textit{Recent Market}$_{t}$: Log relative of the S\&P 500 index over the prior expiration cycle.

\end{description}
We indicate statistical significance at 1\%, 5\%, and 10\% by the superscripts ***, **, and  *, respectively,
where the $p$-values rely on
the \citet*{NeweyWest:87} HAC
estimator (with the lag selected automatically).
The reported put (respectively, call) delta is $-{\cal N}(-d_1)$ (respectively, ${\cal N}(d_1)$), where $d_1=  \frac{1}{ \sigma \sqrt{T_O-t}} \{ - \log k +r (T_O-t) + \frac{1}{2} \sigma^2 (T_O-t)\}$.
SD is the standard deviation, and
$\mathbbm{1}_{\{ q_{t, {T}_O} >0 \}}$ is the proportion (in \%) of option
positions that generate positive returns.
We tabulate the average open interest and trading volume, all observed on the first day of the option expiration cycle.
\begin{center}
\setlength{\tabcolsep}{0.053in}
\begin{tabular}{lll ccc ccc ccc c} \hline
          &           &           &           &           &           &           &           &           &           &           &           &  \\
          & &&\multicolumn{3}{c}{OTM puts on equity} &     & \multicolumn{3}{c}{OTM calls on equity} &  &  \multicolumn{2}{c}{Straddle}\\
          & & &\multicolumn{3}{c}{$\log(k)\times 100$} &   & \multicolumn{3}{c}{$\log(k)\times 100$}  &  &\multicolumn{2}{c}{on equity} \\
          \cline{4-6} \cline{8-10} \cline{12-13}
\multicolumn{1}{l}{Moneyness (\%)} &           &           & -5         & -3         & -1         &           & 1         & 3         & 5         &           & \multicolumn{1}{l}{ATM} & \multicolumn{1}{l}{Crash-} \\
\multicolumn{1}{l}{Delta (\%)} &           &           & -9         & -18         & -35         &           & 41         & 22         & 11         &           & & Neutral\\ \\
\multicolumn{1}{l}{Open interest ($\times 1,000$)} &           &           & 19.3      & 18.2      & 17.0      &           & 16.1      & 15.9      & 14.2      &           &           &  \\
  Volume ($\times 1,000$)       &           &           & 2.6       & 2.5       & 2.9       &           & 2.3       & 2.4       & 1.9       &           &           &  \\
          &           &           &           &           &           &           &           &           &           &           &           &  \\ \hline
          &           &           &           &           &           &           &           &           &           &           &           &  \\
\multicolumn{1}{l}{Dividend Yield} & \multicolumn{1}{l}{H} & \multicolumn{1}{l}{$\mathfrak{s}_{\tiny \mbox{bad}}$ } & \multicolumn{1}{c}{-83***} & \multicolumn{1}{c}{-75***} & \multicolumn{1}{c}{-64***} &           & \multicolumn{1}{c}{7} & \multicolumn{1}{c}{-4} & \multicolumn{1}{c}{-21} &           & \multicolumn{1}{c}{-26***} & \multicolumn{1}{c}{-4***} \\
\multicolumn{1}{l}{} & \multicolumn{1}{l}{M} &   $\mathfrak{s}_{\tiny \mbox{normal}}$         & \multicolumn{1}{c}{-61***} & \multicolumn{1}{c}{-42***} & \multicolumn{1}{c}{-34**} &           & \multicolumn{1}{c}{27} & \multicolumn{1}{c}{12} & \multicolumn{1}{c}{-19} &           & \multicolumn{1}{c}{-7} & \multicolumn{1}{c}{0} \\
\multicolumn{1}{l}{} &      L     &   $\mathfrak{s}_{\tiny \mbox{good}}$         & \multicolumn{1}{c}{-57***} & \multicolumn{1}{c}{-37**} & \multicolumn{1}{c}{-28*} &           & \multicolumn{1}{c}{-21**} & \multicolumn{1}{c}{-38***} & \multicolumn{1}{c}{-53***} &           & \multicolumn{1}{c}{-21***} & \multicolumn{1}{c}{-4***} \\
          &           &           &           &           &           &           &           &           &           &           &           &  \\
\multicolumn{1}{l}{Quadratic Variation} &    H       &  $\mathfrak{s}_{\tiny \mbox{bad}}$         & \multicolumn{1}{c}{-55***} & \multicolumn{1}{c}{-49***} & \multicolumn{1}{c}{-45***} &           & \multicolumn{1}{c}{7} & \multicolumn{1}{c}{13} & \multicolumn{1}{c}{29} &           & \multicolumn{1}{c}{-19***} & \multicolumn{1}{c}{-3*} \\
\multicolumn{1}{l}{} &    M       &   $\mathfrak{s}_{\tiny \mbox{normal}}$        & \multicolumn{1}{c}{-60***} & \multicolumn{1}{c}{-47***} & \multicolumn{1}{c}{-41***} &           & \multicolumn{1}{c}{9} & \multicolumn{1}{c}{-7} & \multicolumn{1}{c}{-22} &           & \multicolumn{1}{c}{-15***} & \multicolumn{1}{c}{-2**} \\
\multicolumn{1}{l}{} &   L        &  $\mathfrak{s}_{\tiny \mbox{good}}$         & \multicolumn{1}{l}{-87***} & \multicolumn{1}{l}{-58***} & \multicolumn{1}{l}{-40***} &           & \multicolumn{1}{c}{-3} & \multicolumn{1}{c}{-37*} & \multicolumn{1}{c}{-100***} &           & \multicolumn{1}{c}{-21***} & \multicolumn{1}{c}{-3**} \\
          &           &           &           &           &           &           &           &           &           &           &           &  \\
\multicolumn{1}{l}{Risk Reversal} &   H        &    $\mathfrak{s}_{\tiny \mbox{bad}}$       & \multicolumn{1}{c}{-71***} & \multicolumn{1}{c}{-50***} & \multicolumn{1}{c}{-39***} &           & \multicolumn{1}{c}{20} & \multicolumn{1}{c}{-4} & \multicolumn{1}{c}{-37} &           & \multicolumn{1}{c}{-14**} & \multicolumn{1}{c}{-1} \\
\multicolumn{1}{l}{} &    M       &  $\mathfrak{s}_{\tiny \mbox{normal}}$         & \multicolumn{1}{l}{-83***} & \multicolumn{1}{l}{-67***} & \multicolumn{1}{l}{-59***} &           & \multicolumn{1}{c}{15} & \multicolumn{1}{c}{7} & \multicolumn{1}{c}{-16} &           & \multicolumn{1}{c}{-19***} & \multicolumn{1}{c}{-2} \\
\multicolumn{1}{l}{} &     L      &     $\mathfrak{s}_{\tiny \mbox{good}}$      & \multicolumn{1}{c}{-45**} & \multicolumn{1}{c}{-35*} & \multicolumn{1}{c}{-26} &           & \multicolumn{1}{c}{-25***} & \multicolumn{1}{c}{-42***} & \multicolumn{1}{c}{-56***} &           & \multicolumn{1}{c}{-21***} & \multicolumn{1}{c}{-4***} \\
          &           &           &           &           &           &           &           &           &           &           &           &  \\
\multicolumn{1}{l}{Change in Volatility} &   H        &   $\mathfrak{s}_{\tiny \mbox{bad}}$        & \multicolumn{1}{c}{-52***} & \multicolumn{1}{c}{-33**} & \multicolumn{1}{c}{-25*} &           & \multicolumn{1}{c}{12} & \multicolumn{1}{c}{10} & \multicolumn{1}{c}{-6} &           & \multicolumn{1}{c}{-8} & \multicolumn{1}{c}{-1} \\
\multicolumn{1}{l}{} &     M      &   $\mathfrak{s}_{\tiny \mbox{normal}}$        & \multicolumn{1}{c}{-83***} & \multicolumn{1}{c}{-62***} & \multicolumn{1}{c}{-53***} &           & \multicolumn{1}{c}{2} & \multicolumn{1}{c}{-18} & \multicolumn{1}{c}{-29} &           & \multicolumn{1}{c}{-25***} & \multicolumn{1}{c}{-3***} \\
\multicolumn{1}{l}{} &    L       &  $\mathfrak{s}_{\tiny \mbox{good}}$         & \multicolumn{1}{l}{-65***} & \multicolumn{1}{l}{-57***} & \multicolumn{1}{l}{-46***} &           & \multicolumn{1}{c}{-4} & \multicolumn{1}{c}{-33*} & \multicolumn{1}{c}{-81***} &           & \multicolumn{1}{c}{-22***} & \multicolumn{1}{c}{-3***} \\
          &           &           &           &           &           &           &           &           &           &           &           &  \\
\multicolumn{1}{l}{Recent Market} &  L         &  $\mathfrak{s}_{\tiny \mbox{bad}}$         & \multicolumn{1}{c}{-58***} & \multicolumn{1}{c}{-48***} & \multicolumn{1}{c}{-43***} &           & \multicolumn{1}{c}{14} & \multicolumn{1}{c}{7} & \multicolumn{1}{c}{8} &           & \multicolumn{1}{c}{-17**} & \multicolumn{1}{c}{-2} \\
\multicolumn{1}{l}{} &    M       & $\mathfrak{s}_{\tiny \mbox{normal}}$          & \multicolumn{1}{c}{-75***} & \multicolumn{1}{c}{-48***} & \multicolumn{1}{c}{-38***} &           & \multicolumn{1}{c}{17} & \multicolumn{1}{c}{11} & \multicolumn{1}{c}{-18} &           & \multicolumn{1}{c}{-13**} & \multicolumn{1}{c}{-2*} \\
\multicolumn{1}{l}{} &     H      &  $\mathfrak{s}_{\tiny \mbox{good}}$         & \multicolumn{1}{c}{-68***} & \multicolumn{1}{c}{-58***} & \multicolumn{1}{c}{-44***} &           & \multicolumn{1}{c}{-18} & \multicolumn{1}{c}{-48***} & \multicolumn{1}{c}{-82***} &           & \multicolumn{1}{c}{-25***} & \multicolumn{1}{c}{-4***} \\
          &           &           &           &           &           &           &           &           &           &           &           &  \\ \hline
          &           &           &           &           &           &           &           &           &           &           &           &  \\
\multicolumn{1}{l}{\textbf{Unconditional}} &           &   Average        & \multicolumn{1}{c}{-67} & \multicolumn{1}{c}{-51} & \multicolumn{1}{c}{-42} &           & \multicolumn{1}{c}{4} & \multicolumn{1}{c}{-10} & \multicolumn{1}{c}{-31} &           & \multicolumn{1}{c}{-19} & \multicolumn{1}{c}{-3} \\
\multicolumn{1}{l}{\textbf{Estimates}} &           &    SD       & \multicolumn{1}{c}{151} & \multicolumn{1}{c}{159} & \multicolumn{1}{c}{146} &           & \multicolumn{1}{c}{151} & \multicolumn{1}{c}{253} & \multicolumn{1}{c}{360} &           & \multicolumn{1}{c}{72} & \multicolumn{1}{c}{14} \\
\multicolumn{1}{l}{} &           &     $\mathbbm{1}_{\{ q_{t, {T}_O} >0 \}}$      & \multicolumn{1}{c}{6\%} & \multicolumn{1}{c}{11\%} & \multicolumn{1}{c}{16\%} &           & \multicolumn{1}{c}{38\%} & \multicolumn{1}{c}{19\%} & \multicolumn{1}{c}{8\%} &           & \multicolumn{1}{c}{30\%} & \multicolumn{1}{c}{38\%} \\
          &           &           &           &           &           &           &           &           &           &           &           &  \\
\hline
\end{tabular}%
\end{center}
\end{table}

\newpage
\begin{table}[h!]
\footnotesize
\caption{\textbf{{
Risk premiums for \emph{28-day} options on the S\&P 500 \emph{futures}}}} \vspace{2mm}
\label{tab:table2}
The sample period is 01/18/1988 to 05/23/2016, with 341 option expiration cycles (28 days to maturity (on average)).
These one-month futures options were discontinued and only the three-month
options were traded after that.
We construct the excess return of OTM puts, OTM calls, and straddles (ATM and crash-neutral) over option expiration cycles. The option settlement price is provided by the CME.
The returns of a crash-neutral straddle combines a long straddle position and a short 5\% OTM put position.
The following is the regression specification (analogously for puts and straddles):
\begin{align*}
&q_{t,{\tiny \mathrm{call}} }^{{T}_O}[k] =
\mu_{\{ {\cal F}_{t} \in \mathfrak{s}_{\tiny \mbox{bad}} \} } \mathbbm{1}_{\{ {\cal F}_{t} \in\mathfrak{s}_{\tiny \mbox{bad}} \}}
+ \mu_{\{ {\cal F}_{t} \in \mathfrak{s}_{\tiny \mbox{normal}} \} }
\mathbbm{1}_{\{ {\cal F}_{t} \in\mathfrak{s}_{\tiny \mbox{normal}} \}}
+ \mu_{\{ {\cal F}_{t} \in \mathfrak{s}_{\tiny \mbox{good}} \} }
\mathbbm{1}_{\{ {\cal F}_{t} \in\mathfrak{s}_{\tiny \mbox{good}} \}} +
\epsilon_{T_{O}}.&
\end{align*}
The proxies for the variable $\mathfrak{s}$, which are known at the beginning of the option expiration cycle, are as described in
the note to Table~\ref{tab:equity_options}.
%
%
%
%
%
%
%
We indicate statistical significance at 1\%, 5\%, and 10\% by the superscripts ***, **, and *, respectively,
where the $p$-values rely on
the \citet*{NeweyWest:87} HAC estimator (with the lag selected automatically).
 The reported put (respectively, call) delta is $-e^{- r (T_O-t)} {\cal N}(-d_1)$ (respectively, $e^{- r (T_O-t)} {\cal N}(d_1)$),
where $d_1=  \frac{1}{ \sigma \sqrt{T_O-t}} \{ - \log k + \frac{1}{2} \sigma^2 (T_O-t)\}$.
SD is the standard deviation,
and $\mathbbm{1}_{\{ q_{t, {T}_O} >0 \}}$ is the proportion (in \%) of option positions
that generate positive returns. 
We tabulate the average open interest and trading volume, all observed on the first day of the option expiration cycle.

\begin{center}
\setlength{\tabcolsep}{0.065in}
\begin{tabular}{lll ccc ccc ccc c} \hline
          &           &           &           &           &           &           &           &           &           &           &           &  \\
          & &&\multicolumn{3}{c}{OTM puts on futures} &     & \multicolumn{3}{c}{OTM calls on futures} &  &  \multicolumn{2}{c}{Straddle}\\
          & & &\multicolumn{3}{c}{$\log(k)\times 100$} &   & \multicolumn{3}{c}{$\log(k)\times 100$}  &  &\multicolumn{2}{c}{on  futures} \\
          \cline{4-6} \cline{8-10} \cline{12-13}
\multicolumn{1}{l}{Moneyness (\%)} &           &           & -5         & -3         & -1         &           & 1        & 3        & 5        &           & \multicolumn{1}{l}{ATM} & \multicolumn{1}{l}{Crash-} \\
\multicolumn{1}{l}{Delta (\%)} &           &           & -9         & -18         & -35         &           & 41        & 22         & 11         &           & & Neutral\\ \\
\multicolumn{1}{l}{Open Interest} &           &          & 1708      & 1544      & 1254      &           & 1560      & 1974      & 1792      &           &           &  \\
Volume          &           &           & 260       & 218       & 204       &           & 142       & 291       & 255       &           &           &  \\
          &           &           &           &           &           &           &           &           &           &           &           &  \\ \hline
          &           &           &           &           &           &           &           &           &           &           &           &  \\
\multicolumn{1}{l}{Dividend Yield } & \multicolumn{1}{l}{H} & \multicolumn{1}{l}{$\mathfrak{s}_{\tiny \mbox{bad}}$} & -77***    & -70***    & -60***    &           & 5         & -15       & -39       &           & -30***    & -4*** \\
\multicolumn{1}{l}{} &     M      &     $\mathfrak{s}_{\tiny \mbox{normal}}$      & -57***    & -41***    & -31**     &           & 23        & 17        & 31        &           & -4        & 1 \\
\multicolumn{1}{l}{} &    L       &     $\mathfrak{s}_{\tiny \mbox{good}}$      & -51***    & -33*      & -25*      &           & -19**     & -35***    & -46***    &           & -18***    & -2** \\
          &           &           &           &           &           &           &           &           &           &           &           &  \\
\multicolumn{1}{l}{Quadratic Variation} &     H      &    $\mathfrak{s}_{\tiny \mbox{bad}}$       & -57***    & -53***    & -45***    &           & 12        & 19        & 25        &           & -16**     & -1 \\
\multicolumn{1}{l}{} &      M     &    $\mathfrak{s}_{\tiny \mbox{normal}}$       & -44**     & -33*      & -32**     &           & 12        & 1         & 10        &           & -10       & -1 \\
\multicolumn{1}{l}{} &      L     &    $\mathfrak{s}_{\tiny \mbox{good}}$       & -84***    & -58***    & -40***    &           & -15       & -52***    & -90***    &           & -25***    & -3*** \\
          &           &           &           &           &           &           &           &           &           &           &           &  \\
\multicolumn{1}{l}{Risk Reversal} &   H        & $\mathfrak{s}_{\tiny \mbox{bad}}$          & -52***    & -31*      & -22       &           & 8         & -4        & -7        &           & -7        & 0 \\
\multicolumn{1}{l}{} &      M     &    $\mathfrak{s}_{\tiny \mbox{normal}}$       & -88***    & -73***    & -60***    &           & 15        & 3         & 7         &           & -21***    & -1 \\
\multicolumn{1}{l}{} &      L     &    $\mathfrak{s}_{\tiny \mbox{good}}$       & -42*      & -38*      & -34*      &           & -18       & -35***    & -63***    &           & -24***    & -4*** \\
          &           &           &           &           &           &           &           &           &           &           &           &  \\
\multicolumn{1}{l}{Change in Volatility} &     H      &  $\mathfrak{s}_{\tiny \mbox{bad}}$         & -60***    & -43***    & -34**     &           & 6         & 15        & 46        &           & -15**     & -1 \\
\multicolumn{1}{l}{} &     M      &     $\mathfrak{s}_{\tiny \mbox{normal}}$      & -67***    & -46***    & -39***    &           & -1        & -11       & -29       &           & -20***    & -2 \\
\multicolumn{1}{l}{} &      L     &    $\mathfrak{s}_{\tiny \mbox{good}}$       & -58***    & -56***    & -44***    &           & 4         & -36**     & -73***    &           & -17***    & -2 \\
          &           &           &           &           &           &           &           &           &           &           &           &  \\
\multicolumn{1}{l}{Recent Market} &   L        &    $\mathfrak{s}_{\tiny \mbox{bad}}$       & -55***    & -42**     & -38***    &           & 22        & 18        & 19        &           & -11       & 0 \\
\multicolumn{1}{l}{} &    M       &   $\mathfrak{s}_{\tiny \mbox{normal}}$        & -71***    & -51***    & -40***    &           & 3         & -1        & 2         &           & -19***    & -2 \\
\multicolumn{1}{l}{} &     H      &   $\mathfrak{s}_{\tiny \mbox{good}}$        & -59***    & -52***    & -39***    &           & -15       & -49***    & -76***    &           & -22***    & -3** \\
          &           &           &           &           &           &           &           &           &           &           &           &  \\ \hline
          &           &           &           &           &           &           &           &           &           &           &           &  \\
\multicolumn{1}{l}{\textbf{Unconditional}} &           &     Average      & -62       & -48       & -39       &           & 3         & -11       & -18       &           & -17       & -2 \\
\multicolumn{1}{l}{\textbf{Estimates}} &           &      SD     & 167       & 166       & 148       &           & 145       & 241       & 441       &           & 74        & 14 \\
\multicolumn{1}{l}{} &           &     $\mathbbm{1}_{\{ q_{t, {T}_O} >0 \}}$       & 6\%         & 12\%        & 17\%        &           & 37\%        & 20\%        & 7\%         &           & 32\%        & 42\% \\
          &           &           &           &           &           &           &           &           &           &           &           &  \\ \hline
\end{tabular}
\end{center}
\end{table}

\newpage
\begin{table}[h!]
\footnotesize
\caption{\textbf{{
Risk premiums for \emph{88-day} options on the S\&P 500 \emph{futures}}}} \vspace{2mm}
\label{tab:opretspfutx}
The sample period is 03/21/1988 to 03/18/2019,
with 125 option expiration cycles (88 days to maturity (on average)). We construct the excess return of OTM puts, OTM calls, and straddles (ATM and crash-neutral) over option expiration cycles.
The option settlement price is provided by the CME.
The returns of a crash-neutral
straddle combines a long straddle position and a short 12\% OTM put position.
The following is the
regression specification (analogously for puts and straddles):
\begin{align*}
&q_{t,{\tiny \mathrm{call}} }^{{T}_O}[k] =
\mu_{\{ {\cal F}_{t} \in \mathfrak{s}_{\tiny \mbox{bad}} \} } \mathbbm{1}_{\{ {\cal F}_{t} \in\mathfrak{s}_{\tiny \mbox{bad}} \}}
+ \mu_{\{ {\cal F}_{t} \in \mathfrak{s}_{\tiny \mbox{normal}} \} }
\mathbbm{1}_{\{ {\cal F}_{t} \in\mathfrak{s}_{\tiny \mbox{normal}} \}}
+ \mu_{\{ {\cal F}_{t} \in \mathfrak{s}_{\tiny \mbox{good}} \} }
\mathbbm{1}_{\{ {\cal F}_{t} \in\mathfrak{s}_{\tiny \mbox{good}} \}} +
\epsilon_{T_{O}}.&
\end{align*}
The proxies for the variable $\mathfrak{s}$, which are known at the beginning of the option expiration cycle, are as described in
the note to Table~\ref{tab:equity_options}.
%
%
%
%
%
%
%
We indicate statistical significance at 1\%, 5\%, and 10\% by the superscripts ***, **, and *, respectively,
where the $p$-values rely on
the \citet*{NeweyWest:87} HAC estimator (with the lag selected automatically).
 The reported put (respectively, call) delta is $-e^{- r (T_O-t)} {\cal N}(-d_1)$ (respectively, $e^{- r (T_O-t)} {\cal N}(d_1)$),
where $d_1=  \frac{1}{ \sigma \sqrt{T_O-t}} \{ - \log k + \frac{1}{2} \sigma^2 (T_O-t)\}$.
SD is the standard deviation, and $\mathbbm{1}_{\{ q_{t, {T}_O} >0 \}}$ is the proportion (in \%) of option positions
that generate positive returns.
We tabulate the average open interest and trading volume, all observed on the first day of the option expiration cycle.
\vspace{-2mm}
\begin{center}
\setlength{\tabcolsep}{0.058in}
\begin{tabular}{lll ccc ccc ccc c} \hline
          &           &           &           &           &           &           &           &           &           &           &           &  \\
          & &&\multicolumn{3}{c}{OTM puts on futures} &     & \multicolumn{3}{c}{OTM calls on futures} &  &  \multicolumn{2}{c}{Straddle}\\
          & & &\multicolumn{3}{c}{$\log(k)\times 100$} &   & \multicolumn{3}{c}{$\log(k)\times 100$}  &  &\multicolumn{2}{c}{on  futures} \\
          \cline{4-6} \cline{8-10} \cline{12-13}
\multicolumn{1}{l}{Moneyness (\%)} &           &           & -12         & -8         & -3         &           & 3         & 8         & 12         &           & \multicolumn{1}{l}{ATM} & \multicolumn{1}{l}{Crash-} \\
\multicolumn{1}{l}{Delta (\%)} &           &           & -5         & -11         & -30         &           & 32         & 13         & 6         &           & & Neutral\\ \\
\multicolumn{1}{l}{Open Interest} &           &           & 969       & 1047      & 959       &           & 839       & 577       & 653       &           &           &  \\
\multicolumn{1}{l}{Volume}          &           &           & 43        & 76        & 78        &           & 50        & 44        & 26        &           &           &  \\
          &           &           &           &           &           &           &           &           &           &           &           &  \\ \hline
          &           &           &           &           &           &           &           &           &           &           &           &  \\
\multicolumn{1}{l}{Dividend Yield} & \multicolumn{1}{l}{H} & \multicolumn{1}{l}{$\mathfrak{s}_{\tiny \mbox{bad}}$} & -73***    & -70***    & -70***    &           & 20        & -48*      & -76***    &           & -22*      & -5 \\
\multicolumn{1}{l}{} &       M    &    $\mathfrak{s}_{\tiny \mbox{normal}}$       & -95***    & -90***    & -76***    &           & 43        & -26       & -34       &           & -14*      & -1 \\
\multicolumn{1}{l}{} &     L      &  $\mathfrak{s}_{\tiny \mbox{good}}$         & -38       & -24       & -21       &           & -41***    & -78***    & -88***    &           & -23*      & -7** \\
\multicolumn{1}{l}{} &           &           &  &      &        &           &        &       &       &           &           &  \\

\multicolumn{1}{l}{Quadratic Variation} &    H       &  $\mathfrak{s}_{\tiny \mbox{bad}}$         & -49**     & -41       & -39       &           & -1        & 1         & -13       &           & -17*      & -3 \\
\multicolumn{1}{l}{} &    M       &    $\mathfrak{s}_{\tiny \mbox{normal}}$       & -83***    & -73***    & -61***    &           & 19        & -58**     & -90***    &           & -21**     & -4 \\
\multicolumn{1}{l}{} &     L      &   $\mathfrak{s}_{\tiny \mbox{good}}$        & -74***    & -71**     & -68***    &           & 4         & -94***    & -94***    &           & -21       & -6 \\
          &           &           &           &           &           &           &           &           &           &           &           &  \\
\multicolumn{1}{l}{Risk Reversal} &   H        &   $\mathfrak{s}_{\tiny \mbox{bad}}$        & -75***    & -72**     & -70***    &           & 18        & -96***    & -101***   &           & -20*      & -5 \\
\multicolumn{1}{l}{} &   M        &   $\mathfrak{s}_{\tiny \mbox{normal}}$        & -85***    & -75***    & -61***    &           & -1        & -39       & -83***    &           & -21*      & -4 \\
\multicolumn{1}{l}{} &   L        &   $\mathfrak{s}_{\tiny \mbox{good}}$        & -45*      & -37       & -37       &           & 6         & -14       & -11       &           & -17       & -4 \\
          &           &           &           &           &           &           &           &           &           &           &           &  \\
\multicolumn{1}{l}{Change in Volatility} &    H       &   $\mathfrak{s}_{\tiny \mbox{bad}}$        & -66***    & -58***    & -52**     &           & 19        & -31       & -53       &           & -15       & -2 \\
\multicolumn{1}{l}{} &     M      &   $\mathfrak{s}_{\tiny \mbox{normal}}$        & -65**     & -56*      & -48*      &           & 2         & -65***    & -82***    &           & -17       & -4 \\
\multicolumn{1}{l}{} &    L       &   $\mathfrak{s}_{\tiny \mbox{good}}$        & -75***    & -71***    & -68***    &           & 2         & -55**     & -62**     &           & -27***    & -6** \\
          &           &           &           &           &           &           &           &           &           &           &           &  \\
\multicolumn{1}{l}{Recent Market} &    L       &  $\mathfrak{s}_{\tiny \mbox{bad}}$         & -66***    & -54**     & -40*      &           & 11        & -29       & -51       &           & -10       & 0 \\
\multicolumn{1}{l}{} &       M    &    $\mathfrak{s}_{\tiny \mbox{normal}}$       & -66***    & -59**     & -55**     &           & -3        & -57**     & -71***    &           & -25**     & -6 \\
\multicolumn{1}{l}{} &     H      &   $\mathfrak{s}_{\tiny \mbox{good}}$        & -74***    & -71**     & -73***    &           & 15        & -64***    & -76***    &           & -23**     & -7* \\
          &           &           &           &           &           &           &           &           &           &           &           &  \\ \hline
          &           &           &           &           &           &           &           &           &           &           &           &  \\
\multicolumn{1}{l}{\textbf{Unconditional}} &           & Average          & -69       & -62       & -56       &           & 7         & -51       & -66       &           & -20       & -5 \\
\multicolumn{1}{l}{\textbf{Estimates}} &           &     SD      & 143       & 158       & 139       &           & 172       & 153       & 171       &           & 68        & 23 \\
\multicolumn{1}{l}{} &           &   $\mathbbm{1}_{\{ q_{t, {T}_O} >0 \}}$         & 6\%         & 6\%         & 11\%        &           & 31\%        & 11\%        & 5\%         &           & 34\%        & 38\% \\
          &           &           &           &           &           &           &           &           &           &           &           &  \\ \hline
\end{tabular}%
\end{center}
\end{table}


\newpage





\newpage
\thispagestyle{empty}
\clearpage
\begin{center}
{\Large{Dark Matter in (Volatility and) Equity Option Risk Premiums}} \\
\vspace{0.04in}
\textbf{\underline{Internet Appendix: Not Intended for Publication}}
\end{center}
\begin{center}
\textbf{Abstract}
\end{center}

Section~\ref{appsec:dispersion} outlines how the risk premium on volatility
uncertainty relates to the risk premiums on local time and jumps crossing the strike.

Section~\ref{app:jumps_across} develops the
analysis that links jump model assumptions under $\mathbb{P}$ and
$\mathbb{Q}$ to the risk premium for jumps crossing
the strike over small $T_{O}-t$. Our focus here is on the setting of a general
semimartingale that admits jumps. Our analysis incorporates the models
of \citet*{Merton:76}, \citet*{Kou:2002}, and \citet*{DuffiePanSingleton:2000}.

Section~\ref{app:var_jumps}
provides the expressions for the local time risk premiums when there are unspanned risks,
dichotomized in the form of diffusive volatility risks and jump volatility risks.

%





\thispagestyle{empty}

\thispagestyle{empty}
\newpage
\setcounter{page}{1}
\renewcommand{\thefootnote}{\arabic{footnote}}
\setcounter{footnote}{0}

\setcounter{section}{0}
\renewcommand{\thesection}{\Roman{section}}
\renewcommand{\thesubsection}{\thesection.\arabic{subsection}}


                                                        \setcounter{equation}{0}
                                                        \renewcommand{\theequation}{IA-\arabic{equation}}
\numberwithin{table}{section}
\numberwithin{theorem}{section}
\numberwithin{figure}{section}

                                                        \setcounter{equation}{0}
                                                        \renewcommand{\theequation}{I\arabic{equation}}


\section{Risk premium for volatility
uncertainty
and its link to risk premiums on (i) local time and (ii)
jumps crossing the strike}
\label{appsec:dispersion}
Consider the time $T_O$ payoff $\{ \log G_{T_O} \}^2= \{\log \frac{F_{T_O}^{T_F}}{F_t^{T_F}}\}^2$. This payoff represents \emph{volatility
uncertainty}.
Define the function
\begin{equation}
\mathfrak{f}[K] ~\equiv~ \frac{2}{K^2} ( 1 - \log \frac{K}{F_{t}^{T_F}}).
\end{equation}
Since  $\{\log \frac{F_{T_O}^{T_F}}{F_t^{T_F}}\}^2 \in {\cal C}^2$,
it may be expressed
as
\begin{eqnarray}
\big\{\log \frac{F_{T_O}^{T_F}}{F_t^{T_F}}\big\}^2 & = &
 \int\limits_0^{F_{t}^{T_F}} \mathfrak{f}[K] \max(K-F_{T_O}^{T_F},0)\,dK  ~+~
 \int\limits_{F_{t}^{T_F}}^\infty \mathfrak{f}[K] \max(F_{T_O}^{T_F}-K,0)\,dK~~\mbox{ \, \, }\label{eq:asb1} \\
 & = &
 \int\limits_0^{1} \omega[k] \max(k - \frac{F_{T_O}^{T_F}}{F_{t}^{T_F}},0)\,dk~+~
\int\limits_{1}^\infty \omega[k] \max(\frac{F_{T_O}^{T_F}}{F_{t}^{T_F}}- k,0)\,dk,
~~\mbox{ \, \, }\label{eq:asb3} \\
\mbox{ where \, }~ \omega[k] & \equiv & \frac{2}{k^2} ( 1 - \log k ), ~~\mbox{ \, with \, }~~k \, = \, \frac{K}{F_{t}^{T_F}},~~\mbox{ \, and \, }~ d k \, = \, \frac{d K}{F_{t}^{T_F}}.~~~\mbox{ \, }
\end{eqnarray}
We can now substitute Tanaka's formula for semimartingales into the expression for $\max(k-G_{T_O},0)$ and
$\max(G_{T_O}- k,0)$ in the right-hand side of
(\ref{eq:asb3}). Therefore, we obtain the following:
\begin{eqnarray}
\big\{\log \frac{F_{T_O}^{T_F}}{F_t^{T_F}}\big\}^2 &=& \int\limits_0^{1} \omega[k] \max(k - G_{T_O},0) dk
+  \int\limits_{1}^\infty \omega[k] \max(G_{T_O}- k,0) dk
~~\mbox{ \, \, }
\nonumber
\\
&=& \int\limits_0^{1} \omega[k] \{ - \int_{t}^{T_O} \mathbbm{1}_{\{G_{\ell-} < k\}} \,dG_{\ell}
~+~ \mathbb{L}^{T_O}_t[k] ~+~ c_t^{T_O}[k] ~+~ d_t^{T_O}[k] \} dk \nonumber \\
&& +  \int\limits_{1}^\infty \omega[k] \{\int_{t}^{T_O} \mathbbm{1}_{\{G_{\ell-} > k\}} \,dG_{\ell}
~+~ \mathbb{L}^{T_O}_t[k] ~+~ a_t^{T_O}[k] ~+~ b_t^{T_O}[k] \} dk
~~\mbox{ \, \, }\label{eq:asb5} \\
&=& \int_{t}^{T_O} \big( \int\limits_{1}^\infty \omega[k] \, \mathbbm{1}_{\{G_{\ell-} > k\}} \, dk - \int\limits_0^{1} \omega[k] \, \mathbbm{1}_{\{G_{\ell-} < k\}} \ dk \big) \, dG_{\ell}
\nonumber \\
&&~+~ \int\limits_{0}^\infty \omega[k]\,\mathbb{L}^{T_O}_t[k]\,dk  \nonumber \\
&& ~+~
\int\limits_{0}^{1} \omega[k]\, ( c_t^{T_O}[k] + d_t^{T_O}[k] ) \,dk
 ~+~
\int\limits_{1}^\infty \omega[k]\, ( a_t^{T_O}[k] + b_t^{T_O}[k]) \,dk.
~\mbox{ \, }
~~\mbox{ \, \, }\label{eq:asb6}
\end{eqnarray}

Using $\mathbb{P}$ and $\mathbb{Q}$ measure expectations, we consequently obtain the following:
\begin{eqnarray}
\mathbb{E}_t^{\mathbb{P}} ( \big\{\log \frac{F_{T_O}^{T_F}}{F_t^{T_F}}\big\}^2 )
&=&
-~\mathbb{E}_t^{\mathbb{P}} ( \int_{t}^{T_O} \big\{ -\int\limits_{1}^\infty \omega[k] \, \mathbbm{1}_{\{G_{\ell-} > k\}} \, dk +
\int\limits_0^{1} \omega[k] \, \mathbbm{1}_{\{G_{\ell-} < k\}} \ dk \big\} \, dG_{\ell} )
~~\mbox{ \, \, \, \, }~~ \nonumber \\
&+&
\int\limits_{0}^\infty \omega[k] \, \mathbb{E}_t^{\mathbb{P}} ( \mathbb{L}^{T_O}_t[k] ) \, dk   \nonumber \\
&+& ~
\int\limits_{0}^1 \omega[k]\, \mathbb{E}_t^{\mathbb{P}}( c_t^{T_O}[k]+ d_t^{T_O}[k] ) \,dk  +
\int\limits_{1}^\infty \omega[k]\, \mathbb{E}_t^{\mathbb{P}}( a_t^{T_O}[k] + b_t^{T_O}[k]) \,dk.
\\
\mathbb{E}_t^{\mathbb{Q}} ( \big\{\log \frac{F_{T_O}^{T_F}}{F_t^{T_F}}\big\}^2 ) &=&
\int\limits_{0}^\infty \omega[k] \, \mathbb{E}_t^{\mathbb{Q}} ( \mathbb{L}^{T_O}_t[k] ) \, dk \nonumber \\
&+&
\int\limits_{0}^1 \omega[k]\, \mathbb{E}_t^{\mathbb{Q}}( c_t^{T_O}[k] + d_t^{T_O}[k] ) \,dk
+ \int\limits_{0}^\infty \omega[k]\, \mathbb{E}_t^{\mathbb{Q}}( a_t^{T_O}[k] + b_t^{T_O}[k] ) \,dk.
\label{eq:asb8}
\end{eqnarray}
This is because
$\mathbb{E}_t^{\mathbb{Q}} ( \int_{t}^{T_O} \big\{\int\limits_{1}^\infty \omega[k] \, \mathbbm{1}_{\{G_{\ell-} > k\}} dk
\big\} dG_{\ell} ) =0$
and $\mathbb{E}_t^{\mathbb{Q}} ( \{\int\limits_0^{1} \omega[k]\, \mathbbm{1}_{\{G_{\ell-} < k\}} \ dk \big\} dG_{\ell} ) =0$.

The expression for the risk premium for volatility uncertainty
is as follows:
\begin{eqnarray}
& & \underbrace{\mathbb{E}_t^{\mathbb{P}} ( \big\{\log \frac{F_{T_O}^{T_F}}{F_t^{T_F}}\big\}^2 )
- \mathbb{E}_t^{\mathbb{Q}} ( \big\{\log \frac{F_{T_O}^{T_F}}{F_t^{T_F}}\big\}^2 )}_{\tiny \mbox{risk~premium~for~volatility~uncertainty}}
~=~  -\mathrm{e}_t^{\mathbb{P}} +
\int\limits_{0}^\infty ~ \omega[k] \, \underbrace{\{ \mathbb{E}_t^{\mathbb{P}} ( \mathbb{L}^{T_O}_t[k] )-\mathbb{E}_t^{\mathbb{Q}} ( \mathbb{L}^{T_O}_t[k] )\}}_{\tiny \mbox{risk~premium~for~local~time}} \,dk   \nonumber \\
&&+~ \int\limits_{0}^1 \omega[k]\, \underbrace{ \{
\mathbb{E}_t^{\mathbb{P}}( c_t^{T_O}[k] ~+~ d_t^{T_O}[k] )
- \mathbb{E}_t^{\mathbb{Q}}( c_t^{T_O}[k] ~+~ d_t^{T_O}[k] ) \}}_{\tiny \mbox{risk~premium~for~jumps~crossing~the~strike}~(k<1)} \, dk \nonumber \\
&&+~ \int\limits_{1}^\infty \omega[k]\, \underbrace{\{
\mathbb{E}_t^{\mathbb{P}}( a_t^{T_O}[k] ~+~ b_t^{T_O}[k] )
- \mathbb{E}_t^{\mathbb{Q}}( a_t^{T_O}[k] ~+~ b_t^{T_O}[k]) \}}_{\tiny \mbox{risk~premium~for~jumps~crossing~the~strike}~(k>1)} \, dk,
\nonumber \\
& &~~~~ \mbox{ where \, \, }~ \mathrm{e}_t^{\mathbb{P}} ~=~ \mathbb{E}_t^{\mathbb{P}} ( \int_{t}^{T_O}
\big\{ -\int\limits_{1}^\infty \omega[k] \, \mathbbm{1}_{\{G_{\ell-} > k\}} \, dk +  \int\limits_0^{1} \omega[k] \, \mathbbm{1}_{\{G_{\ell-} < k\}} \ dk \big\} \, dG_{\ell} ).
\label{eq:DefMuP}
\end{eqnarray}
The term inside the $dG_{\ell}$ integral inside the expectation
in
(\ref{eq:DefMuP}) is the gain/loss from a dynamic trading strategy,
which, at time $\ell$, takes a position in the equity futures
proportional to the quantity $\big( -\int\limits_{1}^\infty \omega[k] \, \mathbbm{1}_{\{G_{\ell-} > k\}} \, dk + \int\limits_0^{1} \omega[k] \, \mathbbm{1}_{\{G_{\ell-} < k\}} \ dk \big)$. In essence, $\mathrm{e}_t^{\mathbb{P}}$ is the expected total gain/loss, over $t$ to $T_O$, from this futures trading strategy.

Finally, 
$\omega[k]>0$ for $0< k<\exp(1)=2.71828$, and $\omega[k]$ is declining for high enough $k$.
$\blacksquare$ \vspace{-3mm}


\section{Option models and risk premiums for jumps crossing the strike}
\label{app:jumps_across}

In this section, we draw on the link between the variations in option risk premiums and modeling ingredients.
Specifically,
we investigate parametric restrictions under which the risk premium for jumps crossing the strike  can be
negative for $k>1$
(i.e., pertaining to OTM calls). Analogous steps apply for $k<1$ (for puts).

We consider the option model based on the price dynamics in (\ref{eq:dou1})--(\ref{eq:dou6}). This
model has price and  volatility jump risks, and spanned and unspanned (diffusive and jump) risks in the volatility dynamics.
The
risk premium adjustments that link $\mathbb{P}$ to $\mathbb{Q}$ are explicit through Girsanov's change of measure
theorem for jump-diffusions (e.g., \citet*{Runggaldier:2003} and \citet*{ContTankov:2004}).

Let ${\bm \lambda}_{\tiny \mbox{jump}}^{\mathbb{P}}$ (${\bm \lambda}_{\tiny \mbox{jump}}^{\mathbb{Q}}$) be the constant
intensity rate of the Poisson process and $\nu^{\mathbb{P}}[\mathbbm{x}_s]$ ($\nu^{\mathbb{Q}}[\mathbbm{x}_s]$) be the density of price jumps under
$\mathbb{P}$ ($\mathbb{Q}$).  The risk premium for jumps crossing the strike $\mathbbm{rp}_{t}^{T_O}[k]$ is
\begin{eqnarray}
\mathbbm{rp}_{t}^{T_O}[k] & \equiv & \mathbb{E}^{\mathbb{P}}_t( a_t^{T_O}[k] + b_t^{T_O}[k] ) ~ - ~ \mathbb{E}^{\mathbb{Q}}_t( a_t^{T_O}[k] + b_t^{T_O}[k] )
~~~~~~~~~~~~~ ~~~~ \mbox{ \, \, }
\nonumber \\
&=& \mathbb{E}^{\mathbb{P}}_t( \sum_{t < \ell \leq T_O} \mathbbm{1}_{\{G_{\ell \, -} \leq k\}} \, \max( G_{\ell} - k, 0 ) ) \, - \, \mathbb{E}^{\mathbb{Q}}_t( \sum_{t < \ell \leq T_O} \mathbbm{1}_{\{G_{\ell \, -} \leq k\}} \, \max( G_{\ell} - k, 0 ) ) ~ \mbox{ \, } \nonumber \\
&+&
\mathbb{E}^{\mathbb{P}}_t( \sum_{t < \ell \leq T_O} \mathbbm{1}_{\{G_{\ell \, -} > k\}} \, \max( k - G_{\ell}, 0 ) ) \, - \, \mathbb{E}^{\mathbb{Q}}_t( \sum_{t < \ell \leq T_O} \mathbbm{1}_{\{G_{\ell \, -} > k\}} \, \max( k - G_{\ell}, 0 ) ). ~ \mbox{ \, \, \, } \label{eq:GeneralDefPMinusQRP}
\end{eqnarray}
Given our focus on the returns of weekly options,
we emphasize analytical tractability and economic insight by developing
our analysis in the limit of small $\Delta T$, where
$\Delta T \equiv T_O - t$.

For small $\Delta T$, the probability, under $\mathbb{P}$ (respectively, $\mathbb{Q}$) of one jump over the time
period $t$ to $t+\Delta T$ approximates to ${\bm \lambda}_{\tiny \mbox{jump}}^{\mathbb{P}}\,\Delta T$ (respectively, ${\bm \lambda}_{\tiny \mbox{jump}}^{\mathbb{Q}} \, \Delta T$). The probability of two or more jumps is negligible for small $\Delta T$.
Therefore, in the limit of small $\Delta T$,
\begin{equation}
G_{\ell-} ~ \mathrm{tends~to} ~ G_t=1 ~ (\mathrm{since}~ G_t = 1 ~\mathrm{(by~construction))}.
~ \mathrm{So} ~ G_{\ell} ~\mathrm{tends~to}~
\underbrace{G_{t}}_{= 1} \, e^{\mathbbm{x}_s} = \ e^{\mathbbm{x}_s}. ~ \mbox{ \, \, \, }
\end{equation}
Simplifying  (\ref{eq:GeneralDefPMinusQRP}), the risk premium for jumps crossing the strike
approximates to
\begin{eqnarray}
\mathbbm{rp}_{t}^{t+\Delta T}[k] & = & {\bm \lambda}_{\tiny \mbox{jump}}^{\mathbb{P}}
\Delta T \int_{-\infty}^{\infty}
\mathbbm{1}_{\{ 1 \leq k\}}
( e^{\mathbbm{x}_s} - k )^{+}  \nu^{\mathbb{P}}[d \mathbbm{x}_s]
- {\bm \lambda}_{\tiny \mbox{jump}}^{\mathbb{Q}}  \Delta T  \int_{-\infty}^{\infty}
\mathbbm{1}_{\{ 1\leq k \}}
( e^{\mathbbm{x}_s} - k )^{+}  \nu^{\mathbb{Q}}[d \mathbbm{x}_s]
 \mbox{ \, \, } \nonumber \\
& +& {\bm \lambda}_{\tiny \mbox{jump}}^{\mathbb{P}}  \Delta T  \int_{-\infty}^{\infty}  \mathbbm{1}_{\{1 > k \}}
( k - e^{\mathbbm{x}_s} )^{+} \, \nu^{\mathbb{P}}[d \mathbbm{x}_s]
-  {\bm \lambda}_{\tiny \mbox{jump}}^{\mathbb{Q}}\Delta T \int_{-\infty}^{\infty}  \mathbbm{1}_{\{ 1> k \}}
( k - e^{\mathbbm{x}_s} )^{+} \nu^{\mathbb{Q}}[d \mathbbm{x}_s],
\mbox{ \, }
\nonumber
\end{eqnarray}
where the error in the approximation is $O[\{\Delta T\}^2]$ and, for brevity, $x^{+} \, \equiv \, \max(x, 0 )$.
Equivalently, since we focus on $k>1$ (pertaining to OTM calls), the task is to compute the following expression:
\begin{eqnarray}
\frac{1}{\Delta T} \mathbbm{rp}_{t}^{t+\Delta T}[k] & = & {\bm \lambda}_{\tiny \mbox{jump}}^{\mathbb{P}} \, \int_{\log(k)}^{\infty}
( e^{\mathbbm{x}_s} - k ) \, \nu^{\mathbb{P}}[\mathbbm{x}_s] \,d \mathbbm{x}_s
~-~ {\bm \lambda}_{\tiny \mbox{jump}}^{\mathbb{Q}} \, \int_{\log(k) }^{\infty}
( e^{\mathbbm{x}_s} - k ) \, \nu^{\mathbb{Q}}[\mathbbm{x}_s] \,d \mathbbm{x}_s.
~ \mbox{ \, \, \, \, }
\label{eq:InterimPMinusQApprox2a}
\end{eqnarray}  \vspace{3mm}
\noindent \textbf{Case~1 (Normally distributed jumps (\citet{Merton:76})).}  For this exercise, we posit
\begin{eqnarray}
\underbrace{\nu^{\mathbb{P}}[\mathbbm{x}_s]}_{\tiny \mbox{density~of~price~jump~under}~\mathbb{P}} & = & \frac{1}{\sqrt{2 \pi {(\sigma^{\mathbb{P}}_{\mathbbm{x}})}^{2}}} \, \,
\exp( -\frac{(\mathbbm{x}_s - \{\mu^{\mathbb{P}}_{\mathbbm{x}} - \frac{1}{2} {(\sigma^{\mathbb{P}}_{\mathbbm{x}}})^2\})^2}{2\, {(\sigma^{\mathbb{P}}_{\mathbbm{x}})}^{2}}) ~~~~~~ ~~ \mbox{ \, } ~~ \mathrm{and} ~~ ~~  \\
\underbrace{\nu^{\mathbb{Q}}[\mathbbm{x}_s]}_{\tiny \mbox{density~of~price~jump~under}~\mathbb{Q}} & = & \frac{1}{\sqrt{2 \pi {(\sigma^{\mathbb{Q}}_{\mathbbm{x}})}^{2}}} \, \,
\exp( -\frac{(\mathbbm{x}_s - \{\mu^{\mathbb{Q}}_{\mathbbm{x}} - \frac{1}{2} {(\sigma^{\mathbb{Q}}_{\mathbbm{x}})}^{2} \})^2}{2\, {(\sigma^{\mathbb{Q}}_{\mathbbm{x}})}^{2}}).
\label{eq:meeerton1}
\end{eqnarray}
Then $\mathbb{E}^{\mathbb{P}}( e^{\mathbbm{x}_s}) =  \exp( \mu^{\mathbb{P}}_{\mathbbm{x}} )$ and
$\mathbb{E}^{\mathbb{Q}}( e^{\mathbbm{x}_s}) = \exp( \mu^{\mathbb{Q}}_{\mathbbm{x}} )$.
It follows from (\ref{eq:InterimPMinusQApprox2a}) that
\begin{equation}
\frac{1}{\Delta T} \mathbbm{rp}_{t}^{t+\Delta T}[k] =
{\bm \lambda}_{\tiny \mbox{jump}}^{\mathbb{P}} \{ e^{\mu^{\mathbb{P}}_{\mathbbm{x}}} {\cal N}(d_1^{\mathbb{P}}[k]) - k \, {\cal N}(d_2^{\mathbb{P}}[k]) \}
~-~
 {\bm \lambda}_{\tiny \mbox{jump}}^{\mathbb{Q}}
 \{ e^{\mu^{\mathbb{Q}}_{\mathbbm{x}}} {\cal N}(d_1^{\mathbb{Q}}[k]) - k \,{\cal N}(d_2^{\mathbb{Q}}[k]) \},
\label{eq:meeerton2}
\end{equation}
where ${\cal N}(.)$ denotes the standard normal cumulative distribution function, and
\begin{eqnarray}
d_1^{\mathbb{P}}[k] &=& \frac{ - \log(k) + \mu^{\mathbb{P}}_{\mathbbm{x}} + \frac{1}{2}{(\sigma^{\mathbb{P}}_{\mathbbm{x}})^{2}}}{\sigma^{\mathbb{P}}_{\mathbbm{x}}},~~ \mbox{ \, \, } ~~ \mathrm{and} ~~~~ d_2^{\mathbb{P}}[k] \, = \, d_1^{\mathbb{P}}[k] \, - \, \sigma^{\mathbb{P}}_{\mathbbm{x}}, ~~  \mbox{ \, \, } ~~ \\
d_1^{\mathbb{Q}}[k] &=& \frac{ - \log(k) + \mu^{\mathbb{Q}}_{\mathbbm{x}} + \frac{1}{2}{(\sigma^{\mathbb{Q}}_{\mathbbm{x}})^{2}}}{\sigma^{\mathbb{Q}}_{\mathbbm{x}}},
 ~~ \mbox{ \, \, } ~~ \mathrm{and} ~~~~ d_2^{\mathbb{Q}}[k] \, = \,  d_1^{\mathbb{Q}}[k] \, - \, \sigma^{\mathbb{Q}}_{\mathbbm{x}}. ~~
 \mbox{ \, \, } ~~
\end{eqnarray}
The ensuing restrictions yield negative risk premiums for jumps crossing the strike and, thus, is an intermediate
step
to supporting negative risk premiums for OTM calls:
\begin{align}
&{\bm \lambda}_{\tiny \mbox{jump}}^{\mathbb{Q}} > {\bm \lambda}_{\tiny \mbox{jump}}^{\mathbb{P}},&
&\mu^{\mathbb{Q}}_{\mathbbm{x}} < \mu^{\mathbb{P}}_{\mathbbm{x}},&
&\mathrm{and}&
&\sigma^{\mathbb{Q}}_{\mathbbm{x}} > \sigma^{\mathbb{P}}_{\mathbbm{x}}.&
&\blacksquare&
\end{align}

\noindent \textbf{Case~2 (Double exponentially distributed jumps (\citet{Kou:2002})).} Under the assumption that the jump distribution under $\mathbb{P}$ and $\mathbb{Q}$ is of the same parametric form, we have
\begin{gather}
\nu^{\mathbb{P}}[\mathbbm{x}_s]~=~
\begin{cases}
p_{+}^{\mathbb{P}} \, \, \eta_{+}^{\mathbb{P}} \, \, e^{- \eta_{+}^{\mathbb{P}} \, \mathbbm{x}_s} \mbox{ \, } &
~ \mbox{ \, } \mathrm{for} ~ ~ \mathbbm{x}_s >0, \\
p_{-}^{\mathbb{P}} \, \, \eta_{-}^{\mathbb{P}} \, \, e^{\eta_{-}^{\mathbb{P}} \, \mathbbm{x}_s} \mbox{ \, }
 & ~ \mbox{ \, } \mathrm{for} ~ ~ \mathbbm{x}_s < 0, ~~~ ~~ \mbox{ \, \, \, where \, $p_{-}^{\mathbb{P}} \, \equiv \, 1 - p_{+}^{\mathbb{P}}$, \, \, \, \, } ~ ~ ~ \mbox{ \, } ~
\end{cases}
\label{kou.1}
\end{gather}
and analogously under $\mathbb{Q}$ (replacing
each superscript $\mathbb{P}$ by a superscript $\mathbb{Q}$ in
equation (\ref{kou.1})).

We assume
that
$0 < p_{+}^{\mathbb{P}} < 1$,
$\eta_{+}^{\mathbb{P}} > 1$,
$\eta_{-}^{\mathbb{P}} > 0$,
$0 < p_{+}^{\mathbb{Q}} < 1$,
$\eta_{+}^{\mathbb{Q}} > 1$,
and $\eta_{-}^{\mathbb{Q}} > 0$.
The mean jump sizes are, respectively, $\frac{1}{\eta_{+}^{\mathbb{P}}}$,  $\frac{1}{\eta_{-}^{\mathbb{P}}}$,
$\frac{1}{\eta_{+}^{\mathbb{Q}}}$, and $\frac{1}{\eta_{-}^{\mathbb{Q}}}$ (\citet[page 1087]{Kou:2002}).

Direct evaluation implies the following expression:
\begin{eqnarray}
\frac{1}{\Delta T} \mathbbm{rp}_{t}^{t+\Delta T}[k] & = &
\frac{{\bm \lambda}_{\tiny \mbox{jump}}^{\mathbb{P}} \, p_{+}^{\mathbb{P}} \, e^{- \log(k) \{\eta_{+}^{\mathbb{P}}-1\} }}{\eta_{+}^{\mathbb{P}}-1}
\, \, - \, \, \frac{{\bm \lambda}_{\tiny \mbox{jump}}^{\mathbb{Q}} \, p_{+}^{\mathbb{Q}} \, e^{- \log(k) \{\eta_{+}^{\mathbb{Q}}-1\}}}{\eta_{+}^{\mathbb{Q}}-1}.
\label{eq:InterimPMinusQApprox2d}
\end{eqnarray}

The following restrictions support negative risk premiums for jumps crossing the strike:
\begin{align}
&{\bm \lambda}_{\tiny \mbox{jump}}^{\mathbb{Q}}
> {\bm \lambda}_{\tiny \mbox{jump}}^{\mathbb{P}},
&
& \frac{1}{\eta_{+}^{\mathbb{P}}} < \frac{1}{\eta_{+}^{\mathbb{Q}}},&
&\mathrm{and}&
& p_{+}^{\mathbb{P}} = p_{+}^{\mathbb{Q}}.~~~~~~\blacksquare&
\label{eq:InterimPMinusQApprox2de}
\end{align}

\noindent \textbf{Case~3 (Normally distributed jumps in equity prices
conditional on exponential jumps in variance (\citet*{DuffiePanSingleton:2000})).} This model admits jumps
in return variance, and the distribution
of price jumps is conditioned on (one-sided) variance jumps.

The consequence is an altered functional form of $\nu^{\mathbb{P}}[\mathbbm{x}_s]$ (and $\nu^{\mathbb{Q}}[\mathbbm{x}_s]$) and
{is amenable to evaluating $\int_{\log(k)}^{\infty}
( e^{\mathbbm{x}_s} - k ) \nu^{\mathbb{P}}[\mathbbm{x}_s] \,d \mathbbm{x}_s$ and $\int_{\log(k)}^{\infty}
( e^{\mathbbm{x}_s} - k ) \nu^{\mathbb{Q}}[\mathbbm{x}_s] \,d \mathbbm{x}_s$.
The model specifies the following:
\begin{eqnarray}
&\mathrm{Jumps}~\mathbbm{x}_\mathrm{v}~\mathrm{in}~\mathrm{v}_t~\mathrm{are~exponentially~and~independently~distributed~with~mean}~ \mu_\mathrm{v}^{\mathbb{P}}, ~ & \mbox{ \quad } \\
&\mathbbm{x}_s \mid \mathbbm{x}_\mathrm{v}  \sim  {\cal N} \left (\beta_{0}^{\mathbb{P}} + \beta_{s,\mathrm{v}}^{\mathbb{P}} \, \mathbbm{x}_\mathrm{v},
(\sigma^{\mathbb{P}}_{s,\mathrm{v}})^2 \right). ~ & \mbox{ \quad }
\label{DPSv-jv.4}
\end{eqnarray}
Equation (\ref{DPSv-jv.4}) allows for simultaneous and correlated jumps in equity price and variance.

Completing the square in the density function of the conditional normal distribution,
we obtain the following density function for price jumps (the form of integral in (\ref{sdff.1}) resembles
\citet*[page 384]{GradshteynRyzhik:1994book}):
\begin{eqnarray}
\nu^{\mathbb{P}}[\mathbbm{x}_s] & = & \int_{0}^{\infty} \, \frac{1}{\sqrt{2 \pi (\sigma^{\mathbb{P}}_{s,\mathrm{v}})^2}} \, \,
\exp( -\frac{(\mathbbm{x}_s - \{ \beta_0^{\mathbb{P}} + \beta_{s,\mathrm{v}}^{\mathbb{P}} \, \mathbbm{x}_\mathrm{v} \})^2}{2 \, (\sigma^{\mathbb{P}}_{s,\mathrm{v}})^2} )
~
\, \, \frac{1}{\mu_\mathrm{v}^{\mathbb{P}}} \, \, e^{- \frac{1}{\mu_\mathrm{v}^{\mathbb{P}}} \mathbbm{x}_\mathrm{v}} \, d \mathbbm{x}_\mathrm{v} ~ \mbox{ \, \, \, \quad } ~ ~ ~ \mbox{ \, \, } ~ \label{sdff.1} \\
& = &\frac{{\cal N}(\frac{\, - \, \sigma^{\mathbb{P}}_{s,\mathrm{v}}}{\mu_\mathrm{v}^{\mathbb{P}} \, \beta_{s,\mathrm{v}}^{\mathbb{P}}} + \frac{(\mathbbm{x}_s - \beta_0^{\mathbb{P}})}{\sigma^{\mathbb{P}}_{s,\mathrm{v}}})}{\mu_\mathrm{v}^{\mathbb{P}} \beta_{s,\mathrm{v}}^{\mathbb{P}}} \, ~ \,
\exp \big( - \, \frac{(\mathbbm{x}_s - \beta_0^{\mathbb{P}})}{\mu_\mathrm{v}^{\mathbb{P}} \, \beta_{s,\mathrm{v}}^{\mathbb{P}}} ~ + ~ \frac{1}{2}  (\frac{\sigma^{\mathbb{P}}_{s,\mathrm{v}}}{\mu_\mathrm{v}^{\mathbb{P}} \, \beta_{s,\mathrm{v}}^{\mathbb{P}}})^2 \big),  \mbox{ \, \, \, \, } ~~
\mbox{ \, }
\label{DPSDensity1.1}
\end{eqnarray}
and analogously under $\mathbb{Q}$ (replacing each superscript $\mathbb{P}$ by a superscript $\mathbb{Q}$ in
(\ref{DPSv-jv.4})--(\ref{DPSDensity1.1})).

The tractability of the jump densities
enables the determination of
the risk premium for jumps crossing the strike in
(\ref{eq:InterimPMinusQApprox2a})
(via numerical integration).
Setting $\sigma^{\mathbb{P}}_{s,\mathrm{v}}=\sigma^{\mathbb{Q}}_{s,\mathrm{v}}$, the following parameter
restrictions facilitate
the outcome of negative risk premiums for jumps crossing the strike:
\begin{align}
&\beta_{0}^{\mathbb{P}} <0,&
& \beta_{s,\mathrm{v}}^{\mathbb{P}} < 0,&
&\beta_{0}^{\mathbb{Q}} < \beta_{0}^{\mathbb{P}},&
& \beta_{s,\mathrm{v}}^{\mathbb{Q}} < \beta_{s,\mathrm{v}}^{\mathbb{P}},&
&\mathrm{and}&
&\mu_\mathrm{v}^{\mathbb{Q}} > \mu_\mathrm{v}^{\mathbb{P}}.&
\end{align}

In other words, the option model imposes inequality restrictions to match the empirical patterns.
These parametric restrictions
have not, to our knowledge, been tested, and may be difficult
to validate.
Similar to the emphasis in \citet*{Chen_Dou_Kogan:JF2020} and \citet*{Cheng_Dou_Liao:ECMTA2021}, these
restrictions highlight the dark matter property of option models.
$\blacksquare$ \vspace{-4mm}




\section{Option models and local time risk premiums for moneyness $k$}
\label{app:var_jumps}

We outline restrictions that generate negative local time risk premiums,
in the context of the model in
(\ref{eq:dou1})--(\ref{eq:dou6}).
{By the definition of covariance, and using
$\mathbb{E}_{t}^{\mathbb{Q}}( \frac{M_{t}}{M_{T_{O}} e^{r ({T}_O - t)}} ) \, = \, 1$, we have \small
\begin{eqnarray}
\mathrm{cov}_t^{\mathbb{Q}}( \frac{M_{t}}{M_{{T}_O} e^{r ({T}_O - t)}}, \mathbb{L}^{T_O}_t[k] )
& = &
\mathbb{E}_{t}^{\mathbb{Q}}( \frac{M_{t}}{M_{{T}_O} e^{r ({T}_O - t)}} \, \mathbb{L}^{T_O}_t[k] ) ~ - ~
\overbrace{\mathbb{E}_{t}^{\mathbb{Q}}( \frac{M_{t}}{M_{{T}_O} e^{r ({T}_O - t)}} )}^{= \, 1} \,
\mathbb{E}_{t}^{\mathbb{Q}}( \mathbb{L}^{T_O}_t[k] ) ~~ \mbox{ \, \, \, }  \nonumber \\
&=& \mathbb{E}_{t}^{\mathbb{Q}}( \frac{M_{t}}{M_{{T}_O} e^{r ({T}_O - t)}} \mathbb{L}^{T_O}_t[k] ) ~-~
\mathbb{E}_{t}^{\mathbb{Q}}( \mathbb{L}^{T_O}_t[k] ) ~ \mbox{ \, } \label{c5.2} \\
&=& \underbrace{\mathbb{E}_{t}^{\mathbb{P}}( \mathbb{L}^{T_O}_t[k] ) ~-~ \mathbb{E}_{t}^{\mathbb{Q}}( \mathbb{L}^{T_O}_t[k] ).}_{\tiny \mbox{local~time~risk~premium}} ~~~ \mbox{ \, }
\label{eq:covqgsbstatement}
\end{eqnarray} \normalsize
To evaluate $\mathrm{cov}_t^{\mathbb{Q}}( \frac{M_{t}}{M_{{T}_O} e^{r ({T}_O - t)}}, \mathbb{L}^{T_O}_t[k] )$,
we consider the dynamics of $\frac{M_{t}}{M_{{T}_O} e^{r ({T}_O - t)}}$ under $\mathbb{Q}$
as well as those of local time $\mathbb{L}^{T_O}_t[k]$. \vspace{-4mm}

\subsection{Expression for $\frac{M_{t}}{M_{{T}_O} e^{r ({T}_O - t)}}$ dynamics under $\mathbb{Q}$}
\label{app:OptionModelsAndOptionRiskPremiums}
Using equation (\ref{eq:dou1}), we have the following representation:\footnote{In light of Girsanov's theorem, ${z}^{\mathbb{Q}}_t$ and ${u}^{\mathbb{Q}}_t$ are independent standard Brownian motions under the probability measure $\mathbb{Q}$, linked to ${z}^{\mathbb{P}}_t$ and ${u}^{\mathbb{P}}_t$, by
$d {z}^{\mathbb{P}}_t-d {z}^{\mathbb{Q}}_t = {\eta}[t,\mathrm{v}_t] \,dt$ and
$d {u}^{\mathbb{P}}_t-d {u}^{\mathbb{Q}}_t =  {\theta}[t,\mathrm{v}_t] \, dt$.}
\begin{eqnarray}
\frac{M_{t}}{M_{{T}_O} e^{r ({T}_O - t)}}
&=& \overbrace{e^{ \int_{t}^{{T}_O} \{ -\frac{1}{2} (\eta[s,\mathrm{v}_s])^2 ds - \eta[s,\mathrm{v}_s] d z_s^{\mathbb{Q}} \}}}^{\tiny \mbox{spanned~diffusive~component}} ~ \times ~ \, \,
\overbrace{e^{ \int_{t}^{{T}_O} \{
- \frac{1}{2} (\theta[s,\mathrm{v}_s])^2 ds - \theta[s,\mathrm{v}_s]
d u^{\mathbb{Q}}_s \}}}^{\tiny \mbox{unspanned~diffusive~component}} \, \, ~ \times \mbox{ \, \, } ~ ~ \mbox{ \, \, } \nonumber \\
& & ~ \mbox{ \, \, } ~ \mbox{ \, \, } \underbrace{e^{
\{
\sum_{t < \ell \leq T_O} (-\mathbbm{x}_m)
~ - \, \int_{t}^{{T}_O} {{\bm \lambda}^{\mathbb{Q}}_{\tiny \mbox{jump}}} \, \mathbb{E}^{\mathbb{Q}}( e^{-\mathbbm{x}_m} - 1 ) \, ds \}}.}_{\tiny \mbox{unspanned~jump~component}}
~ ~ ~ \mbox{ \, \, \, \, } \label{eq:RecipPK5TermsDPSLT1}
\end{eqnarray}
For compactness of equation presentation, define as follows:
\begin{eqnarray}
\mathcal{R}_{T_O}^{\tiny \mbox{span~diffusive}} &\equiv&
e^{ \int_{t}^{{T}_O} \{ -\frac{1}{2} (\eta[s,\mathrm{v}_s])^2 ds - \eta[s,\mathrm{v}_s] d z_s^{\mathbb{Q}} \}}, \label{ah.1}\\
\mathcal{R}_{T_O}^{\tiny \mbox{unspan~diffusive}} &\equiv&
e^{ \int_{t}^{{T}_O} \{
- \frac{1}{2} (\theta[s,\mathrm{v}_s])^2 ds - \theta[s,\mathrm{v}_s]
d u^{\mathbb{Q}}_s \}}, ~ ~ \mbox{ \, \, } ~~~ \mathrm{and} \mbox{ \, \, } \\
\mathcal{R}_{T_O}^{\tiny \mbox{unspan~jump}} & \equiv &
e^{
\{
\sum_{t < \ell \leq T_O} (-\mathbbm{x}_m)
~ - \, \int_{t}^{{T}_O} {{\bm \lambda}^{\mathbb{Q}}_{\tiny \mbox{jump}}} \, \mathbb{E}^{\mathbb{Q}}( e^{-\mathbbm{x}_m} - 1 ) \, ds \}}. ~ \mbox{ \, \, }
\end{eqnarray}
Then, we can write the reciprocal of the Radon-Nikodym derivative as  follows:
\begin{eqnarray}
\frac{M_{t}}{M_{{T}_O} e^{r ({T}_O - t)}} \, \, \, = \, \,
\mathcal{R}_{T_O}^{\tiny \mbox{span~diffusive}} \, \,
\times \, \mathcal{R}_{T_O}^{\tiny \mbox{unspan~diffusive}} \, \,
\times \, \mathcal{R}_{T_O}^{\tiny \mbox{unspan~jump}}. ~ \mbox{ \quad \, \, } ~
\label{eq:RecipPK5TermsDPSLT1cv}
\end{eqnarray}
Thus, $\frac{M_{t}}{M_{{T}_O} e^{r ({T}_O - t)}}$ is multiplicative in three positive (orthogonal) martingales under $\mathbb{Q}$.

\subsection{Characterizing the
sign of the local time risk premiums}

For the results that follow, we define the
following. \small
\begin{equation}
\mathrm{Let~} \, \mathcal{I}_{s} ~ \mbox{be~the~sub-filtration~of} ~ \mathcal{F}_{s}~\mathrm{generated~by} ~ \mathcal{R}_{s}^{\tiny \mbox{span~diffusive}},\mathrm{~that~is,}~\mathrm{by}~\eta[s,\mathrm{v}_s]~\mathrm{and}~\eta[s,\mathrm{v}_s] d z_s^{\mathbb{Q}}. ~ \mbox{ \, \, \, } ~
 \label{eq:SubFiltrationDefinition}
\end{equation} \normalsize
Exploiting the law of total covariance, the risk premium for local time, with moneyness $k$, is \small
\begin{eqnarray}
\mathrm{cov}_t^{\mathbb{Q}}( \overbrace{\frac{M_{t}}{M_{{T}_O} e^{r ({T}_O - t)}} }^{\mathrm{from~\tiny(\ref{eq:RecipPK5TermsDPSLT1}})}, \, \mathbb{L}_t^{{T}_O}[k] )&=&  \mathbb{E}_{t}^{\mathbb{Q}}( \mathrm{cov}_t^{\mathbb{Q}}( \frac{M_{t}}{M_{{T}_O} e^{r ({T}_O - t)}},
\, \mathbb{L}_t^{{T}_O}[k] {\Big |} \, \mathcal{I}_{T_O} ) ) \nonumber \\
&&  + \, \mathrm{cov}_t^{\mathbb{Q}}( \mathbb{E}_{t}^{\mathbb{Q}}( \frac{M_{t}}{M_{{T}_O} e^{r ({T}_O - t)}} \,  {\Big |} \mathcal{I}_{T_O} ), \, \, \mathbb{E}_{t}^{\mathbb{Q}}( \mathbb{L}_t^{{T}_O}[k] \,{\Big |} \, \mathcal{I}_{T_O} ) ) ~ \mbox{ \, \, \, }
\nonumber
\\
&=& \mathbb{E}_{t}^{\mathbb{Q}}( \mathrm{cov}_t^{\mathbb{Q}}(
\mathcal{R}_{T_O}^{\tiny \mbox{span~diffusive}}
\times \mathcal{R}_{T_O}^{\tiny \mbox{unspan~diffusive}}
\times \mathcal{R}_{T_O}^{\tiny \mbox{unspan~jump}},
\, \mathbb{L}_t^{{T}_O}[k] {\Big |} \, \mathcal{I}_{T_O} ) ) ~~ \mbox{ \, \, } ~~~ \nonumber \\
& & \, + \, \mathrm{cov}_t^{\mathbb{Q}}( \mathbb{E}_{t}^{\mathbb{Q}}( \frac{M_{t}}{M_{{T}_O}} e^{-r ({T}_O - t)} \, \, {\Big |} \mathcal{I}_{T_O} ), \, \, \mathbb{E}_{t}^{\mathbb{Q}}( \mathbb{L}_t^{{T}_O}[k] \, {\Big |}\, \mathcal{I}_{T_O} ) )
\nonumber
\\
&=&\mathbb{E}_{t}^{\mathbb{Q}}( \mathcal{R}_{T_O}^{\tiny \mbox{span~diffusive}} \times
\mathrm{cov}_t^{\mathbb{Q}}(
\mathcal{R}_{T_O}^{\tiny \mbox{unspan~diffusive}}
\times \mathcal{R}_{T_O}^{\tiny \mbox{unspan~jump}},
\, \mathbb{L}_t^{{T}_O}[k] {\Big |} \, \mathcal{I}_{T_O} ) ) ~~ \mbox{ \, \, } ~~~ \nonumber \\
& &
\, +\,
\mathrm{cov}_t^{\mathbb{Q}}( \mathbb{E}_{t}^{\mathbb{Q}}( \frac{M_{t}}{M_{{T}_O}} e^{-r ({T}_O - t)} \, \, {\Big |} \mathcal{I}_{T_O} ), \, \, \mathbb{E}_{t}^{\mathbb{Q}}( \mathbb{L}_t^{{T}_O}[k] \, {\Big |} \, \mathcal{I}_{T_O} ) ). \mbox{ \, }~\mbox{ \, \, }~
\label{eq:ConditCovarSpanning2Linesa}
\end{eqnarray} \normalsize

To reproduce the empirical finding of negative risk premiums of OTM calls, one
may require negative local time
risk premiums, which in view of equation (\ref{eq:ConditCovarSpanning2Linesa}) leads us to assess when the
two terms appearing in (\ref{eq:ConditCovarSpanning2Linesa}) can be negative.

In particular, examining (\ref{eq:ConditCovarSpanning2Linesa}), we are interested in when the term involving unspanned risks, specifically,
\begin{eqnarray}
\mathrm{cov}_t^{\mathbb{Q}}( \mathcal{R}_{T_O}^{\tiny \mbox{unspan~diffusive}} \times \mathcal{R}_{T_O}^{\tiny \mbox{unspan~jump}},\, \mathbb{L}_t^{{T}_O}[k] {\Big |} \,
\mathcal{I}_{T_O} ) ~ ~ \mbox{ \quad \, is negative. \quad \, } ~ ~~ \label{eq:CovarianceTermWhoseSign}
\end{eqnarray}

To keep the analysis contained, our approach is twofold,
as follows:
\begin{enumerate}

\item Assess the economic implications of the sign of $\mathrm{cov}_t^{\mathbb{Q}}( \mathcal{R}_{T_O}^{\tiny \mbox{unspan~diffusive}} \times \mathcal{R}_{T_O}^{\tiny \mbox{unspan~jump}},\, \mathbb{L}_t^{{T}_O}[k] {\Big |} \,
\mathcal{I}_{T_O} )$ in Section~\ref{app:unspannedDiffusiveVolRisk}
(see (\ref{eq:FinalCovarianceTerm3DPS})--(\ref{eq:FinalCovarianceTermisNegIfThetaNeg4DPS})
and in Section~\ref{app:unspannedVolJumpRisk}
(see
(\ref{c5.10DPS3})--(\ref{eq:FinalCovarianceTermisNegIfThetaNeg4DPSSecondTerm})).

\item Then, assess the economic implications of the sign of
$\mathrm{cov}_t^{\mathbb{Q}}( \mathbb{E}_{t}^{\mathbb{Q}}( \frac{M_{t}}{M_{{T}_O}} e^{-r ({T}_O - t)} \,  {\Big |} \mathcal{I}_{T_O} ), \, \, \mathbb{E}_{t}^{\mathbb{Q}}( \mathbb{L}_t^{{T}_O}[k] \, {\Big |} \mathcal{I}_{T_O} ) )$ in
Section~\ref{app:LTRPSpannedDiffusiveVolRisk}.

\end{enumerate}
We turn to these tasks in turn. \vspace{-4mm}

\subsection{Evolution of  diffusive component of the futures return
under $\mathbb{Q}$}

First, we note that the evolution of $\mathrm{v}_{\ell}$
under $\mathbb{Q}$ is
\begin{eqnarray}
\mathrm{v}_{\ell} & =  & \mathrm{v}_{t}\, e^{ \kappa_{\mathrm{vol}}^{\mathbb{Q}} ( t - \ell ) }
~+~
\int_{t}^{\ell} \phi_{\tiny \mbox{vol}}^{\mathbb{Q}} e^{ \kappa_{\tiny \mbox{vol}}^{\mathbb{Q}} ( s - \ell ) } ds
~+~ \sigma_{\tiny \mbox{vol}}\,\rho_{\tiny \mbox{vol}} \overbrace{\int_{t}^{\ell}  e^{ \kappa_{\tiny \mbox{vol}}^{\mathbb{Q}} ( s - \ell ) } \sqrt{\mathrm{v}_{s}} \,  dz_s^{\mathbb{Q}}}^{\tiny \mbox{spanned~diffusive~volatility~risk}} \nonumber \\
&+&\sigma_{\tiny \mbox{vol}} \, \sqrt{1-\rho^2_{\tiny \mbox{vol}}}
\underbrace{\int_{t}^{\ell}  e^{ \kappa_{\tiny \mbox{vol}}^{\mathbb{Q}} ( s - \ell ) } \sqrt{\mathrm{v}_{s}} \,  du_s^{\mathbb{Q}}}_{\tiny \mbox{unspanned~diffusive~volatility~risk}}
\, + \,
\underbrace{\int_{t}^{\ell}  e^{ \kappa_{\tiny \mbox{vol}}^{\mathbb{Q}} ( s - \ell ) }\,
\mathbbm{x}_{\mathrm{v}} \,
d \mathbb{N}^{\mathbb{Q}}_{s}}_{\tiny \mbox{unspanned~volatility~jump~risk}}~~~~~~\mbox{for $\ell \geq t$.\, \, \, }~~
\mbox{ \, \, } ~
\label{cv.sDPS}
\end{eqnarray}

To obtain an expression for $\mathbb{L}^{T_O}_t[k]$, we note that the path-by-path continuous part of the quadratic variation $[ G^\mathrm{c}, G^\mathrm{c} ]_{s} =
\int_{t}^s \{ \sqrt{\mathrm{v}_{\ell}} \, G_{\ell} \}^2 \, d\ell = \int_{t}^s \, \mathrm{v}_{\ell} \,G_\ell^2 \,d\ell$.
We deduce the form of $\mathbb{L}^{T_O}_t[k]$ as
\begin{eqnarray}
\mathbb{L}^{T_O}_t[k] & = & \frac{1}{2} \int_{t}^{T_O} \delta_{\{G_\ell- k\}} d [ G^\mathrm{c}, G^\mathrm{c} ]_{\ell}
~=~ \frac{1}{2} \, \int_{t}^{{T}_O} \, \delta_{\{G_\ell - k\}} \, \mathrm{v}_{\ell} \, G_\ell^2 \, d\ell ~ \mbox{ \, } ~ ~~ ~ \nonumber \\
& = & \frac{1}{2} \int_{t}^{T_O} \delta_{\{G_\ell - k\}}
\big\{ \, \overbrace{\mathrm{v}_{t}\, e^{ \kappa_{\tiny \mbox{vol}}^{\mathbb{Q}} ( t - \ell ) } + \int_{t}^{\ell} \phi_{\tiny \mbox{vol}}^{\mathbb{Q}} e^{ \kappa_{\tiny \mbox{vol}}^{\mathbb{Q}} ( s - \ell ) } ds }^{\tiny \mbox{irrelevant~for~conditional~covariance~in~(\ref{eq:CovarianceTermWhoseSign})}}
~ + ~ \sigma_{\tiny \mbox{vol}}\,\rho_{\tiny \mbox{vol}} \int_{t}^{\ell}  e^{ \kappa_{\tiny \mbox{vol}}^{\mathbb{Q}} ( s - \ell ) } \sqrt{\mathrm{v}_{s}} \,  dz_s^{\mathbb{Q}} ~
\nonumber \\
& + & \underbrace{\sigma_{\tiny \mbox{vol}} \, \sqrt{1-\rho^2_{\tiny \mbox{vol}}} \int_{t}^{\ell}  e^{ \kappa_{\tiny \mbox{vol}}^{\mathbb{Q}} ( s - \ell ) } \sqrt{\mathrm{v}_{s}} \,  du_s^{\mathbb{Q}}}_{\tiny \mbox{covaries~(relevant~in~(\ref{eq:CovarianceTermWhoseSign}))}}
\, + \,
\underbrace{\int_{t}^{\ell}  e^{ \kappa_{\tiny \mbox{vol}}^{\mathbb{Q}} ( s - \ell ) }\,
\mathbbm{x}_{\mathrm{v}} \,
d \mathbb{N}^{\mathbb{Q}}_{s}}_{\tiny \mbox{covaries~(relevant~in~(\ref{eq:CovarianceTermWhoseSign}))}} \big\}
\, G_\ell^2 \, d\ell.
\label{eq:longLDPS1}
\end{eqnarray}
Using (\ref{eq:longLDPS1}), we now substitute
for $\mathbb{L}^{T_O}_t[k]$ into
(\ref{eq:CovarianceTermWhoseSign}). Recognizing that some terms are irrelevant in the computation
of
$\mathrm{cov}_t^{\mathbb{Q}}( \mathcal{R}_{T_O}^{\tiny \mbox{unspan~diffusive}} \times \mathcal{R}_{T_O}^{\tiny \mbox{unspan~jump}},\, \mathbb{L}_t^{{T}_O}[k] {\Big |} \,
\mathcal{I}_{T_O} )$, we determine as follows:
\begin{eqnarray}
& &
\mathrm{cov}_t^{\mathbb{Q}}( \mathcal{R}_{T_O}^{\tiny \mbox{unspan~diffusive}} \times \mathcal{R}_{T_O}^{\tiny \mbox{unspan~jump}}, \, \mathbb{L}_t^{{T}_O}[k] {\Big |} \,
\mathcal{I}_{T_O} ) ~
\label{asssd.1}
\nonumber \\
& & ~ = \, \mathrm{cov}_t^{\mathbb{Q}}( \mathcal{R}_{T_O}^{\tiny \mbox{unspan~diffusive}} ~
\times ~ \mathcal{R}_{T_O}^{\tiny \mbox{unspan~jump}},
\nonumber \\
& & ~ \mbox{ \, \, } ~~~~ ~ (\frac{1}{2} \int_{t}^{T_O} \delta_{\{G_\ell - k\}} \sigma_{\mathrm{vol}} \, \sqrt{1-\rho^2_{\tiny \mbox{vol}}} \int_{t}^{\ell}  e^{ \kappa_{\tiny \mbox{vol}}^{\mathbb{Q}} ( s - \ell ) } \sqrt{\mathrm{v}_{s}} \,  du_s^{\mathbb{Q}} \, G_\ell^2 \, d\ell \, \, + \, \nonumber \\
& & ~ \mbox{ \, \, \, \, } ~~~~ ~ ~ \frac{1}{2} \int_{t}^{T_O} \delta_{\{G_\ell- k\}} \int_{t}^{\ell}  e^{ \kappa_{\mathrm{vol}}^{\mathbb{Q}} ( s - \ell ) }
\mathbbm{x}_{\mathrm{v}} \,
d \mathbb{N}^{\mathbb{Q}}_{s} \, G_\ell^2 \, d\ell \, \big)
\, \, {\Big |} \mathcal{I}_{T_O} ) ~
~ \mbox{ \, \, \, \, \, }
\nonumber \\
& & ~ = \, \mathrm{cov}_t^{\mathbb{Q}}(
\mathcal{R}_{T_O}^{\tiny \mbox{unspan~diffusive}},
\frac{1}{2} \int_{t}^{T_O} \delta_{\{G_\ell- k\}} \sigma_{\tiny \mbox{vol}} \sqrt{1-\rho^2_{\tiny \mbox{vol}}} \int_{t}^{\ell}  e^{ \kappa_{\tiny \mbox{vol}}^{\mathbb{Q}} ( s - \ell ) } \sqrt{\mathrm{v}_{s}} du_s^{\mathbb{Q}} G_\ell^2 \, d\ell \, {\Big |} \mathcal{I}_{T_O} )
\mbox{ \, \, } \,
\nonumber \\
& & ~ \mbox{ \, \, }
~ + ~ \mathrm{cov}_t^{\mathbb{Q}}(
\mathcal{R}_{T_O}^{\tiny \mbox{unspan~jump}}, \frac{1}{2} \int_{t}^{T_O} \delta_{\{G_\ell- k\}} \int_{t}^{\ell}  e^{ \kappa_{\mathrm{vol}}^{\mathbb{Q}} ( s - \ell ) }
\mathbbm{x}_{\mathrm{v}} \,
d \mathbb{N}^{\mathbb{Q}}_{s}
\, G_\ell^2 \, d\ell \, \, {\Big |} \mathcal{I}_{T_O} ),
~ ~ \mbox{ \, \, \, \, \, } \label{eq:JumpRelatedLT2}
\end{eqnarray}
where we have exploited independence between $d u^{\mathbb{Q}}_s$ and $d \mathbb{N}^{\mathbb{Q}}_{s}$.

Thus, the conditional covariance in
(\ref{eq:JumpRelatedLT2}) consists of two
parts: (i) an unspanned diffusion-related term and
(ii) an unspanned jump-related term.

Next, we elaborate the economic rationale under which these derived terms can be signed. \vspace{-4mm}


\subsection{Negative local time risk premium for unspanned diffusive volatility risk}
\label{app:unspannedDiffusiveVolRisk}

The sign of the first term in (\ref{eq:JumpRelatedLT2}) (after substituting from
(\ref{eq:RecipPK5TermsDPSLT1})) is the sign of
{\small \begin{equation}
\mathrm{cov}_t^{\mathbb{Q}}( e^{ \int_{t}^{{T}_O} \{
- \frac{1}{2} (\theta[s,\mathrm{v}_s])^2 ds - \theta[s,\mathrm{v}_s]
d u^{\mathbb{Q}}_s \}},
\frac{1}{2} \int_{t}^{T_O} \delta_{\{G_\ell-k\}} \sigma_{\mathrm{vol}} \, \int_{t}^{\ell}  e^{ \kappa_{\mathrm{vol}}^{\mathbb{Q}} ( s - \ell ) } \sqrt{\mathrm{v}_{s}} \,  du_s^{\mathbb{Q}} \, G_\ell^2 \, d\ell \, \, {\Big |} \mathcal{I}_{T_O}  ), ~ ~
\end{equation}}
which is the same (in light of Stein's lemma) as the sign of
\small
\begin{eqnarray}
& & \mathrm{cov}_t^{\mathbb{Q}}( \int_{t}^{{T}_O} - \theta[s,\mathrm{v}_s]
d u^{\mathbb{Q}}_s, \, \frac{1}{2} \int_{t}^{T_O} \delta_{\{G_\ell- k\}} \sigma_{\mathrm{vol}} \, \int_{t}^{\ell}  e^{ \kappa_{\mathrm{vol}}^{\mathbb{Q}} ( s - \ell ) } \sqrt{\mathrm{v}_{s}} \,  du_s^{\mathbb{Q}} \, G_\ell^2 \, d\ell \, \, {\Big |} \mathcal{I}_{T_O} ) \nonumber \\
& & \, \, = ~ \mathrm{cov}_t^{\mathbb{Q}}( \int_{t}^{{T}_O} -
\{-\theta_{\mathrm{LT}} \, \sqrt{\mathrm{v}_s} \, \, d u^{\mathbb{Q}}_{s}\}, \, \frac{1}{2} \int_{t}^{T_O} \delta_{\{G_\ell - k\}}
\sigma_{\tiny \mbox{vol}} \, \int_{t}^{\ell}  e^{ \kappa_{\tiny \mbox{vol}}^{\mathbb{Q}} ( s - \ell ) } \sqrt{\mathrm{v}_{s}} \,  du_s^{\mathbb{Q}} \, G_\ell^2 \, d\ell
\, {\Big |} \mathcal{I}_{T_O} )  ~ \mbox{ \, \, \, } ~ \nonumber \\
& & \, \, = ~ \mathrm{cov}_t^{\mathbb{Q}}( \int_{t}^{{T}_O} \theta_{\mathrm{LT}} \, \sqrt{\mathrm{v}_s} \, \, d u^{\mathbb{Q}}_{s},
\, \int_{t}^{{T}_O} \sqrt{\mathrm{v}_{s}} \, \{\ \int_{s}^{{T}_O} \frac{\sigma_{\tiny \mbox{vol}}}{2} e^{\kappa_{\tiny \mbox{vol}}^{\mathbb{Q}} ( s - \ell ) } \, \delta_{\{G_\ell - k\}} \, G_\ell^2 \, d\ell \} \, du^{\mathbb{Q}}_{s} \,{\Big |} \mathcal{I}_{T_O} )  \, \, \nonumber \\
& & \, \, = ~ \mathbb{E}_{t}^{\mathbb{Q}}( \int_{t}^{{T}_O} \theta_{\mathrm{LT}} \, \sqrt{\mathrm{v}_s} \,  \sqrt{\mathrm{v}_s} \,
\, \{\ \int_{s}^{{T}_O} \frac{\sigma_{\tiny \mbox{vol}} }{2} e^{\kappa_{\tiny \mbox{vol}}^{\mathbb{Q}} ( s - \ell ) } \,
\delta_{\{G_\ell- k\}}
\, G_\ell^2 \, d\ell \} \,  ds \, {\Big |} \mathcal{I}_{T_O})  \, \nonumber \\
& & \, \, = ~ \theta_{\mathrm{LT}} \, \, \,
\underbrace{\mathbb{E}_{t}^{\mathbb{Q}}(  \int_{t}^{{T}_O} \, \mathrm{v}_s \, \, \{\ \int_{s}^{{T}_O} \frac{\sigma_{\tiny \mbox{vol}}}{2}
e^{ \kappa_{\tiny \mbox{vol}}^{\mathbb{Q}} ( s - \ell ) } \, \delta_{\{G_\ell - k\}}
\, G_\ell^2 \, d\ell \} \, ds \,{\Big |} \mathcal{I}_{T_O} ) }_{~\geq ~0}.
\label{eq:FinalCovarianceTerm3DPS}
\end{eqnarray} \normalsize

Inspection of (\ref{eq:FinalCovarianceTerm3DPS}) shows that  \vspace{-3mm}
\begin{eqnarray}
\mbox{ the diffusion-related term in (\ref{eq:JumpRelatedLT2}) is negative if ~$\theta_{\mathrm{LT}} \, <  \, 0$. \, \, } ~ ~~ ~ ~~
\mbox{ \, \quad \quad \quad } 
\label{eq:FinalCovarianceTermisNegIfThetaNeg4DPS}
\end{eqnarray}
This is the restriction required for negative local time risk premiums for unspanned diffusive volatility risks. $\blacksquare$ \vspace{-4mm}

\subsection{Negative local time risk premiums due to unspanned volatility jump risks}
\label{app:unspannedVolJumpRisk}

With the term $\int_{t}^{{T}_O} {{\bm \lambda}^{\mathbb{Q}}_{\tiny \mbox{jump}}} \, \mathbb{E}^{\mathbb{Q}}( e^{-\mathbbm{x}_m} - 1 ) \, ds$
not relevant for the conditional covariance, the second term in (\ref{eq:JumpRelatedLT2}) is \vspace{-3mm}
\begin{eqnarray}
& & \mathrm{cov}_t^{\mathbb{Q}}(
e^{ \{ \sum_{t < \ell \leq T_O} (-\mathbbm{x}_m) \} },
\frac{1}{2} \int_{t}^{T_O} \delta_{\{G_\ell-k\}}
\int_{t}^{\ell}  e^{ \kappa_{\tiny \mbox{vol}}^{\mathbb{Q}} ( s - \ell ) }
\mathbbm{x}_{\mathrm{v}} \,
d \mathbb{N}^{\mathbb{Q}}_{s} \, G_\ell^2 \, d\ell \, {\Big |} \mathcal{I}_{T_O} )
\nonumber \\
& & = \mathrm{cov}_t^{\mathbb{Q}}(
e^{ \{ \sum_{t < \ell \leq T_O} (-\mathbbm{x}_m) \}}, \frac{1}{2} \int_{t}^{T_O} \{\ \int_{s}^{T_O} \delta_{\{G_\ell- k\}}
e^{ \kappa_{\tiny \mbox{vol}}^{\mathbb{Q}} ( s - \ell ) }\,
\mathbbm{x}_{\mathrm{v}} \,
G_\ell^2 \, d\ell \} \, ~ d \mathbb{N}^{\mathbb{Q}}_{s}
\, {\Big |} \mathcal{I}_{T_O} )
 ~ ~ \mbox{ \, \quad } ~
\nonumber \\
& &  = ~ \mathrm{cov}_t^{\mathbb{Q}}(
e^{ \{ \sum_{t < \ell \leq T_O} (-\mathbbm{x}_m)  \}},~\frac{1}{2} \sum_{t < \ell \leq T_O} \{\ \int_{s}^{T_O}
\delta_{\{G_\ell - k\}}
e^{ \kappa_{\tiny \mbox{vol}}^{\mathbb{Q}} ( s - \ell ) }\,
\mathbbm{x}_{\mathrm{v}} \,
G_\ell^2 \, d\ell \}
\, {\Big |} \mathcal{I}_{T_O}).
 ~ ~ \mbox{ \, \quad } ~
\label{c5.10DPS3}
\end{eqnarray}
Among the determinants of the sign of equation (\ref{c5.10DPS3}) and, thus, of
(\ref{eq:JumpRelatedLT2}) is the sign of $\mathrm{cov}^{\mathbb{Q}}( e^{-\mathbbm{x}_m},
\mathbbm{x}_{\mathrm{v}} )$.
In particular, for a
negative contribution to the local time risk premium, one is
led
to postulate the following restriction: \vspace{-3mm}
\begin{eqnarray}
\mathrm{cov}^{\mathbb{Q}}( e^{-\mathbbm{x}_m},
\mathbbm{x}_{\mathrm{v}} ) \, < \, 0.
\label{eq:FinalCovarianceTermisNegIfThetaNeg4DPSSecondTerm}
\end{eqnarray}
Equation (\ref{eq:FinalCovarianceTermisNegIfThetaNeg4DPSSecondTerm}) holds
when model parameters under $\mathbb{Q}$ are such that
large jumps in volatility associate with large up jumps in the pricing kernel.
$\blacksquare$ \vspace{-4mm}

\subsection{Local time risk premiums due to spanned risks}
\label{app:LTRPSpannedDiffusiveVolRisk}

In light of the fact that
\begin{align}
&\mathbb{E}_{t}^{\mathbb{Q}}( \mathcal{R}_{T_O}^{\tiny \mbox{unspan~diffusive}} \, {\Big |}\, \mathcal{I}_{T_O} ) = 1&
&\mathrm{and}&
&\mathbb{E}_{t}^{\mathbb{Q}}( \mathcal{R}_{T_O}^{\tiny \mbox{unspan~jump}} \, {\Big |}\, \mathcal{I}_{T_O} ) = 1, &
\end{align}
we consider the final term $\mathrm{cov}_t^{\mathbb{Q}}( \mathbb{E}_{t}^{\mathbb{Q}}( \frac{M_{t}}{M_{{T}_O}} e^{-r ({T}_O - t)} \, \, {\Big |} \mathcal{I}_{T_O} ), \, \, \mathbb{E}_{t}^{\mathbb{Q}}( \mathbb{L}_t^{{T}_O}[k] \, {\Big |} \mathcal{I}_{T_O} ) )$ in equation (\ref{eq:ConditCovarSpanning2Linesa}).

Direct evaluation of the covariance is unrevealing.
Therefore,  we cast this final term in terms of economic variables, specifically, expectations of option payoffs.

To see our rationale, we work through the covariance as follows:
\begin{eqnarray}
& &
\mathrm{cov}_t^{\mathbb{Q}}( \mathbb{E}_{t}^{\mathbb{Q}}( \frac{M_{t}}{M_{{T}_O}} e^{-r ({T}_O - t)} \, \, {\Big |} \mathcal{I}_{T_O} ), \, \, \mathbb{E}_{t}^{\mathbb{Q}}( \mathbb{L}_t^{{T}_O}[k] \, {\Big |} \mathcal{I}_{T_O} ) )  \nonumber \\
&& ~ = ~
\mathrm{cov}_t^{\mathbb{Q}}( \mathbb{E}_{t}^{\mathbb{Q}}( e^{ \int_{t}^{{T}_O} \{ -\frac{1}{2} (\eta[s,\mathrm{v}_s])^2 ds - \eta[s,\mathrm{v}_s] d z_s^{\mathbb{Q}} \}} \, \, {\Big |} \mathcal{I}_{T_O} ), \, \, \mathbb{E}_{t}^{\mathbb{Q}}( \mathbb{L}_t^{{T}_O}[k] \, {\Big |}\, \mathcal{I}_{T_O} ) ) ~ \mbox{ \, \, \quad } ~~ \label{eq:CovarianceForSpanned1} \\
& & ~ = \, \mathbb{E}_{t}^{\mathbb{Q}}( \mathbb{E}_{t}^{\mathbb{Q}}( e^{ \int_{t}^{{T}_O} \{ -\frac{1}{2} (\eta[s,\mathrm{v}_s])^2 ds - \eta[s,\mathrm{v}_s] d z_s^{\mathbb{Q}} \}} \, {\Big |} \mathcal{I}_{T_O} ) \, \, \mathbb{E}_{t}^{\mathbb{Q}}( \mathbb{L}_t^{{T}_O}[k] \, {\Big |} \mathcal{I}_{T_O} ) )
\nonumber \\
& &~~~~~~-~~
\overbrace{\mathbb{E}_{t}^{\mathbb{Q}}( \mathbb{E}_{t}^{\mathbb{Q}}( e^{ \int_{t}^{{T}_O} \{ -\frac{1}{2} (\eta[s,\mathrm{v}_s])^2 ds - \eta[s,\mathrm{v}_s] d z_s^{\mathbb{Q}} \}} \, {\Big |} \mathcal{I}_{T_O} ) )}^{= \, \, \mathbb{E}_{t}^{\mathbb{Q}}( e^{ \int_{t}^{{T}_O} \{ -\frac{1}{2} (\eta[s,\mathrm{v}_s])^2 ds - \eta[s,\mathrm{v}_s] d z_s^{\mathbb{Q}} \}} ) \, ~ = ~ 1} ~
\overbrace{\mathbb{E}_{t}^{\mathbb{Q}}( \mathbb{E}_{t}^{\mathbb{Q}}( \mathbb{L}_t^{{T}_O}[k] \,{\Big |} \mathcal{I}_{T_O} ) )}^{= \, \mathbb{E}_{t}^{\mathbb{Q}}( \mathbb{L}_t^{{T}_O}[k] )} ~ \mbox{ \, \, \quad } ~
\nonumber \\
& & ~ = \, \mathbb{E}_{t}^{\mathbb{Q}}( \underbrace{ ~ \mathbb{E}_{t}^{\mathbb{Q}}( e^{ \int_{t}^{{T}_O} \{ -\frac{1}{2} (\eta[s,\mathrm{v}_s])^2 ds - \eta[s,\mathrm{v}_s] d z_s^{\mathbb{Q}} \}} \, {\Big |} \mathcal{I}_{T_O} ) ~ }_{= \, \, e^{ \int_{t}^{{T}_O} \{ -\frac{1}{2} (\eta[s,\mathrm{v}_s])^2 ds - \eta[s,\mathrm{v}_s] d z_s^{\mathbb{Q}} \}}} \, \, \mathbb{E}_{t}^{\mathbb{Q}}( \mathbb{L}_t^{{T}_O}[k] \, {\Big |} \mathcal{I}_{T_O} ) ) ~ - ~ \mathbb{E}_{t}^{\mathbb{Q}}( \mathbb{L}_t^{{T}_O}[k] ) ~ \mbox{ \, \, \, \, \, \, } ~ ~ \nonumber \\
& & ~ = \, \mathbb{E}_{t}^{\mathbb{Q}}( \mathbb{E}_{t}^{\mathbb{Q}}( e^{ \int_{t}^{{T}_O} \{ -\frac{1}{2} (\eta[s,\mathrm{v}_s])^2 ds - \eta[s,\mathrm{v}_s] d z_s^{\mathbb{Q}} \}} \, \, \mathbb{L}_t^{{T}_O}[k] \,{\Big |} \mathcal{I}_{T_O} ) ) ~ - ~ \mathbb{E}_{t}^{\mathbb{Q}}( \mathbb{L}_t^{{T}_O}[k] ) ~ \mbox{ \, \, \, \, \, \, } ~ ~
\nonumber \\
& & ~ = \, \mathbb{E}_{t}^{\mathbb{Q}}( e^{ \int_{t}^{{T}_O} \{ -\frac{1}{2} (\eta[s,\mathrm{v}_s])^2 ds - \eta[s,\mathrm{v}_s] d z_s^{\mathbb{Q}} \}} \, \, \mathbb{L}_t^{{T}_O}[k] ) ~ - ~ \mathbb{E}_{t}^{\mathbb{Q}}( \mathbb{L}_t^{{T}_O}[k] ) ~ ~ ~ \mbox{ \, \, \, \, \, \, (now use (\ref{ah.1})) \, \, \, } \nonumber \\
& & ~ = \, \mathbb{E}_{t}^{\mathbb{Q}}( \mathcal{R}_{T_O}^{\tiny \mbox{span~diffusive}} \, \, \mathbb{L}_t^{{T}_O}[k] ) ~ - ~ \mathbb{E}_{t}^{\mathbb{Q}}( \mathbb{L}_t^{{T}_O}[k] ) ~ \mbox{ \, \, \, \, \, \, } \label{eq:Interim1} \\
& & ~ = \,
\mathrm{cov}_t^{\mathbb{Q}}(
e^{ \int_{t}^{{T}_O} \{ -\frac{1}{2} (\eta[s,\mathrm{v}_s])^2 ds - \eta[s,\mathrm{v}_s] d z_s^{\mathbb{Q}} \}}, \mathbb{L}_t^{{T}_O}[k] ).
\label{eqa.contt0}
\end{eqnarray}
We now use Tanaka's formula in equation (\ref{eq:Interim1}) to
substitute out $\mathbb{L}_t^{{T}_O}[k]$ and
re-express our quantity of interest in terms of option payoffs.
From the definition of expected call returns in equation (\ref{eq:ExpectedHoldingReturn1GneralDynamics}), we note that the
expected \emph{excess} return of holding a call option over $t$ to ${T}_O$ is
\begin{eqnarray}
\underbrace{1 + \mu^{{T}_O}_{t,{\tiny \mathrm{call}}}[k] - e^{r ({T}_O - t)}}_{\tiny \mbox{expected~excess~return~of~calls}} ~ = ~ e^{r ({T}_O - t)} ~ \, \frac{\mathbb{E}_{t}^{\mathbb{P}}( \max(G_{{T}_O} - k, 0) ) \, - \, \mathbb{E}_{t}^{\mathbb{Q}}( \max(G_{{T}_O} - k,0) )}
{ \mathbb{E}_{t}^{\mathbb{Q}}(
\max(G_{{T}_O} - k, 0)  )}. ~ \mbox{ \, \, \, } \label{eq:ExpectedExcessHoldingReturn1GeneralDynamics}
\end{eqnarray} } 
Thus, the call option risk premium inherits the sign of (using Tanaka's formula)
\begin{eqnarray}
\mathbb{E}_{t}^{\mathbb{P}}( \max( G_{T_O} - k, 0 ) ) -
\mathbb{E}_{t}^{\mathbb{Q}}( \max( G_{T_O} - k, 0 ) )
&=& \mathbb{E}_{t}^{\mathbb{P}}( \int_{t+}^{T_O} \mathbbm{1}_{\{G_{\ell  -} > k\}} dG_{\ell})
~~ \nonumber \\
&+& ~ \mathbb{E}_{t}^{\mathbb{P}}( \mathbb{L}^{T_O}_t[k]) ~ -
~ \mathbb{E}_{t}^{\mathbb{Q}}( \mathbb{L}^{T_O}_t[k])  \nonumber \\
&+&
\underbrace{\mathbb{E}_{t}^{\mathbb{P}}( a_t^{T_O}[k] + b_t^{T_O}[k] ) - \mathbb{E}_{t}^{\mathbb{Q}}( a_t^{T_O}[k]
+b_t^{T_O}[k]).}_{\text{ \tiny \, (risk premium for jumps crossing the strike (already signed in
Section~\ref{app:jumps_across})) }} ~~~ \mbox{ \, \, \, \, \, \, }. \nonumber
\end{eqnarray}
To further reduce the problem to what we have already derived based on conditioning on $\mathcal{I}_{T_O}$, note the following simplification
steps:
 \begin{eqnarray}
& & \mathbb{E}_{t}^{\mathbb{P}}( \max( G_{T_O} - k, 0 ) ) ~ - ~
\mathbb{E}_{t}^{\mathbb{Q}}( \max( G_{T_O} - k, 0 ) )  \nonumber \\
& & ~ ~ = ~ \mathbb{E}_{t}^{\mathbb{P}}( \int_{t+}^{T_O} \mathbbm{1}_{\{G_{\ell  -} > k\}} dG_{\ell}) ~ \nonumber \\
&&~~~ ~ ~ \mbox{ \, \, \, \, }
+\overbrace{\mathbb{E}_{t}^{\mathbb{Q}}( \mathcal{R}_{T_O}^{\tiny \mbox{span~diffusive}} \, \, \mathbb{L}_t^{{T}_O}[k] ) ~ - ~ \mathbb{E}_{t}^{\mathbb{Q}}( \mathbb{L}_t^{{T}_O}[k] )}^{\text{ from
(\ref{eq:Interim1})}}  ~ ~
\nonumber \\
& & ~~~ ~ ~ \mbox{ \, \, \, \, } ~ + ~ \overbrace{\mathbb{E}_{t}^{\mathbb{Q}}( \mathcal{R}_{T_O}^{\tiny \mbox{span~diffusive}} \times
\mathrm{cov}_t^{\mathbb{Q}}(
\mathcal{R}_{T_O}^{\tiny \mbox{unspan~diffusive}}
\times \mathcal{R}_{T_O}^{\tiny \mbox{unspan~jump}},
\, \mathbb{L}_t^{{T}_O}[k] {\Big |} \, \mathcal{I}_{T_O} ) )}^{\text{ from
(\ref{eq:ConditCovarSpanning2Linesa}) }} ~ \mbox{ \, \, \, \, } ~ \mbox{ \, \, } \nonumber \\
& & ~~~ ~ ~ \mbox{ \, \, \, \, } ~ + ~
\mathbb{E}_{t}^{\mathbb{P}}( a_t^{T_O}[k]  +  b_t^{T_O}[k] )
- \mathbb{E}_{t}^{\mathbb{Q}}( a_t^{T_O}[k] + b_t^{T_O}[k]).
~~~ \mbox{ \, \, \, \, \, \, } \label{eq:ReexpressCallRiskPremiumMinus2}
\end{eqnarray}

Rearranging for clarity and to see the term that is left to be signed, we then have
 \begin{eqnarray}
& & \mathbb{E}_{t}^{\mathbb{P}}( \max( G_{T_O} - k, 0 ) ) ~ - ~
\mathbb{E}_{t}^{\mathbb{Q}}( \max( G_{T_O} - k, 0 ) )  \nonumber \\
& & ~ = ~
\overbrace{\mathbb{E}_{t}^{\mathbb{P}}( a_t^{T_O}[k] ) + b_t^{T_O}[k] )
- \mathbb{E}_{t}^{\mathbb{Q}}( a_t^{T_O}[k] + b_t^{T_O}[k])}^{\text{ \tiny \, (already signed in
Section~\ref{app:jumps_across})}} ~~~ \mbox{ \, \, \, \, \, \, } \nonumber \\
& & ~ ~ ~ + ~ \overbrace{\mathbb{E}_{t}^{\mathbb{Q}}( \mathcal{R}_{T_O}^{\tiny \mbox{span~diffusive}} \times
\mathrm{cov}_t^{\mathbb{Q}}(
\mathcal{R}_{T_O}^{\tiny \mbox{unspan~diffusive}}
\times \mathcal{R}_{T_O}^{\tiny \mbox{unspan~jump}},
\, \mathbb{L}_t^{{T}_O}[k] {\Big |} \, \mathcal{I}_{T_O} ) )}^{\text{ \tiny \, (already signed by equations (\ref{eq:ConditCovarSpanning2Linesa}) and
(\ref{eq:JumpRelatedLT2})) }} ~ \mbox{ \, \, \, \, }
\nonumber \\
& & ~ ~ ~ + ~ \mathbb{E}_{t}^{\mathbb{P}}( \int_{t+}^{T_O} \mathbbm{1}_{\{G_{\ell  -} > k\}} dG_{\ell} ) ~ - ~ \overbrace{\mathbb{E}_{t}^{\mathbb{P}}( \frac{1}{\mathcal{R}_{T_O}^{\tiny \mbox{unspan~diffusive}} \, \mathcal{R}_{T_O}^{\tiny \mbox{unspan~jump}}} \, \, \int_{t+}^{T_O} \mathbbm{1}_{\{G_{\ell  -} > k\}} dG_{\ell} )}^{= ~ \mathbb{E}_{t}^{\mathbb{Q}}(  \mathcal{R}_{T_O}^{\tiny \mbox{span~diffusive}} \, \int_{t+}^{T_O} \mathbbm{1}_{\{G_{\ell  -} > k\}} dG_{\ell} )} ~ ~  \mbox{ \, } ~
\nonumber
\\
& & ~ ~ ~ + ~
\underbrace{ ~ \mathbb{E}_{t}^{\mathbb{Q}}( \mathcal{R}_{T_O}^{\tiny \mbox{span~diffusive}} \, \max( G_{T_O} - k, 0 ) ) ~ - ~
\mathbb{E}_{t}^{\mathbb{Q}}( \max( G_{T_O} - k, 0 ) ).}_{\text{ \tiny
$= \, \mathbb{E}_{t}^{\mathbb{Q}}(  \mathcal{R}_{T_O}^{\tiny \mbox{span~diffusive}} \, \int_{t+}^{T_O} \mathbbm{1}_{\{G_{\ell  -} > k\}} dG_{\ell} ) ~ + ~ \mathbb{E}_{t}^{\mathbb{Q}}( \mathcal{R}_{T_O}^{\tiny \mbox{span~diffusive}} \, \, \mathbb{L}_t^{{T}_O}[k] ) ~ - ~ \mathbb{E}_{t}^{\mathbb{Q}}( \mathbb{L}_t^{{T}_O}[k] )$ }}
\label{eq:ReexpressCallRiskPremium}
\end{eqnarray}
The last two terms in (\ref{eq:ReexpressCallRiskPremium}) are adding and subtracting the same quantity. This
a consequence of using Tanaka's formula to
reverse engineer
the local time $\mathbb{L}_t^{{T}_O}[k]$ in terms of the call payoff.
Further recognize that
$\mathbb{E}_{t}^{\mathbb{Q}}(  \mathcal{R}_{T_O}^{\tiny \mbox{span~diffusive}} \, \int_{t+}^{T_O} \mathbbm{1}_{\{G_{\ell  -} > k\}} dG_{\ell} )$
and $\mathbb{E}_{t}^{\mathbb{P}}( \frac{1}{\mathcal{R}_{T_O}^{\tiny \mbox{unspan~diffusive}} \, \mathcal{R}_{T_O}^{\tiny \mbox{unspan~jump}}} \, \, \int_{t+}^{T_O} \mathbbm{1}_{\{G_{\ell  -} > k\}} dG_{\ell} )$ are identical (by Girsanov's Theorem and the definitions in equations (\ref{eq:RecipPK5TermsDPSLT1}) and (\ref{ah.1})).


\vspace{1mm}

The following feature is
evident from equation (\ref{eq:ReexpressCallRiskPremium}):
\begin{itemize}

\item In the special case that \emph{there are no unspanned risks in the pricing kernel}, we would have
(i) $\mathcal{R}_{T_O}^{\tiny \mbox{unspan~diffusive}} \equiv 1$ and (ii) $\mathcal{R}_{T_O}^{\tiny \mbox{unspan~jump}} \equiv 1$
(state-by-state). Hence,
\emph{the call option risk premium would inherit the same sign as that of the final line}, specifically
of $\mathbb{E}_{t}^{\mathbb{Q}}( \mathcal{R}_{T_O}^{\tiny \mbox{span~diffusive}} \max( G_{T_O} - k, 0 ) ) -
\mathbb{E}_{t}^{\mathbb{Q}}( \max( G_{T_O} - k, 0 ) )$, since the first three lines of equation (\ref{eq:ReexpressCallRiskPremium})
would vanish.

\end{itemize}
\vspace{1mm}
We will now show that the sign of the final line of equation (\ref{eq:ReexpressCallRiskPremium}) is
positive regardless of whether or not there are unspanned risks in the
pricing kernel. \vspace{1mm}

\noindent \textbf{Result.} The following result is true:
\begin{eqnarray}
\mathbb{E}_{t}^{\mathbb{Q}}( \mathcal{R}_{T_O}^{\tiny \mbox{span~diffusive}} \, \max( G_{T_O} - k, 0 ) ) ~ - ~
\mathbb{E}_{t}^{\mathbb{Q}}( \max( G_{T_O} - k, 0 ) ) ~ > ~ 0.
\label{eq:InequalityInBlackScholesTypeFormula3}
\end{eqnarray}

\noindent \textbf{Proof.} The proof of this result is tedious and presented
next in Section~\ref{appsec:cccv1}.
$\blacksquare$ \vspace{-4mm}


\subsection{Proof that equation (\ref{eq:InequalityInBlackScholesTypeFormula3}) holds}
\label{appsec:cccv1}


By conditioning
on the jump component of the equity futures and its variance,
and exploiting independence from the diffusive
components,
one can see that, for the purpose of determining
the \emph{sign} of $\mathbb{E}_{t}^{\mathbb{Q}}( \mathcal{R}_{T_O}^{\tiny \mbox{span~diffusive}}  \max( G_{T_O} - k, 0 ) ) -
\mathbb{E}_{t}^{\mathbb{Q}}( \max( G_{T_O} - k, 0 ) )$, one can reduce the problem
to computing this quantity when there are \emph{no jumps}.

In other words, with no loss of generality, we are justified in working
with the following $\mathbb{Q}$ dynamics:
\begin{eqnarray}
\frac{d G_{t}}{G_{t}} & = & \sqrt{\mathrm{v}_t} \,d z^{\mathbb{Q}}_t ~~ \mbox{ \, \, \, \, \, \, \, \quad \quad } ~~ ~ ~\mathrm{and}~ ~ ~~~~ \label{eqa:spp1a}
\\
d \mathrm{v}_t & = &  ( \phi_{\tiny \mbox{vol}}^{\mathbb{Q}} - \kappa_{\tiny \mbox{vol}}^{\mathbb{Q}} \,\mathrm{v}_t )\, dt ~ + ~
\sigma_{\tiny \mbox{vol}} \, \sqrt{\mathrm{v}_t} \,\rho_{\tiny \mbox{vol}} \,d z_t^{\mathbb{Q}}
~+~\sigma_{\tiny \mbox{vol}} \sqrt{\mathrm{v}_t} \, \sqrt{1-\rho^2_{\tiny \mbox{vol}}  }
\, du_t^{\mathbb{Q}}.
~ ~ \mbox{ \, \, } ~ \mbox{ \, \, \, \, \, \, }
\label{eqa:spp1}
\end{eqnarray}
\noindent \textbf{Step 1.} For the purpose of the proof, we recast the Brownian motions by introducing
independent Brownian
motions $w_t^{(\mathbb{Q},1)}$ and $w_t^{(\mathbb{Q},2)}$, under
$\mathbb{Q}$, as follows:
\begin{eqnarray}
w_t^{(\mathbb{Q},1)} \, = \, \sqrt{1-\rho^2_{\tiny \mbox{vol}}} \, z_t^{\mathbb{Q}} \, - \, \rho_{\tiny \mbox{vol}} \,  u_t^{\mathbb{Q}} ~ \mbox{ \, \, \quad and \, \, \, \, } ~
w_t^{(\mathbb{Q},2)} \, = \, \rho_{\tiny \mbox{vol}} \, z_t^{\mathbb{Q}} \, + \,
\sqrt{1-\rho^2_{\tiny \mbox{vol}}} \, u_t^{\mathbb{Q}}. ~ \mbox{ \, \, \, \, \, \, \, \, } \label{eq:CholeskyTypeDef}
\end{eqnarray}
Hence, we have
\begin{equation}
z_t^{\mathbb{Q}} \, = \, \sqrt{1-\rho^2_{\tiny \mbox{vol}}} \, w_t^{(\mathbb{Q},1)} \, + \, \rho_{\tiny \mbox{vol}} \,  w_t^{(\mathbb{Q},2)}.
\end{equation}

\noindent \textbf{Step 2.} The variance process $(\mathrm{v}_s)$
is driven only by $(w_s^{(\mathbb{Q},2)})$.
Next, we proceed as follows:
\begin{eqnarray}
G_{T_O} & = & \overbrace{G_t}^{= \, 1} \, \, e^{ \int_{t}^{{T}_O} \{ -\frac{1}{2} \mathrm{v}_s\, ds + \sqrt{\mathrm{v}_s} \, d z_s^{\mathbb{Q}} \}} ~ \, \nonumber \\
&=&  e^{ \int_{t}^{{T}_O} \{ -\frac{1}{2} \mathrm{v}_s \, ds + \sqrt{\mathrm{v}_s} \,
[\sqrt{1-\rho^2_{\tiny \mbox{vol}}} \, d w_s^{(\mathbb{Q},1)} \, + \, \rho_{\tiny \mbox{vol}} \, d w_s^{(\mathbb{Q},2)}] \}}
~ \mbox{ \, \, }\\
& = & {G}_{T_O}^{\perp} \, e^{ \int_{t}^{{T}_O} \{ -\frac{1}{2} (\sqrt{\mathrm{v}_s})^2 \, (1-\rho^2_{\tiny \mbox{vol}}) \, ds
+ (\sqrt{\mathrm{v}_s}) \, \sqrt{1-\rho^2_{\tiny \mbox{vol}}} \, d w_s^{(\mathbb{Q},1)} \}}.
\end{eqnarray}
Additionally,
\begin{eqnarray}
\mathcal{R}_{T_O}^{\tiny \mbox{span~diffusive}} & = &
e^{ \int_{t}^{{T}_O} \{ -\frac{1}{2} (\eta[s,\mathrm{v}_s])^2 \, ds - \eta[s,\mathrm{v}_s] \, d z_s^{\mathbb{Q}} \}} \mbox{ \, } \mbox{ \, \, \, }
\nonumber \\
& = &
\mathcal{R}_{T_O}^{\perp} \, e^{ \int_{t}^{{T}_O} \{ -\frac{1}{2} (\eta[s,\mathrm{v}_s])^2 (1-\rho^2_{\tiny \mbox{vol}}) \, ds + \eta[s,\mathrm{v}_s] \sqrt{1-\rho^2_{\tiny \mbox{vol}}}\, d w_s^{(\mathbb{Q},1)} \}}. ~
~ \mbox{ \, \, } ~~
\end{eqnarray}
Finally,
\begin{eqnarray}
\mathcal{R}_{T_O}^{\tiny \mbox{span~diffusive}} \, G_{T_O}
& = & e^{ \int_{t}^{{T}_O} \{ -\frac{1}{2} (\eta[s,\mathrm{v}_s])^2 ds - \eta[s,\mathrm{v}_s] (\sqrt{1-\rho^2_{\tiny \mbox{vol}}} d w_s^{(\mathbb{Q},1)} \, + \, \rho_{\tiny \mbox{vol}} d w_s^{(\mathbb{Q},2)}) \}} ~ \mbox{ \, } \times \mbox{ \, \, \, } \nonumber \\
& & ~ \mbox{ \, \, \, \, } ~ ~ \overbrace{G_t}^{= \, 1} \, \, e^{ \int_{t}^{{T}_O} \{ -\frac{1}{2} \mathrm{v}_s ds + \sqrt{\mathrm{v}_s} (\sqrt{1-\rho^2_{\tiny \mbox{vol}}} d w_s^{(\mathbb{Q},1)} \, + \, \rho_{\tiny \mbox{vol}} d w_s^{(\mathbb{Q},2)}) \}} ~ \mbox{ \, } \mbox{ \, \, \, } \nonumber \\
& = & \mathcal{R}_{T_O}^{\perp}  \, {G}_{T_O}^{\perp} \,
{\cal V}_{T_{O}}^{\bullet}
 \, e^{ \int_{t}^{{T}_O} \{ -\frac{1}{2} (\sqrt{\mathrm{v}_s}-\eta[s,\mathrm{v}_s])^2 (1-\rho^2_{\tiny \mbox{vol}}) \,ds + (\sqrt{\mathrm{v}_s}-\eta[s,\mathrm{v}_s]) \sqrt{1-\rho^2_{\tiny \mbox{vol}}} \, d w_s^{(\mathbb{Q},1)} \}}. \nonumber
~ \mbox{ \, \, \, } ~
\end{eqnarray}
We have defined, for compactness of presentation, the following quantities:
\begin{eqnarray}
\mathcal{R}_{T_O}^{\perp} & \equiv & e^{ \int_{t}^{{T}_O} \{ -\frac{1}{2} (-\eta[s,\mathrm{v}_s])^2 \, \rho^2_{\tiny \mbox{vol}} \, ds + (-\eta[s,\mathrm{v}_s]) \, \rho_{\tiny \mbox{vol}} \, d w_s^{(\mathbb{Q},2)} \}},
~ ~ ~ \mbox{ \, \, \, \, \, \, \, \, } ~ ~ ~ ~
\label{cvg.21} \\
{G}_{T_O}^{\perp} & \equiv & e^{ \int_{t}^{{T}_O} \{ -\frac{1}{2} (\sqrt{\mathrm{v}_s})^2 \rho^2_{\tiny \mbox{vol}} ds + (\sqrt{\mathrm{v}_s}) \, \rho_{\tiny \mbox{vol}}\, d w_s^{(\mathbb{Q},2)} \}}, ~~ \mbox{ \, \, \, \, \, \quad \quad \quad \quad } ~ \mbox{ \, and } ~ ~ ~
\label{cvg.22}\\
{\cal V}_{T_{O}} & \equiv & {\cal V}_{T_{O}}^{\bullet} \, \, {\cal V}_{T_{O}}^{\perp}, ~~ ~~~ \mbox{ \, \, \, \, \quad \quad \quad \quad } \mbox{ \, \, \, \, \quad \quad \quad \quad \quad \quad \, \, \, where } ~~~
\\
{\cal V}_{T_{O}}^{\bullet} & \equiv & e^{ \int_{t}^{{T}_O}
\{ (1-\rho^2_{\tiny \mbox{vol}}) \sqrt{\mathrm{v}_s} ( - \eta[s,\mathrm{v}_s] )\, ds \}}
~ \mbox{ \, \, \, \, \, and \, \, \, \,  } ~ {\cal V}_{T_{O}}^{\perp} ~ \, \equiv
~ \, e^{ \int_{t}^{{T}_O}
\{ \rho^2_{\tiny \mbox{vol}} \sqrt{\mathrm{v}_s} ( - \eta[s,\mathrm{v}_s] ) \, ds \}}. ~ \mbox{ \, \, \, \, \, \, } ~
~ \label{eq:DefinitionOfA-TO}
\end{eqnarray}
\noindent \textbf{Step 3.} With these substitutions,
we have decomposed
$G_{T_O}$,
$\mathcal{R}_{T_O}^{\tiny \mbox{span~diffusive}}$, and
$\mathcal{R}_{T_O}^{\tiny \mbox{span~diffusive}} \, G_{T_O}$
into the product of terms whose increments are (instantaneously)
perfectly correlated with $(w_s^{(\mathbb{Q},1)})$ and terms
(i.e., ${G}_{T_O}^{\perp}$ and $\mathcal{R}_{T_O}^{\perp}$),
whose
increments are independent of $(w_s^{(\mathbb{Q},1)})$,
as well as a term ${\cal V}_{T_{O}}^{\bullet}$ which is
informative about the sign of the equity premium. Furthermore,
\begin{equation}
\mbox{the variance process $(\mathrm{v}_s)$ is independent of the Brownian motion} ~ (w_s^{(\mathbb{Q},1)}). ~ ~ ~ ~ \mbox{ \, \, \, \, \, \, }
\end{equation}

Hence, the distribution of
(i) $\log( \frac{\mathcal{R}_{T_O}^{\tiny \mbox{span~diffusive}} \, G_{T_O}}{ \mathcal{R}_{T_O}^{\perp} \,{G}_{T_O}^{\perp} \, {\cal V}_{T_{O}}^{\bullet}
} )$,
(ii) $\log( \frac{G_{T_O}}{{G}_{T_O}^{\perp}} )$,
and (iii) $\log( \frac{\mathcal{R}_{T_O}^{\tiny \mbox{span~diffusive}}}{\mathcal{R}_{T_O}^{\perp}} )$, conditional on
the path of variance $\lbrace \mathrm{v}_s, t \leq s \leq T_{O} \rbrace$
and on $\mathcal{F}_t$, is jointly normal
with
\begin{eqnarray}
\log( \frac{\mathcal{R}_{T_O}^{\tiny \mbox{span~diffusive}} \, G_{T_O}}{\mathcal{R}_{T_O}^{\perp} \,{G}_{T_O}^{\perp} \, {\cal V}_{T_{O}}^{\bullet} } ) {\Big |} \lbrace \mathrm{v}_s, t \leq s \leq T_{O} \rbrace, \mathcal{F}_t ~ & \sim & {\cal N}( - \frac{1}{2}
\mathbbm{d}_{t,T_O}^2, \mathbbm{d}_{t,T_O}^2 ), ~ ~ \mbox{ \, \, \, \quad } ~ ~ ~ \mbox{ \, \, } \nonumber \\
\log( \frac{G_{T_O}}{{G}_{T_O}^{\perp}} ) {\Big |} \lbrace \mathrm{v}_s, t \leq s \leq T_{O} \rbrace, \mathcal{F}_t ~ & \sim & {\cal N}( - \frac{1}{2} \mathbbm{v}_{t,T_O}^2, \mathbbm{v}_{t,T_O}^2 ), ~ \mbox{ \, \, \, } ~ \mbox{ \, \, \, and \, \, \, }
\nonumber \\
\log( \frac{\mathcal{R}_{T_O}^{\tiny \mbox{span~diffusive}}}{\mathcal{R}_{T_O}^{\perp}} )
{\Big |} \lbrace \mathrm{v}_s, t \leq s \leq T_{O} \rbrace, \mathcal{F}_t ~ & \sim & {\cal N}( - \frac{1}{2} \mathbbm{e}_{t,T_O}^2, \mathbbm{e}_{t,T_O}^2 ), ~ ~ \mbox{ \, \, \, \quad } ~ ~ ~ \mbox{ \, \, }
\end{eqnarray}
where
\begin{eqnarray}
\mathbbm{d}_{t,T_O}^2 &\equiv&  \int_{t}^{{T}_O} (\sqrt{\mathrm{v}_s}-\eta[s,\mathrm{v}_s])^2 (1-\rho^2_{\tiny \mbox{vol}}) ds, \\
\mathbbm{v}_{t,T_O}^2 &\equiv& \int_{t}^{{T}_O} \mathrm{v}_s (1-\rho^2_{\tiny \mbox{vol}}) ds, ~ ~ ~ \mbox{ \, \, \, \, \, \, \, \, \, \, and \, \, \, \, \, \, \, \, \, \, } ~ ~ ~ \\
\mathbbm{e}_{t,T_O}^2  &\equiv& \int_{t}^{{T}_O} (\eta[s,\mathrm{v}_s])^2 (1-\rho^2_{\tiny \mbox{vol}}) \,ds.
\end{eqnarray}

\noindent \textbf{Step 4.} Using a technique
that
is a variant of \citet*{HullWhite:87},
we condition on the path of variance
$\lbrace \mathrm{v}_s, t \leq s \leq T_{O} \rbrace$ and on $\mathcal{F}_t$, to
derive as follows:
\small
\begin{eqnarray}
& & \mathbb{E}_{t}^{\mathbb{Q}}( \mathcal{R}_{T_O}^{\tiny \mbox{span~diffusive}} \,
\max( G_{T_O} - k, 0 ) )
\nonumber \\
& & ~ ~ \mbox{ \, \quad } = \,
\mathbb{E}_{t}^{\mathbb{Q}}( \mathcal{R}_{T_O}^{\perp} \, \{
{G}_{T_O}^{\perp} \,{\cal V}_{T_{O}}^{\bullet}  {\cal N}\big( \frac{\log(\frac{
{G}_{T_O}^{\perp} \, {\cal V}_{T_{O}}^{\bullet}   }{k})
+ \frac{1}{2} \mathbbm{v}_{t,T_O}^2 }{ \mathbbm{v}_{t,T_O}} \big)
 \, - \, k {\cal N}\big( \frac{\log(\frac{ {G}_{T_O}^{\perp}  \, {\cal V}_{T_{O}}^{\bullet}  }{k}) - \frac{1}{2} \mathbbm{v}_{t,T_O}^2 }{\mathbbm{v}_{t,T_O}} \big) \} \, ), ~ \mbox{ \, \, \, \quad \quad } ~
\end{eqnarray} \normalsize
where ${\cal N}(.)$ is the standard normal cumulative distribution function.
Similarly,
\small
\begin{eqnarray}
\mathbb{E}_{t}^{\mathbb{Q}}( \max( G_{T_O} - k, 0 ) ) &= &
 \mathbb{E}_{t}^{\mathbb{Q}}( \underbrace{ {G}_{T_O}^{\perp} {\cal N}\big( \frac{\log(\frac{ {G}_{T_O}^{\perp}}{k}) + \frac{1}{2} \mathbbm{v}_{t,T_O}^2
 }{ \mathbbm{v}_{t,T_O}} \big) \, - \, k {\cal N}\big( \frac{\log(\frac{ {G}_{T_O}^{\perp}}{k}) - \frac{1}{2} \mathbbm{v}_{t,T_O}^2 }{ \mathbbm{v}_{t,T_O}} \big)}_{\equiv ~ \mathrm{call}_t^{\tiny \mbox{BS}}[
{G}_{T_O}^{\perp}, k ]} ). \mbox{ \, \, \, \, \, \, \, \, } ~ \mbox{ \, \, \quad }  \label{eq:BlackScholesTypeFormula2}
\end{eqnarray} \normalsize
\noindent \textbf{Step 5.} We ask the following question:
\begin{eqnarray}
\mathrm{When~is}~\mathbb{E}_{t}^{\mathbb{Q}}( \mathcal{R}_{T_O}^{\tiny \mbox{span~diffusive}} \, \max( G_{T_O} - k, 0 ) ) ~ - ~
\mathbb{E}_{t}^{\mathbb{Q}}( \max( G_{T_O} - k, 0 ) ) ~ > ~ 0  \mbox{?} ~~~ \mbox{ \, } \mbox{ \, \, \, }
\end{eqnarray}
A few observations are in order. First, the equity premium is positive when $\alpha_{\tiny \mbox{vol}} > 0$ and $\lambda_{\tiny \mbox{vol}} > 0$. This
implies that $\eta[s,\mathrm{v}_s]= - \frac{1}{ \sqrt{\mathrm{v}_s}}(
\alpha_{\tiny \mbox{vol}} +\lambda_{\tiny \mbox{vol}} \, \mathrm{v}_s) < 0$.

Hence, by equation (\ref{eq:DefinitionOfA-TO}) and $\eta[s,\mathrm{v}_s]<0$, it holds that
\begin{equation}
{\cal V}_{T_{O}} \, > \, 1 ~~ \mbox{ \, \, \, } ~ \mathrm{and} ~~ \mbox{ \, \, \, } ~ {\cal V}_{T_{O}}^{\bullet} \, > \, 1. ~~~ \mbox{ \, \, \, \, \, \, }
\end{equation}
Since call option prices are monotonically increasing
in the price of the underlying and since
${G}_{T_O}^{\perp} \mathcal{V}_{T_{O}}^{\bullet} \, > \, {G}_{T_O}^{\perp}$, we have
\begin{equation}
\mathrm{call}_t^{\tiny \mbox{BS}}[ {G}_{T_O}^{\perp} \mathcal{V}_{T_{O}}^{\bullet} , k]
\, > \,
\mathrm{call}_t^{\tiny \mbox{BS}}[ {G}_{T_O}^{\perp}, k ]. ~ ~ ~ \label{eq:FirstIneq1}
\end{equation}
We note that the covariance between $\mathcal{R}_{T_O}^{\perp}$
and
${G}_{T_O}^{\perp}$ under $\mathbb{Q}$ is positive (by
(\ref{cvg.21})--(\ref{cvg.22})
and since $(-\eta[s,\mathrm{v}_s]) \sqrt{\mathrm{v}_s} >0$).
Furthermore, call options have a nonnegative delta. The upshot is
that $\mathcal{R}_{T_O}^{\perp}$ and $\mathrm{call}_t^{\tiny \mbox{BS}}[ {G}_{T_O}^{\perp}, k ]$ have a positive
covariance under $\mathbb{Q}$. With
$\mathbb{E}_{t}^{\mathbb{Q}}( \mathcal{R}_{T_O}^{\perp})=1$, it holds that
\begin{equation}
\mathbb{E}_{t}^{\mathbb{Q}}( \mathcal{R}_{T_O}^{\perp} \, \{
\mathrm{call}_t^{\tiny \mbox{BS}}[ {G}_{T_O}^{\perp}, k ] \} )
\, - \, \mathbb{E}_{t}^{\mathbb{Q}}( \mathrm{call}_t^{\tiny \mbox{BS}}[ {G}_{T_O}^{\perp}, k ] ) ~ = ~
\mathrm{cov}_t^{\mathbb{Q}}( \mathcal{R}_{T_O}^{\perp},  \mathrm{call}_t^{\tiny \mbox{BS}}[ {G}_{T_O}^{\perp}, k ]  ) \, > \, 0.
~ \mbox{ \, } ~ \label{eq:SecondIneq2}
\end{equation}
Therefore,
combining (\ref{eq:FirstIneq1})
and
(\ref{eq:SecondIneq2}), we have $\mathbb{E}_{t}^{\mathbb{Q}}( \mathcal{R}_{T_O}^{\perp} \, \{
\mathrm{call}_t^{\tiny \mbox{BS}}[ {G}_{T_O}^{\perp} \mathcal{V}_{T_{O}}^{\bullet}, k ] \} )
 >  \mathbb{E}_{t}^{\mathbb{Q}}( \mathrm{call}_t^{\tiny \mbox{BS}}[ {G}_{T_O}^{\perp}, k ] )$.
The consequence is that
\begin{eqnarray}
\mathbb{E}_{t}^{\mathbb{Q}}( \mathcal{R}_{T_O}^{\tiny \mbox{span~diffusive}} \, \max( G_{T_O} - k, 0 ) ) ~ - ~
\mathbb{E}_{t}^{\mathbb{Q}}( \max( G_{T_O} - k, 0 ) ) ~~ > ~ 0.
~ \mbox{ \, } \mbox{ \, \, \, }
\label{app:caaal1}
\end{eqnarray}
We have the proof. $\blacksquare$ \vspace{-4mm}

\subsection{No unspanned risks in the pricing kernel imply zero straddle risk premium}
\label{app:jumps_acrossPutsStraddles}

The statement to prove is the following: When there are no unspanned risks in the
pricing kernel, the straddle risk premium (corresponding to $k=1$) is zero.


For the proof, we
first state the following companion result corresponding to equation (\ref{app:caaal1}) for OTM puts (steps are similar and omitted):

\noindent \textbf{Result.} The following result is true:
\begin{eqnarray}
\mathbb{E}_{t}^{\mathbb{Q}}( \mathcal{R}_{T_O}^{\tiny \mbox{span~diffusive}} \, \max( k - G_{T_O}, 0 ) ) ~ - ~
\mathbb{E}_{t}^{\mathbb{Q}}( \max( k - G_{T_O}, 0 ) ) ~ < ~ 0.
\label{eq:InequalityInBlackScholesTypeFormula3Put}
\end{eqnarray}
Move next to our object of interest, specifically the straddle risk premium.

Recall from Appendix~\ref{appsec:jumppps} (part III) that
\begin{eqnarray*}
\mathbb{A}_t^{T_O}[1] ~ \equiv ~  \sum_{t < \ell \leq T_O} \underbrace{\{
 \mathbbm{1}_{\{G_{\ell \, -} < 1\}} \, \max(  G_{\ell}- 1, 0 ) \, + \,
 \mathbbm{1}_{\{G_{\ell  -} > 1\}} \, \max( 1 - G_{\ell}, 0 )\}}_{\tiny \mbox{jumps~crossing~the~strike~from~below~and~above,}~ k = 1}. ~ \mbox{ \, \, \, \, } ~ &
\end{eqnarray*} 
{Using
equations (\ref{eq:ReexpressCallRiskPremium}) and
(\ref{eq:BlackScholesTypeFormula2}) (as well as the analogous
(but, for brevity, not presented)
equations for put options)
the sign of the risk premium on ATM straddles is the same as the sign of
\begin{eqnarray}
& & 2 \overbrace{\mathbb{E}_{t}^{\mathbb{Q}}( \mathcal{R}_{T_O}^{\tiny \mbox{span~diffusive}} \times
\mathrm{cov}_t^{\mathbb{Q}}(
\mathcal{R}_{T_O}^{\tiny \mbox{unspan~diffusive}}
\times \mathcal{R}_{T_O}^{\tiny \mbox{unspan~jump}},
\mathbb{L}_t^{{T}_O}[1] {\Big |}  \mathcal{I}_{T_O} ) )}^{\text{ \tiny (already signed by equations (\ref{eq:ConditCovarSpanning2Linesa}) and
(\ref{eq:JumpRelatedLT2})) }} \nonumber \\
&&~~
+2
\{\mathbb{E}_{t}^{\mathbb{P}}( \mathbb{A}_t^{T_O}[1] ) - \mathbb{E}_{t}^{\mathbb{Q}}( \mathbb{A}_t^{T_O}[1] )\} ~
\nonumber \\
& & ~ ~ ~ + ~ \mathbb{E}_{t}^{\mathbb{P}}( \int_{t+}^{T_O} \mathbbm{1}_{\{G_{\ell  -} > 1\}} dG_{\ell} ) ~ \nonumber \\
&&- ~ \overbrace{\mathbb{E}_{t}^{\mathbb{P}}( \frac{1}{\mathcal{R}_{T_O}^{\tiny \mbox{unspan~diffusive}} \, \mathcal{R}_{T_O}^{\tiny \mbox{unspan~jump}}} \, \, \int_{t+}^{T_O} \mathbbm{1}_{\{G_{\ell  -} > 1\}} dG_{\ell} )}^{= ~ \mathbb{E}_{t}^{\mathbb{Q}}(  \mathcal{R}_{T_O}^{\tiny \mbox{span~diffusive}} \, \int_{t+}^{T_O} \mathbbm{1}_{\{G_{\ell  -} > 1\}} dG_{\ell} )} ~ ~  \mbox{ \, } ~
\nonumber \\
& & ~ ~ ~ + ~
\underbrace{ ~ \mathbb{E}_{t}^{\mathbb{Q}}( \mathcal{R}_{T_O}^{\tiny \mbox{span~diffusive}} \, \max( G_{T_O} - 1, 0 ) )}_{\text{ \tiny
$= \, \mathbb{E}_{t}^{\mathbb{Q}}( \mathcal{R}_{T_O}^{\perp} \,
\{ \mathrm{call}_t^{\tiny \mbox{BS}}[ {G}_{T_O}^{\perp} \mathcal{V}_{T_{O}}^{\bullet} , 1] \} )$ }} ~ - ~
\underbrace{ ~ \mathbb{E}_{t}^{\mathbb{Q}}( \max( G_{T_O} - 1, 0 ) )}_{\text{ \tiny
$= \, \mathbb{E}_{t}^{\mathbb{Q}}( \mathrm{call}_t^{\tiny \mbox{BS}}[ {G}_{T_O}^{\perp}, 1 ] )$ }} \nonumber \\
& & ~  - \mathbb{E}_{t}^{\mathbb{P}}( \int_{t+}^{T_O} \mathbbm{1}_{\{G_{\ell  -} < 1\}} dG_{\ell} ) \nonumber \\
&& + \overbrace{\mathbb{E}_{t}^{\mathbb{P}}( \frac{1}{\mathcal{R}_{T_O}^{\tiny \mbox{unspan~diffusive}} \, \mathcal{R}_{T_O}^{\tiny \mbox{unspan~jump}}} \,\int_{t+}^{T_O} \mathbbm{1}_{\{G_{\ell  -} < 1\}} dG_{\ell} )}^{= \mathbb{E}_{t}^{\mathbb{Q}}(  \mathcal{R}_{T_O}^{\tiny \mbox{span~diffusive}} \, \int_{t+}^{T_O} \mathbbm{1}_{\{G_{\ell  -} < 1\}} dG_{\ell} )}
 ~ ~  \mbox{ \, } ~
\nonumber \\
& & +
\underbrace{ ~ \mathbb{E}_{t}^{\mathbb{Q}}( \mathcal{R}_{T_O}^{\tiny \mbox{span~diffusive}} \, \max( 1 - G_{T_O}, 0 ) )}_{\text{ \tiny
$= \mathbb{E}_{t}^{\mathbb{Q}}( \mathcal{R}_{T_O}^{\perp} \{ \mathrm{put}_t^{\tiny \mbox{BS}}[ {G}_{T_O}^{\perp} \mathcal{V}_{T_{O}}^{\bullet} , 1] \} )$ }} -
\underbrace{ ~ \mathbb{E}_{t}^{\mathbb{Q}}( \max( 1 - G_{T_O}, 0 ) ).}_{\text{ \tiny
$=\mathbb{E}_{t}^{\mathbb{Q}}( \mathrm{put}_t^{\tiny \mbox{BS}}[ {G}_{T_O}^{\perp}, 1 ] )$ }}
\label{eq:ReexpressStraddleRiskPremium}
\end{eqnarray}
Simplifying equation
(\ref{eq:ReexpressStraddleRiskPremium}), the sign
of the risk premium on straddles is the same as the sign of
\begin{eqnarray}
& & 2 ~ \mathbb{E}_{t}^{\mathbb{Q}}( \mathcal{R}_{T_O}^{\tiny \mbox{span~diffusive}} \times
\mathrm{cov}_t^{\mathbb{Q}}(
\mathcal{R}_{T_O}^{\tiny \mbox{unspan~diffusive}}
\times \mathcal{R}_{T_O}^{\tiny \mbox{unspan~jump}},
\, \mathbb{L}_t^{{T}_O}[1] {\Big |} \, \mathcal{I}_{T_O} ) )
\nonumber \\
& &
~+~  2 \{\mathbb{E}_{t}^{\mathbb{P}}( \mathbb{A}_t^{T_O}[1] ) - \mathbb{E}_{t}^{\mathbb{Q}}( \mathbb{A}_t^{T_O}[1] )\} \nonumber \\
& & ~ + ~ \mathbb{E}_{t}^{\mathbb{P}}( \int_{t+}^{T_O} \mathbbm{1}_{\{G_{\ell  -} > 1\}} dG_{\ell} ) ~ - ~ \mathbb{E}_{t}^{\mathbb{P}}( \frac{1}{\mathcal{R}_{T_O}^{\tiny \mbox{unspan~diffusive}} \, \mathcal{R}_{T_O}^{\tiny \mbox{unspan~jump}}} \, \, \int_{t+}^{T_O} \mathbbm{1}_{\{G_{\ell  -} > 1\}} dG_{\ell} ) ~ ~  \mbox{ \, \, \, } ~
\nonumber \\
& & ~ - ~ \mathbb{E}_{t}^{\mathbb{P}}( \int_{t+}^{T_O} \mathbbm{1}_{\{G_{\ell  -} < 1\}} dG_{\ell} ) ~ + ~ \mathbb{E}_{t}^{\mathbb{P}}( \frac{1}{\mathcal{R}_{T_O}^{\tiny \mbox{unspan~diffusive}} \, \mathcal{R}_{T_O}^{\tiny \mbox{unspan~jump}}} \, \, \int_{t+}^{T_O} \mathbbm{1}_{\{G_{\ell  -} < 1\}} dG_{\ell} ) ~ ~  \mbox{ \, \, \, } ~
\nonumber \\
&&
~+~\mathbb{E}_{t}^{\mathbb{Q}}( \mathcal{R}_{T_O}^{\perp} \,
\{ \mathrm{straddle}_t^{\tiny \mbox{BS}}[ {G}_{T_O}^{\perp} \mathcal{V}_{T_{O}}^{\bullet} , 1] \} ) ~ - ~
\mathbb{E}_{t}^{\mathbb{Q}}( \mathrm{straddle}_t^{\tiny \mbox{BS}}[ {G}_{T_O}^{\perp}, 1 ] ),
\nonumber
\end{eqnarray}
where
\begin{eqnarray}
\mathrm{straddle}_t^{\tiny \mbox{BS}}[ {G}_{T_O}^{\perp} \mathcal{V}_{T_{O}}^{\bullet} , 1]
&\equiv&
\mathrm{call}_t^{\tiny \mbox{BS}}[ {G}_{T_O}^{\perp} \mathcal{V}_{T_{O}}^{\bullet} , 1] \, + \, \mathrm{put}_t^{\tiny \mbox{BS}}[ {G}_{T_O}^{\perp} \mathcal{V}_{T_{O}}^{\bullet} , 1] ~~~ ~ ~ \mbox{ \, \, \, and \, \, \, } ~ ~
\\
\mathrm{straddle}_t^{\tiny \mbox{BS}}[ {G}_{T_O}^{\perp}, 1 ]
& \equiv &
\mathrm{call}_t^{\tiny \mbox{BS}}[ {G}_{T_O}^{\perp}, 1 ] \, + \,  \mathrm{put}_t^{\tiny \mbox{BS}}[ {G}_{T_O}^{\perp}, 1 ]. ~
\mbox{ \, }
\end{eqnarray}

Now, we assume that
\begin{eqnarray}
& & \mathrm{straddle}_t^{\tiny \mbox{BS}}[ {G}_{T_O}^{\perp} \mathcal{V}_{T_{O}}^{\bullet} , 1] ~ \, \approx ~
\mathrm{straddle}_t^{\tiny \mbox{BS}}[ {G}_{T_O}^{\perp}, 1 ], ~~
~ ~
\mbox{ \, \, \, \, \quad \quad and that \, \, \quad \, } ~ \label{eq:straddleFirstAssumption-a} \\
& & \mathbb{E}_{t}^{\mathbb{Q}}( \mathcal{R}_{T_O}^{\perp} \,
\{ \mathrm{straddle}_t^{\tiny \mbox{BS}}[ {G}_{T_O}^{\perp} , 1] \} )
~ \, \approx ~
\mathbb{E}_{t}^{\mathbb{Q}}( \mathrm{straddle}_t^{\tiny \mbox{BS}}[ {G}_{T_O}^{\perp} , 1] ).
\mbox{ \, \, \, \, \, \, \quad } ~~ \label{eq:straddleSecondAssumption-b}
\end{eqnarray}
The first condition in (\ref{eq:straddleFirstAssumption-a}) is consistent with straddles being approximately delta-neutral. Next,
$\mathcal{R}_{T_O}^{\perp}$ is a term which comes from the \emph{spanned} component of the pricing kernel
and so the correlation between this quantity and the (delta-neutral) straddle is (approximately)
zero, leading to (\ref{eq:straddleSecondAssumption-b}).

It follows that the sign of the risk premium on straddles is
that of
\begin{eqnarray}
& & 2 ~ \mathbb{E}_{t}^{\mathbb{Q}}( \mathcal{R}_{T_O}^{\tiny \mbox{span~diffusive}} \times
\mathrm{cov}_t^{\mathbb{Q}}(
\mathcal{R}_{T_O}^{\tiny \mbox{unspan~diffusive}}
\times \mathcal{R}_{T_O}^{\tiny \mbox{unspan~jump}},
\, \mathbb{L}_t^{{T}_O}[1] {\Big |} \, \mathcal{I}_{T_O} ) )  \nonumber \\
& & ~ ~ + ~ 2 \{\mathbb{E}_{t}^{\mathbb{P}}( \mathbb{A}_t^{T_O}[1] ) - \mathbb{E}_{t}^{\mathbb{Q}}( \mathbb{A}_t^{T_O}[1] )\} ~ \mbox{ \, \, \, } \nonumber \\
& & ~ ~ + ~ \mathbb{E}_{t}^{\mathbb{P}}( \int_{t+}^{T_O} \mathbbm{1}_{\{G_{\ell  -} > 1\}} dG_{\ell} ) ~ - ~ \mathbb{E}_{t}^{\mathbb{P}}( \frac{1}{\mathcal{R}_{T_O}^{\tiny \mbox{unspan~diffusive}} \, \mathcal{R}_{T_O}^{\tiny \mbox{unspan~jump}}} \, \, \int_{t+}^{T_O} \mathbbm{1}_{\{G_{\ell  -} > 1\}} dG_{\ell} ) ~ ~  \mbox{ \, \, \, } ~
\nonumber \\
& & ~ ~ - ~ \mathbb{E}_{t}^{\mathbb{P}}( \int_{t+}^{T_O} \mathbbm{1}_{\{G_{\ell  -} < 1\}} dG_{\ell} ) ~ + ~ \mathbb{E}_{t}^{\mathbb{P}}( \frac{1}{\mathcal{R}_{T_O}^{\tiny \mbox{unspan~diffusive}} \, \mathcal{R}_{T_O}^{\tiny \mbox{unspan~jump}}} \, \, \int_{t+}^{T_O} \mathbbm{1}_{\{G_{\ell  -} < 1\}} dG_{\ell} ). ~ ~  \mbox{ \, \, \, } ~
\label{eq:ReexpressStraddleRiskPremiumSimplifiedAgain}
\end{eqnarray}
In particular,
if there were no unspanned risks in the pricing kernel; that is, if
$\mathcal{R}_{T_O}^{\tiny \mbox{unspan~diffusive}} \equiv 1$ and $\mathcal{R}_{T_O}^{\tiny \mbox{unspan~jump}} \equiv 1$,
equation (\ref{eq:ReexpressStraddleRiskPremiumSimplifiedAgain}) would
evaluate to zero. Our rationale is:
\begin{itemize}

\item The
covariance term would then be identically zero.

\item Additionally, $\mathbb{E}_{t}^{\mathbb{P}}( \mathbb{A}_t^{T_O}[1] ) - \mathbb{E}_{t}^{\mathbb{Q}}( \mathbb{A}_t^{T_O}[1] ) =  0$ (if there were no jumps).
    \item Finally, the third and fourth lines would cancel.

\end{itemize}
Thus, to support our empirical findings
of a negative risk premium on straddles, there must be
unspanned risks in the pricing} kernel. $\blacksquare$ 

\setcounter{table}{0}
\renewcommand{\thetable}{IA-\arabic{table}}

\newpage
\begin{table}[h!]
\footnotesize
\caption{\textbf{{
Risk premiums for holding weekly options over 2-day and 3-day windows}}}
\vspace{2mm}
\label{tab:2and3day}
The sample period is 01/13/2011 to 12/20/2018, with 415 weekly option expiration cycles. The weekly options data on S\&P 500 index is from the CBOE. We construct the excess return of OTM puts, OTM calls, and straddles (ATM and crash-neutral).  These calculations are done at the ask option price. The returns of a crash-neutral
straddle combines a long straddle position and a short 3\% OTM put position.
\begin{description}

\item[$-$] Panel A computes option returns from Wednesday to Friday (2-day to maturity (on average)).
\item[$-$] Panel B computes option returns from Tuesday to Friday (3-day to maturity (on average)).
\end{description}
We indicate statistical significance at 1\%, 5\%, and 10\% by the superscripts ***, **, and *, respectively,
where the $p$-values rely on the \citet*{NeweyWest:87} HAC estimator (with the lag selected automatically).
The reported
put (respectively, call) delta is $-{\cal N}(-d_1)$ (respectively, ${\cal N}(d_1)$), where $d_1=  \frac{1}{ \sigma \sqrt{T_O-t}} \{ - \log k + r (T_O-t) + \frac{1}{2} \sigma^2 (T_O-t)\}$. 
SD is the standard deviation, and $\mathbbm{1}_{\{ q_{t, {T}_O} >0 \}}$ is the proportion (in \%) of
option positions that generate positive returns.
\begin{center}
\setlength{\tabcolsep}{0.08in}
\begin{tabular}{lll ccc c ccc c cc} \hline
       &   & &           &           &           &           &           &           &  \\
&&&\multicolumn{10}{c}{\textbf{Panel A: \emph{2-day} holding period returns}} \\ \\
       &   & &\multicolumn{3}{c}{OTM puts on equity} &           &  \multicolumn{3}{c}{OTM calls on equity}& &\multicolumn{2}{c}{Straddle} \\
       &   & &\multicolumn{3}{c}{ $\log(k)\times 100$} &        & \multicolumn{3}{c}{$\log(k)\times 100$} &    &\multicolumn{2}{c}{on equity} \\
          \cline{4-6} \cline{8-10} \cline{12-13}
Moneyness (\%)   &  &      & -3         & -2         & -1         &          & 1         & 2         & 3 &    &  ATM  &     Crash-Neutral\\
Delta (\%)       &  & & -2         & -5         & -17         &           & 17         & 5         & 2 &    &   &     Neutral\\
 &         &       &    &           &           &           &                      &  &    &    &\\ \hline
           &           &           &           &           &           &           &           &           & &    &    & \\
\multicolumn{1}{l}{\textbf{Unconditional}} &           & \multicolumn{1}{l}{Average} &-82   & -44   & -31   &       & -17   & -43   & -79   &       & -12   & -1 \\
\multicolumn{1}{l}{\textbf{Estimates}} &           & \multicolumn{1}{l}{SD} & 170   & 367   & 296   &       & 281   & 402   & 310   &       & 94    & 5 \\
\multicolumn{1}{l}{ } &           & \multicolumn{1}{l}{$\mathbbm{1}_{\{ q_{t, {T}_O} >0 \}}$} & 2\%     & 5\%     & 10\%    &       & 13\%    & 4\%     & 1\%     &       & 35\%    & 36\% \\
          &           &           &           &           &           &           &           &           & &    &    & \\ \hline \\
&&&\multicolumn{10}{c}{\textbf{Panel B: \emph{3-day} holding period returns}} \\ \\
                &   & &\multicolumn{3}{c}{OTM puts on equity} &           &  \multicolumn{3}{c}{OTM calls on equity}& &\multicolumn{2}{c}{Straddle} \\
       &   & &\multicolumn{3}{c}{ $\log(k)\times 100$} &        & \multicolumn{3}{c}{$\log(k)\times 100$} &    &\multicolumn{2}{c}{on equity} \\
          \cline{4-6} \cline{8-10} \cline{12-13}
Moneyness (\%)   &  &      & -3         & -2         & -1         &          & 1         & 2         & 3 &    &  ATM  &     Crash-Neutral\\
Delta (\%)       &  & & -1         & -3         & -13         &           & 13         & 3         & 1 &    &   &     Neutral\\
 &         &       &    &           &           &           &                      &  &    &    &\\ \hline
           &           &           &           &           &           &           &           &           & &    &    & \\
\multicolumn{1}{l}{\textbf{Unconditional}} &           & \multicolumn{1}{l}{Average} & -72   & -47   & -27   &       & 18    & -28   & -57   &       & -6    & -0 \\
\multicolumn{1}{l}{\textbf{Estimates}} &           & \multicolumn{1}{l}{SD} & 205   & 252   & 227   &       & 546   & 471   & 443   &       & 87    & 6  \\
\multicolumn{1}{l}{ } &           & \multicolumn{1}{l}{$\mathbbm{1}_{\{ q_{t, {T}_O} >0 \}}$}
& 3\%     & 7\%     & 15\%    &
& 19\%    & 7\%     & 3\%     &
& 40\%    & 42\% \\
          &           &           &           &           &           &           &           &           & &    &    & \\ \hline
\end{tabular}%
\end{center}
\end{table}

\newpage
\begin{table}[h!]
\footnotesize
\caption{\textbf{{
Risk premiums for \emph{weekly} OTM calls on the S\&P 500 index, deeper than 3\% OTM}}}
\vspace{2mm}
\label{tab:deep_weekly}
This table complements Table~\ref{tab:weekly} by presenting results on call option excess returns deeper than 3\% OTM. These calculations are done at the ask option price. The sample period is 01/13/2011 to 12/20/2018, with 415 weekly option expiration cycles (8 days to maturity (on average)).
The weekly options data on
the S\&P 500 index is from the CBOE.
The following is the regression specification (analogously for puts and straddles):
\begin{align*}
&q_{t,{\tiny \mathrm{call}} }^{{T}_O}[k] =
\mu_{\{ {\cal F}_{t} \in \mathfrak{s}_{\tiny \mbox{bad}} \} } \mathbbm{1}_{\{ {\cal F}_{t} \in\mathfrak{s}_{\tiny \mbox{bad}} \}}
+ \mu_{\{ {\cal F}_{t} \in \mathfrak{s}_{\tiny \mbox{normal}} \} }
\mathbbm{1}_{\{ {\cal F}_{t} \in\mathfrak{s}_{\tiny \mbox{normal}} \}}
+ \mu_{\{ {\cal F}_{t} \in \mathfrak{s}_{\tiny \mbox{good}} \} }
\mathbbm{1}_{\{ {\cal F}_{t} \in\mathfrak{s}_{\tiny \mbox{good}} \}} ~+~ \underbrace{\epsilon_{T_{O}}.}_{\mathrm{error~term}}&
\end{align*}
We use proxies for the variable $\mathfrak{s}$, known at the beginning of the expiration cycle.
The variable construction for this weekly exercise is described
in the text.
For example, WEI is the weekly economic index. 
\\ 

We indicate statistical significance at 1\%, 5\%, and 10\% by the superscripts ***, **, and *, respectively,
where the $p$-values rely on the \citet*{NeweyWest:87} HAC estimator (with the lag selected automatically).
The reported
put (respectively, call) delta is $-{\cal N}(-d_1)$ (respectively, ${\cal N}(d_1)$), where $d_1=  \frac{1}{ \sigma \sqrt{T_O-t}} \{ - \log k + r (T_O-t) + \frac{1}{2} \sigma^2 (T_O-t)\}$. 
SD is the standard deviation, and $\mathbbm{1}_{\{ q_{t, {T}_O} >0 \}}$ is the proportion (in \%) of
option positions that generate positive returns.
We tabulate the average open interest and trading volume, all observed on the first day of the weekly option expiration cycle.
\begin{center}
\setlength{\tabcolsep}{0.14in}
\begin{tabular}{lll ccc} \hline
       &   & & \multicolumn{3}{c}{OTM calls on equity}\\
       &   & & \multicolumn{3}{c}{$\log(k)\times 100$} \\
          \cline{4-6}
Moneyness (\%)                    &  &      & 4         & 5         & 6 \\
Delta (\%)                        &  &      & 3         & 2         & 1 \\ \\
Open Interest ($\times 1,000$)    &  &      &  5.6   & 5.3   & 3.9    \\
Volume ($\times 1,000)$           &  &      &  0.9   & 0.9   & 0.6    \\ \hline
                   &  &      &          &          &  \\
\multicolumn{1}{l}{Change in WEI} & \multicolumn{1}{l}{L} & \multicolumn{1}{l}{$\mathfrak{s}_{\tiny \mbox{bad}}$} & -73*** & -92*** & -100*** \\
\multicolumn{1}{l}{ } & \multicolumn{1}{l}{M} & \multicolumn{1}{l}{$\mathfrak{s}_{\tiny \mbox{normal}}$} &  -91*** & -96*** & -97***  \\
\multicolumn{1}{l}{ } & \multicolumn{1}{l}{H} & \multicolumn{1}{l}{$\mathfrak{s}_{\tiny \mbox{good}}$} & -90*** & -99*** & -100*** \\ \hline
\multicolumn{1}{l}{Quadratic Variation} & \multicolumn{1}{l}{H} & \multicolumn{1}{l}{$\mathfrak{s}_{\tiny \mbox{bad}}$} & \-65*** & -87*** & -97*** \\
\multicolumn{1}{l}{ } & \multicolumn{1}{l}{M} & \multicolumn{1}{l}{$\mathfrak{s}_{\tiny \mbox{normal}}$} & -88*** & -100*** & -100*** \\
\multicolumn{1}{l}{ } & \multicolumn{1}{l}{L} & \multicolumn{1}{l}{$\mathfrak{s}_{\tiny \mbox{good}}$} & -100*** & -100*** & -100***\\
\hline
\multicolumn{1}{l}{Risk Reversal} & \multicolumn{1}{l}{H} & \multicolumn{1}{l}{$\mathfrak{s}_{\tiny \mbox{bad}}$} & -88*** & -94*** & -100*** \\
\multicolumn{1}{l}{ } & \multicolumn{1}{l}{M} & \multicolumn{1}{l}{$\mathfrak{s}_{\tiny \mbox{normal}}$} &  -88*** & -100*** & -100*** \\
\multicolumn{1}{l}{ } & \multicolumn{1}{l}{L} & \multicolumn{1}{l}{$\mathfrak{s}_{\tiny \mbox{good}}$} & -78*** & -93*** & -97***\\
\hline
\multicolumn{1}{l}{Change in Volatility} & \multicolumn{1}{l}{H} & \multicolumn{1}{l}{$\mathfrak{s}_{\tiny \mbox{bad}}$} & -79*** & -91*** & -100***  \\
\multicolumn{1}{l}{ } & \multicolumn{1}{l}{M} & \multicolumn{1}{l}{$\mathfrak{s}_{\tiny \mbox{normal}}$} & -88*** & -96*** & -97*** \\
\multicolumn{1}{l}{ } & \multicolumn{1}{l}{L} & \multicolumn{1}{l}{$\mathfrak{s}_{\tiny \mbox{good}}$} & -87*** & -100*** & -100***\\
\hline
\multicolumn{1}{l}{Recent Market} & \multicolumn{1}{l}{L} & \multicolumn{1}{l}{$\mathfrak{s}_{\tiny \mbox{bad}}$} & -76*** & -88*** & -97***  \\
\multicolumn{1}{l}{ } & \multicolumn{1}{l}{M} & \multicolumn{1}{l}{$\mathfrak{s}_{\tiny \mbox{normal}}$} & -81*** & -100*** & -100*** \\
\multicolumn{1}{l}{ } & \multicolumn{1}{l}{H} & \multicolumn{1}{l}{$\mathfrak{s}_{\tiny \mbox{good}}$} & -96*** & -98*** & -100***  \\
\hline
\\
\multicolumn{1}{l}{\textbf{Unconditional}} &           & \multicolumn{1}{l}{Average} & -85   & -96   & -99 \\
\multicolumn{1}{l}{\textbf{Estimates}} &           & \multicolumn{1}{l}{SD} & 131   & 52    & 21\\
\multicolumn{1}{l}{ } &           & \multicolumn{1}{l}{$\mathbbm{1}_{\{ q_{t, {T}_O} >0 \}}$} & 2\%     & 1\%     & 0\%\\

\\ \hline

\end{tabular}%
\end{center}
\end{table}

\newpage
\begin{table}[h!]
\footnotesize
\caption{\textbf{Option risk premiums
based on the midpoint of bid and ask prices}}
\vspace{2mm}
\label{tab:bid-ask}
All option return calculations are done at the \emph{midpoint} of bid and ask option prices.
The sample period of this exercise for S\&P 500 index options is as follows:
\begin{description}
\item[$-$] Weekly options: 01/13/2011 to 12/20/2018, with 415 weekly expiration cycles (8 days to maturity (on average)).
\item[$-$] 28-day options: 01/22/1990 to 12/24/2018, with 348 expiration cycles (28 days to maturity (on average)).
\end{description}
We construct the excess return of OTM puts, OTM calls, and straddles (ATM and crash-neutral) over expiration cycles.
Presented are the results from the following regression specification (analogously for puts and straddles):
\begin{align*}
&q_{t,{\tiny \mathrm{call}} }^{{T}_O}[k] =
\mu_{\{ {\cal F}_{t} \in \mathfrak{s}_{\tiny \mbox{bad}} \} } \mathbbm{1}_{\{ {\cal F}_{t} \in\mathfrak{s}_{\tiny \mbox{bad}} \}}
+ \mu_{\{ {\cal F}_{t} \in \mathfrak{s}_{\tiny \mbox{normal}} \} }
\mathbbm{1}_{\{ {\cal F}_{t} \in\mathfrak{s}_{\tiny \mbox{normal}} \}}
+ \mu_{\{ {\cal F}_{t} \in \mathfrak{s}_{\tiny \mbox{good}} \} }
\mathbbm{1}_{\{ {\cal F}_{t} \in\mathfrak{s}_{\tiny \mbox{good}} \}} +
\epsilon_{T_{O}}.&
\end{align*}
We use proxies for the variable $\mathfrak{s}$, known at the beginning of the option expiration cycle.
The variable construction for the weekly and monthly exercise is described
in the text. We indicate statistical significance at 1\%, 5\%, and 10\% by the superscripts ***, **, and *, respectively,
where the $p$-values rely on the \citet*{NeweyWest:87} HAC estimator (with the lag selected automatically).
SD is the standard deviation, and $\mathbbm{1}_{\{ q_{t, {T}_O} >0 \}}$ is the proportion (in \%) of
option positions that generate positive returns.
\begin{center}
\setlength{\tabcolsep}{0.05in}
\begin{tabular}{lll ccc c ccc c cc} \hline
       &   & &           &           &           &           &           &           &  \\
&&&\multicolumn{10}{c}{\textbf{Panel A: Weekly options}} \\ \\
&   & &\multicolumn{3}{c}{OTM puts on equity} &           &  \multicolumn{3}{c}{OTM calls on equity}& &\multicolumn{2}{c}{Straddle} \\
       &   & &\multicolumn{3}{c}{ $\log(k)\times 100$} &        & \multicolumn{3}{c}{$\log(k)\times 100$} &    &\multicolumn{2}{c}{on equity} \\
          \cline{4-6} \cline{8-10} \cline{12-13}
Moneyness (\%)   &  &      & -3         & -2         & -1         &          & 1         & 2         & 3 &    &  ATM  &     Crash-Neutral\\
 &         &       &    &           &           &           &                      &  &    &    &\\ \hline
 &         &       &    &           &           &           &                      &  &    &    &\\
\multicolumn{1}{l}{Risk Reversal} & \multicolumn{1}{l}{H} & \multicolumn{1}{l}{$\mathfrak{s}_{\tiny \mbox{bad}}$} & -68*** & -49*** & -32** &       & 32    & -46** & -92*** &       & -7    & 0  \\
\multicolumn{1}{l}{ } & \multicolumn{1}{l}{M} & \multicolumn{1}{l}{$\mathfrak{s}_{\tiny \mbox{normal}}$} & -93*** & -73*** & -53*** &       & 25    & 13    & -22   &       & -10*  & 0    \\
\multicolumn{1}{l}{ } & \multicolumn{1}{l}{L} & \multicolumn{1}{l}{$\mathfrak{s}_{\tiny \mbox{good}}$} & -14   & -5    & -4    &       & -1    & -13   & -43** &       & -3    & 0 \\
          &           &           &           &           &           &           &           &           &  &    &    &\\
\multicolumn{1}{l}{Change in Volatility} & \multicolumn{1}{l}{H} & \multicolumn{1}{l}{$\mathfrak{s}_{\tiny \mbox{bad}}$} & -39   & -35   & -35** &       & 7     & -1    & -50*** &       & -12*  & -1\\
\multicolumn{1}{l}{ } & \multicolumn{1}{l}{M} & \multicolumn{1}{l}{$\mathfrak{s}_{\tiny \mbox{normal}}$} & -69*** & -48*** & -27   &       & 60    & 10    & -36   &       & -1    & 0 \\
\multicolumn{1}{l}{ } & \multicolumn{1}{l}{L} & \multicolumn{1}{l}{$\mathfrak{s}_{\tiny \mbox{good}}$}  & -68*** & -44*** & -28*  &       & -9    & -54*** & -71*** &       & -6    & 0 \\
          &           &           &           &           &           &           &           &           &  &    &    & \\ \hline
          &           &           &           &           &           &           &           &           &  &    &    &\\
          \multicolumn{1}{l}{\textbf{Unconditional}} &           & \multicolumn{1}{l}{Average} &-59   & -43   & -30   &       & 19    & -15   & -52   &       & -7    & -0 \\
\multicolumn{1}{l}{\textbf{Estimates}} &           & \multicolumn{1}{l}{SD} & 250   & 226   & 189   &       & 294   & 292   & 250   &       & 80    & 7 \\
\multicolumn{1}{l}{ } &           & \multicolumn{1}{l}{$\mathbbm{1}_{\{ q_{t, {T}_O} >0 \}}$} & 7\%     & 10\%    & 17\%    &       & 27\%    & 12\%    & 5\%     &       & 42\%    & 45\%\\
          &           &           &           &           &           &           &           &           & &    &    & \\ \hline \\
&&&\multicolumn{10}{c}{\textbf{Panel B: 28-day options}} \\ \\

                          &   & &\multicolumn{3}{c}{OTM puts on equity} &           &  \multicolumn{3}{c}{OTM calls on equity}& &\multicolumn{2}{c}{Straddle} \\
       &   & &\multicolumn{3}{c}{ $\log(k)\times 100$} &        & \multicolumn{3}{c}{$\log(k)\times 100$} &    &\multicolumn{2}{c}{on equity} \\
          \cline{4-6} \cline{8-10} \cline{12-13}
Moneyness (\%)   &  &      & -5         & -3         & -1         &          & 1         & 3         & 5 &    &  ATM  &     Crash-Neutral\\
 &         &       &    &           &           &           &                      &  &    &    &\\ \hline
 &         &       &    &           &           &           &                      &  &    &    &\\
\multicolumn{1}{l}{Risk Reversal} & \multicolumn{1}{l}{H} & \multicolumn{1}{l}{$\mathfrak{s}_{\tiny \mbox{bad}}$} & -69*** & -47*** & -37*** &       & 25    & 5     & -25   &       & -11   & 0 \\
\multicolumn{1}{l}{ } & \multicolumn{1}{l}{M} & \multicolumn{1}{l}{$\mathfrak{s}_{\tiny \mbox{normal}}$} &  -83*** & -66*** & -58*** &       & 19    & 14    & -8    &       & -16*** & -1 \\
\multicolumn{1}{l}{ } & \multicolumn{1}{l}{L} & \multicolumn{1}{l}{$\mathfrak{s}_{\tiny \mbox{good}}$} & -42** & -32   & -23   &       & -23** & -40*** & -54*** &       & -19** & -3**  \\
          &           &           &           &           &           &           &           &           &  &    &    &\\
\multicolumn{1}{l}{Change in Volatility} & \multicolumn{1}{l}{H} & \multicolumn{1}{l}{$\mathfrak{s}_{\tiny \mbox{bad}}$} & -49*** & -29*  & -23   &       & 16    & 17    & 5     &       & -5    & 1 \\
\multicolumn{1}{l}{ } & \multicolumn{1}{l}{M} & \multicolumn{1}{l}{$\mathfrak{s}_{\tiny \mbox{normal}}$} &-81*** & -59*** & -51*** &       & 6     & -11   & -19   &       & -22*** & -3* \\
\multicolumn{1}{l}{ } & \multicolumn{1}{l}{L} & \multicolumn{1}{l}{$\mathfrak{s}_{\tiny \mbox{good}}$} & -63*** & -55*** & -43*** &       & -1    & -30   & -80*** &       & -19*** & -2** \\
          &           &           &           &           &           &           &           &           &  &    &    & \\ \hline
          &           &           &           &           &           &           &           &           &  &    &    &\\
\multicolumn{1}{l}{\textbf{Unconditional}} &           & \multicolumn{1}{l}{Average} & -65   & -49   & -40   &       & 8     & -4    & -22   &       & -16   & -2 \\
\multicolumn{1}{l}{\textbf{Estimates}} &           & \multicolumn{1}{l}{SD} & 161   & 167   & 152   &       & 157   & 273   & 420   &       & 74    & 14 \\
\multicolumn{1}{l}{ } &           & \multicolumn{1}{l}{$\mathbbm{1}_{\{ q_{t, {T}_O} >0 \}}$} & 6\%     & 11\%    & 16\%    &       & 38\%    & 20\%    & 8\%     &       & 32\%    & 41\%\\
          &           &           &           &           &           &           &           &           & &    &    & \\ \hline
\end{tabular}%
\end{center}
\end{table}

\end{document}